\documentclass[a4paper,11pt]{article}

\usepackage{jheppub} 
\usepackage{subfigure}
\usepackage[utf8x]{inputenc}
\usepackage{amsthm}
\usepackage{array}
\usepackage{multirow}
\usepackage{booktabs}
\usepackage{shuffle}
\usepackage{tikz}
\usepackage{graphicx}

\usetikzlibrary{decorations.pathreplacing}
\usepackage{expl3}
\usetikzlibrary{decorations.pathreplacing,calligraphy}

\usepackage{graphicx}
\usepackage{amssymb}
\usepackage{mathrsfs}
\usepackage{bm}

\usepackage{mathtools}

\usetikzlibrary{calc}
\allowdisplaybreaks[4]
\usepackage[all]{xy}

\makeatletter
\protected\def\xvcenter{%
  \hbox\bgroup$\everyvbox{\everyvbox{}\aftergroup\m@th\aftergroup$\aftergroup\egroup}%
  \vcenter
}
\DeclareRobustCommand{\midscript}[1]{
  \mathchoice{\mid@script\scriptstyle{#1}}
    {\mid@script\scriptstyle{#1}}
    {\mid@script\scriptscriptstyle{#1}}
    {\mid@script\scriptscriptstyle{#1}}
}
\newcommand{\mid@script}[2]{
  \vcenter{\hbox{$\m@th#1#2$}}
}

\makeatletter

\newcommand{\bea}{\begin{eqnarray}}
\newcommand{\eea}{\end{eqnarray}}
\newcommand{\bean}{\begin{eqnarray*}}
\newcommand{\eean}{\end{eqnarray*}}
\newcommand{\nn}{\nonumber \\}

\def\IC{\mathbb{C}}
\def\IP{\mathbb{P}}
\def\O #1{\overline{#1}}

\def\W #1{\widetilde{#1}}

\def\braket#1{\left\langle #1 \right\rangle}

\def\eref#1{(\ref{#1})}

\def\a{{\alpha}}

\def\b{{\beta}}

\def\la{\lambda}
\def\eps{\epsilon}

\def\vev{\braket}

\def\Spaa{\vev}

\def\Label#1{\label{#1}%
  \smash{\hbox to0pt{\raise1ex\hbox{\tiny[#1]}\hss}}}

\title{Note on solutions of scattering equations}
\author{ Bo Feng$^{abcd}$, Chang Hu$^{a}$, Yaobo Zhang$^{a}$
\footnote{Emails:  fengbo@zju.edu.cn, isiahalbert@126.com,
 yaobozhang@zju.edu.cn. } \\~\\
{$^a$\small Zhejiang Institute of Modern Physics, Zhejiang University, Hangzhou, 310027, P. R. China \\
$^b$ Center of Mathematical Science, Zhejiang University, Hangzhou, 310027, P. R. China\\
$^c$ Peng Huanwu Center for Fundamental Theory, Hefei, Anhui 230026, China\\
$^d$ Beijing Computational Science Research Center, Beijing 100084, China}}
\date{\today}
\abstract{
  In the CHY-frame for the amplitudes, there are two kinds of singularities we need to deal with. The first one is the pole singularities when the kinematics is not general, such that some of $S_A\to 0$. The second one is the collapse of locations of points after solving scattering equations (i.e., the singular solutions). These two types of singularities are tightly related to each other, but the exact mapping is not well understood. In this paper, we have initiated the systematic study of the mapping. We have demonstrated the different mapping patterns using three typical situations, i.e., the factorization limit, the soft limit and the forward limit.
 }
\keywords{CHY, scattering equations}

\begin{document}
\maketitle
\flushbottom

\section{Motivation}

In past years, there are huge processes in the effective computation and deep understanding of scattering amplitudes using the so-called "on-shell program"\footnote{See for example, the tree-level on-shell recursion relation \cite{Britto:2004ap,Britto:2005fq} and one-loop unitarity cut method \cite{Bern:1994cg,Bern:1994zx,Britto:2004nc}. }. Among these on-shell methods, the CHY-formalism \cite{Cachazo:2013gna,Cachazo:2013hca,Cachazo:2013iea,Cachazo:2014nsa,Cachazo:2014xea} provides a fantastic angle to study scattering amplitudes. In the CHY-formalism, the tree-level amplitude is given by
\bea
    {\cal A}_n& = & \int {\left(\prod_{i=1}^n dz_i\right)\over {\rm
vol}(SL(2,\IC))} \Omega({\cal E}) {\cal F}(z)
    =  \int {\left(\prod_{i=1}^n dz_i\right)\over d\omega}  \Omega({\cal E}) {\cal F} \ ,
    ~~~~\label{gen-A}
\eea
where the integration is done over $z_i$'s, which describe the locations of
$n$-external particles  living on $\IC\IP^1$. Although there are $n$ variables, because the M\"obius $SL(2,\IC)$ symmetry, three variables should be fixed, thus we need to divide the $d\omega={d z_r d z_s d z_t\over z_{rs} z_{st} z_{tr}}$ with $z_{ij}=z_i-z_j$ by the gauge fixing. The $\Omega({\cal E})$ is given by
\bea \Omega({\cal E})\equiv z_{ij}z_{jk}z_{ki}
\prod_{a\neq i,j,k}\delta\left( {\cal E}_a\right) \ ,
~~~\label{measure-Omega} \eea
where ${\cal E}_a$'s are the scattering equations defined as
\bea
    {\cal E}_a\equiv \sum_{b\neq a} {S_{ab}\over z_a-z_b}=0,~~~~~a=1,2,...,n \ ,
    ~~~~\label{SE-def}
\eea
with $S_{ab}=(k_a+k_b)^2= 2k_a\cdot k_b$, and $k_a$, $a=1,2,...,n$ as $n$ massless momenta for $n$-external particles. The ${\cal F}$ is called the CHY-integrand, which defines the particular theory we are considering. Since there are $(n-3)$ delta-functions for $(n-3)$ variables, there is, in fact, no integration to be done in \eref{gen-A} and amplitudes are calculated as
\bea
    {\cal A}_n=\sum_{sol} {z_{ij} z_{jk}z_{ki} z_{rs} z_{st} z_{tr}\over (-)^{i+j+k+r+s+t} |\Phi|_{ijk}^{rst}}{\cal F} \ ,~~~
    ~~~~\label{gen-A-2-1}
    \eea
where three arbitrary indices $i,j,k$ correspond to three removed scattering equations while three arbitrary indices $r,s,t$ correspond to three fixed locations mentioned above. The sum is over the solution set of the scattering equations, which is generically a discrete set of points. Furthermore, in the above, the Jacobi matrix $\Phi$ is calculated as ($a$ for rows and $b$ for column) 
\bea \Phi_{ab}= {\partial {\cal E}_a\over \partial z_b}=
\left\{
\begin{array}{ll}
  {S_{ab}\over z_{ab}^2} & a\neq b\\
  -\sum\limits_{c\neq a}{S_{ac}\over z_{ac}^2}~~~~&  a=b\end{array}\right. \ ,
 ~~~~ \label{Phi-Jacobi}
\eea
and $|\Phi|_{ijk}^{rst}$ is the determinant of $\Phi$ after removing the $i$-th, $j$-th and $k$-th rows and $r$-th, $s$-th and $t$-th columns.

From the above brief introduction, one can see that unlike the familiar off-shell Lagrangian formalism where the Feynman diagrams give a very intuitive picture to see the interaction by vertexes and propagators, all physical information in the CHY-formalism is coded by universal scattering equations and special rational  CHY-integrand of a given theory. Thus it is crucial to understand how these two total different descriptions are mapped to each other. For example, the general tree-level amplitudes have the factorization property, easily seen from the Feynman diagrams when putting a particular propagator on-shell. How do we see this in the CHY-formalism? Since it is a general property, it can not depend on ${\cal F}$, and thus from \eref{gen-A-2-1}, the information must be coded in the solutions of scattering equations.

There are many works on studying  scattering equations, and their solutions \cite{Kalousios:2013eca,Weinzierl:2014vwa,Dolan:2014ega,Lam:2014tga,Casali:2014hfa,Kalousios:2015fya,Cachazo:2015nwa,Huang:2015yka,Sogaard:2015dba, Cardona:2015eba,Cardona:2015ouc,Dolan:2015iln,Lam:2015sqb,Bosma:2016ttj,Zlotnikov:2016wtk,Cachazo:2016ror,Chen:2016fgi,Chen:2017edo,Liu:2018brz,DeLaurentis:2019vkf}. 
By these studies, now we have an obvious picture about the factorization property in the CHY-formalism. As shown in \cite{Dolan:2014ega}, if all $S_A\neq 0$, the values of the $z_a$'s are distinct (i.e, $z_i\neq z_j, \forall i,j$) in every solution for all $(n-3)!$ solutions of scattering equations. However, suppose there is one and only one $S_A=0$, which corresponds to a particular factorization channel. In that case, we will find that there are $(n_A-2)!\times (n-n_A-2)!$ solutions \cite{Cachazo:2013iea}, where $z_i=z_0,\forall i\in A$\footnote{In fact, because the gauge fixing of three $z_i$'s, between the two subsets of $A,\O A$, it is the one containing at most one gauge fixing point having $z_i=z_0$. }.  For later convenience, we will call a solution "regular" if all $z_i$'s are distinct and a solution to be "singular" if some $z_i$ collapse to the same value.
Using this terminology, one can say the solution singularity reflects that pole singularities. The understanding of factorization in the CHY-formalism is crucial because, through its study, the "integration rule" for simple poles and its generalization for higher poles have been proposed \cite{Baadsgaard:2015voa,Baadsgaard:2015ifa,Cardona:2016gon} and we can read out the analytic expression for any CHY-integrand without solving the scattering equations. One crucial concept coming out from the "integration rule" is the {\bf pole index}
\bea \chi(A):=\mathbb{L}[A]-2(|A|-1)~~~~\label{pole-degree}\eea
where $\mathbb{L}[A]$ be the number\footnote{Again more accurately it is the difference of number between solid lines and dashed lines, which represent the factor $z_{ij}=z_i-z_j$ in the denominator or the numerator respectively. } of lines connecting these nodes
inside $A$ and $|A|$ is the number of nodes. The condition
\bea \chi(A)\geq 0~~~~\label{Pole-cond}\eea
will be called the {\bf pole condition} for a given subset. More
explicitly, each subset gives a possible nonzero pole contribution when and only when $\chi(A)\geq 0$, and  the pole will be ${1\over
S_A^{\chi(A)+1}}$, where $S_{A}=(p_{a_1}+p_{a_2}+\cdots+p_{a_m})^2$
for massless momentum $p_i$\footnote{ Because the momentum conservation, we will treat the subset $A$ to be equivalent to its complement $\O A=\{1,2,\ldots, n\}-A$ when taking the pole. However, as we will show, the behaviour of solutions will be very different between $A, \O A$.}.

Factorization singularity is just one type of singularity of tree-level amplitudes. There are other types of singularities, for example, soft singularity and forward singularity. 
In fact,forward singularity appearances naturally in the construction of loop-level  CHY-formalism and dealing with it is important for the whole construction
\cite{Adamo:2013tsa,Casali:2014hfa,Adamo:2015hoa,Geyer:2015bja,
Geyer:2015jch,Baadsgaard:2015hia,He:2015yua,Cachazo:2015aol,
Zlotnikov:2016wtk,Cardona:2016wcr,He:2016mzd,Gomez:2016cqb,
Gomez:2017lhy,Gomez:2017cpe,Geyer:2017ela,Ahmadiniaz:2018nvr,Agerskov:2019ryp,Feng:2016nrf,
Geyer:2016wjx,Geyer:2018xwu,Geyer:2019hnn,Farrow:2020voh}.
Different from factorization singularity, where we can arrange kinematics so there is one and only one $S_A=0$, both soft and forward limits will have multiple poles going to zero simultaneously. For example, if $k_n\to 0$, we will have all $S_{ni}\to 0, i=1,...,n-1$. Similarly, for forward limit $k_+=-k_-$, we will have all $(k_+ +k_-+k_i)^2\to 0$. It is natural to ask what is the solution behaviour under these limits. For forward limit, beautiful analysis has been done in \cite{He:2015yua}. Motivated by these observations, in this note, we will start a systematical discussion about the relation of the following two things: {\bf the kinematic singularities} and {\bf the singular solutions of scattering equations}. More explicitly, at one side, starting from the singular kinematic configuration with one or more singular poles, we will ask:
\begin{itemize}

\item (A1) If there will be singular solutions of scattering equations?

\item (A2) If there are, how many singular solutions?

\item (A3) What is the exact behaviour of these singular solutions? Are all of them have similar
behaviour or different patterns?

\item (A4) What are the contributions of these singular solutions when putting them into CHY-integrands?

\item (A5) Do regular solutions contribute to a singular part of scattering amplitudes?

\end{itemize}
On another side, starting from a given singular solution, we will ask:
\begin{itemize}

\item (B1) Is there a singular kinematic configuration implied by the solution?

\item (B2) If there are, how many different singular kinematic configurations?
Are these poles compatible or not compatible?

\end{itemize}
Although we can not answer the above questions in this note, we will illustrate cases where two side problems can be theoretically analyzed. Besides theoretical analysis, we also provide numerical checking to support our arguments.

The plan of the note is following. In the second section, we have discussed from singular solutions what is the kinematic configuration we can infer. From the second three to section five, we use factorization limit, soft limit and forward limit to demonstrate how to get pieces of information of singular solutions from the singular kinematics. Finally, a summary is given in section six. In this paper, to better understand the theoretical analysis, we have offered several numerical computations. In the appendix, how to implement the numerical kinematic configuration has been carefully discussed.

\section{From singular solutions to kinematics}

As discussed in the previous section, we will discuss relations between singular solutions and kinematics in this note. This section will assume the behaviour of singular solutions and derive the corresponding consequence of the kinematics, especially the pole structure. Let us consider an asymptotic behaviour of the kinematics $k_i$'s such that when $\eps\to 0$, the kinematics will reach the one we want:
\begin{equation}
    k_i=\sum_{a=0} \eps^a k_i^{(a)}
    ~~~~~\label{kin-limit}
\end{equation}
With this asymptotic behaviour, solutions $z_i$ have also  asymptotic behaviour like:
\begin{equation}
    z_i=\sum_{a=0} \eps^a z_i^{(a)}
    ~~~~~\label{sol-limit}
\end{equation}
Singular solutions will correspond to $z_i^{(0)} \to z_S$ for some $i\in S$ where the subset $S$
has   at least two elements.

A first important observation made in \cite{Dolan:2014ega} is that {\sl if there is at least one singular solution, then at least one kinematic pole goes to zero}. The argument in \cite{Dolan:2014ega} is following.
Assuming there is at least one singular solution in \eref{sol-limit}
satisfying  following condition:
\bea {\rm Cond~~A}:~~~ z_i^{(0)}= z_S,~~~~a\in S;~~~~z_{ab}^{(0)}\neq 0,~~z_{b\W b}^{(0)}\neq 0,~~~\forall a\in S,~ b,\W b\not\in S~~~\label{Cond-A} \eea
Putting this back, the scattering equations \eref{SE-def} will be factorized into two parts.
For $a\in S$, we will have
\bea {\cal E}_a & = & \sum_{t\neq a} {k_a\cdot k_t\over z_a-z_t}
= \sum_{t\neq a, t\in S} {k_a\cdot k_t\over z_a-z_t}+\sum_{t\not\in S} {k_a\cdot k_t\over z_a-z_t} ={1\over \eps} \left\{ \sum_{t\neq a, t\in S} {{\cal K}_{at}^{(00)}\over z_{at}^{(1)}}\right\}
\nn & & +\left\{\sum_{t\neq a, t\in S} {{\cal K}_{at}^{(01)}+{\cal K}_{at}^{(10)}\over z_{at}^{(1)}}+ \sum_{t\neq a, t\in S} {-(z_{at}^{(2)}){\cal K}_{at}^{(00)}\over (z_{at}^{(1)})^2}+\sum_{t\not\in S} {{\cal K}_{at}^{(00)}\over z_{at}^{(0)}}\right\}+{\cal O}(\eps)~~~~ \label{dolan-1-1}\eea
where for simplicity we have defined $z_{ij}^{(a)}\equiv z_{i}^{(a)}-z_{j}^{(a)}$ and
${\cal K}_{ij}^{(ab)}\equiv k_i^{(a)}\cdot k_j^{(b)}$, while for $a\not \in S$, we have
\bea {\cal E}_a & = & \sum_{t\neq a,t\not\in S} {k_a\cdot k_t\over z_a-z_t}
 +\sum_{t\in S} {k_a\cdot k_t\over z_a-z_t}=\sum_{t\neq a,t\not\in S} {{\cal K}_{at}^{(00)}\over z_{at}^{(0)}}+ \sum_{t\in S} {{\cal K}_{at}^{(00)}\over z_a^{(0)}-z_S} +{\cal O}(\eps)~~~~ \label{dolan-1-2}\eea
An important point is that scattering equation should hold for each order of $\eps$, thus
from the leading order of \eref{dolan-1-1}, we get the relation
\bea {\cal E}_{a;\eps^{-1}}\equiv \sum_{t\neq a, t\in S} {{\cal K}_{at}^{(00)}\over z_{at}^{(1)}}=0~~~~\label{dolan-1-3}\eea
Using this result, we can see
\bea
      (K^{(0)}_S)^{2}&\equiv& (\sum_{a\in S} k_a^{(0)})^2=  \sum_{\substack{a,b\in S\\a<b}}2 k_a^{(0)}\cdot k_b^{(0)}
      =\sum_{\substack{a,b\in S\\a<b}}\frac{z_a^{(1)}-z_b^{(1)}}{z_a^{(1)}-z_b^{(1)}}2 k_a^{(0)}\cdot k_b^{(0)}\nn &= &\sum_{\substack{a,b\in S\\a\neq b}}\frac{z_a^{(1)}}{z_a^{(1)}-z_b^{(1)}}2 k_a^{(0)}\cdot k_b^{(0)}=2\sum_{\substack{a\in S}}z_a^{(1)}\sum_{\substack{b\in S\\ b\neq a}}\frac{k_a^{(0)}\cdot k_b^{(0)}}{z_a^{(1)}-z_b^{(1)}}=2\sum_{\substack{a\in S}}z^{(1)}_a {\cal E}_{a;\eps^{-1}}=0
   ~~~~~ \label{dolan-1-4}
\eea
In other words, assuming the singular solution with behaviour \eref{Cond-A} we have
derived the kinematic information $(K^{(0)}_S)^{2}=0$.

Above derivation is beautiful, but if one check it carefully, one can find a hidden assumption having been made, i.e., all $z^{(1)}_a-z^{(1)}_b\neq 0$ for $a,b\in S$. In other words, points in the subset $S$ going to the same location with the same order of speed, i.e., there is no subset such that points of this subset collapse to each other faster. If there is a subset violating the assumption, the leading term in \eref{dolan-1-1} will be at least the order of ${1\over \eps^2}$ and the conclusion will be modified. With this in mind,  we consider various generalizations of the above derivation with different singular solution behaviours.

\subsection{The first case}

Let us start with the simplest generalization. Again, let us assume there is at least one singular solution satisfying the condition A given in \eref{Cond-A}. However, for these points going together, a subset of points collapses to each other faster, i.e., they satisfy one more condition.
\bea {\rm Cond~~B}: & ~~~ &  z_i^{(1)}= z_{S_1}(i.e,z_{a\W a}^{(1)}=0),~~~~z_{a\W a}^{(2)}\neq 0~~~\forall~~ a,\W a\in S_1\subset S;\nn
&~~~ & z_{ab}^{(1)}\neq 0,~~z_{b\W b}^{(1)}\neq 0,~~~\forall~~a\in S_1,~b,\W b\in S\setminus S_1~~~\label{Cond-B} \eea
With these two conditions, the scattering equation \eref{dolan-1-2} is not modified, but equation \eref{dolan-1-1} should be refined as
\bea {\cal E}_a & = &  \sum_{t\neq a, t\in S_1} {k_a\cdot k_t\over z_a-z_t}+\sum_{t\in S\setminus S_1} {k_a\cdot k_t\over z_a-z_t}+\sum_{t\not\in S} {k_a\cdot k_t\over z_a-z_t}={1\over \eps^2} \left\{ \sum_{t\neq a, t\in S_1} {{\cal K}_{at}^{(00)}\over z_{at}^{(2)}}\right\} \nn
&  &+ {1\over \eps} \left\{\sum_{t\neq a, t\in S_1} {{\cal K}_{at}^{(01)}+{\cal K}_{at}^{(10)}\over z_{at}^{(2)}}+ \sum_{t\neq a, t\in S_1} {-(z_{at}^{(3)}){\cal K}_{at}^{(00)}\over (z_{at}^{(2)})^2}+\sum_{t\in S\setminus S_1} {{\cal K}_{at}^{(00)}\over z_{at}^{(1)}}\right\}+{\cal O}(\eps^0)~~~~ \label{dolan-gen1-1}\eea
when $a\in S_1$ and
\bea {\cal E}_a & = &  \sum_{t\neq a, t\in S\setminus S_1} {k_a\cdot k_t\over z_a-z_t}+\sum_{t\in S_1} {k_a\cdot k_t\over z_a-z_t}+\sum_{t\not\in S} {k_a\cdot k_t\over z_a-z_t} \nn
& = &{1\over \eps} \left\{ \sum_{t\neq a, t\in S} {{\cal K}_{at}^{(00)}\over z_{at}^{(1)}}\right\}
+\left\{\sum_{t\neq a, t\in S} {{\cal K}_{at}^{(01)}+{\cal K}_{at}^{(10)}\over z_{at}^{(1)}}+ \sum_{t\neq a, t\in S} {-(z_{at}^{(2)}){\cal K}_{at}^{(00)}\over (z_{at}^{(1)})^2}+\sum_{t\not\in S} {{\cal K}_{at}^{(00)}\over z_{at}^{(0)}}\right\}+{\cal O}(\eps)~~~~ \label{dolan-gen1-2}\eea
when $a\in S\setminus S_1$. From the refined scattering equations \eref{dolan-gen1-1} and
\eref{dolan-gen1-2}, we obtain following identities
\bea {\cal E}_{a\in S_1; \eps^{-2}} & = & \sum_{t\neq a, t\in S_1} {{\cal K}_{at}^{(00)}\over z_{at}^{(2)}}=0~~~~ \label{dolan-gen1-3-1}\\
{\cal E}_{a\in S_1; \eps^{-1}} & = & \sum_{t\neq a, t\in S_1} {{\cal K}_{at}^{(01)}+{\cal K}_{at}^{(10)}\over z_{at}^{(2)}}+ \sum_{t\neq a, t\in S_1} {-(z_{at}^{(3)}){\cal K}_{at}^{(00)}\over (z_{at}^{(2)})^2}+\sum_{t\in S\setminus S_1} {{\cal K}_{at}^{(00)}\over z_{at}^{(1)}}~~~~ \label{dolan-gen1-3-2}\\
{\cal E}_{a\in S\setminus S_1; \eps^{-1}} & = & \sum_{t\neq a, t\in S} {{\cal K}_{at}^{(00)}\over z_{at}^{(1)}}=\sum_{t\in S_1} {{\cal K}_{at}^{(00)}\over z_{at}^{(1)}}+\sum_{t\neq a, t\in S\setminus S_1} {{\cal K}_{at}^{(00)}\over z_{at}^{(1)}}~~~~ \label{dolan-gen1-3-3}\\
{\cal E}_{a\in S\setminus S_1; \eps^{0}} & = & \sum_{t\neq a, t\in S} {{\cal K}_{at}^{(01)}+{\cal K}_{at}^{(10)}\over z_{at}^{(1)}}+ \sum_{t\neq a, t\in S} {-(z_{at}^{(2)}){\cal K}_{at}^{(00)}\over (z_{at}^{(1)})^2}+\sum_{t\not\in S} {{\cal K}_{at}^{(00)}\over z_{at}^{(0)}}~~~~ \label{dolan-gen1-3-4}\eea
Now mimic previous argument, first suing \eref{dolan-gen1-3-1} one can show
\bea
      (k^{(0)}_{S_1})^{2}&\equiv& (\sum_{a\in S_1} k_a^{(0)})^2=  \sum_{\substack{a,b\in S_1\\a<b}}2 k_a^{(0)}\cdot k_b^{(0)}
      =\sum_{\substack{a,b\in S_1\\a<b}}\frac{z_a^{(2)}-z_b^{(2)}}{z_a^{(2)}-z_b^{(2)}}2 k_a^{(0)}\cdot k_b^{(0)}\nn & = & \sum_{\substack{a,b\in S_1\\a\neq b}}\frac{z_a^{(2)}}{z_a^{(2)}-z_b^{(2)}}2 k_a^{(0)}\cdot k_b^{(0)}=2\sum_{\substack{a\in S_1}}z_a^{(2)} {\cal E}_{a\in S_1;\eps^{-2}}=0
   ~~~~ \label{dolan-gen1-4-1}
\eea
Secondly, let us consider following combination
\bea  & & 2K_{S_1}\cdot K_{S\setminus S_1}+ K^2_{S\setminus S_1}=\sum_{a\in S_1, b\in S\setminus S_1} 2k_a^{(0)} \cdot k_{b}^{(0)}+ \sum_{b,\W b\in S\setminus S_1,b<\W b} 2k_b^{(0)} \cdot k_{\W b}^{(0)}\nn
& = & \sum_{a\in S_1, b\in S\setminus S_1} {z_{ab}^{(1)}\over z_{ab}^{(1)}}2k_a^{(0)} \cdot k_{b}^{(0)}+ \sum_{b,\W b\in S\setminus S_1,b<\W b} {(z_{b}^{(1)}-z_{S_1})-(z_{\W b}^{(1)}-z_{S_1})\over z_{b}^{(1)}-z_{\W b}^{(1)}}2k_b^{(0)} \cdot k_{\W b}^{(0)}\nn
& = & \sum_{a\in S_1, b\in S\setminus S_1} {(z_{b}^{(1)}-z_{S_1})\over (z_{b}^{(1)}-z_{S_1})}2k_a^{(0)} \cdot k_{b}^{(0)}+ \sum_{b,\W b\in S\setminus S_1,b\neq \W b} {(z_{b}^{(1)}-z_{S_1})\over z_{b}^{(1)}-z_{\W b}^{(1)}}2k_b^{(0)} \cdot k_{\W b}^{(0)}\nn
& = & \sum_{b\in S\setminus S_1} 2(z_{b}^{(1)}-z_{S_1})\left( \sum_{a\in S_1} {1\over (z_{b}^{(1)}-z_{S_1})}k_a^{(0)} \cdot k_{b}^{(0)}+\sum_{\W b\in S\setminus S_1,b\neq \W b} {1\over z_{b}^{(1)}-z_{\W b}^{(1)}}k_b^{(0)} \cdot k_{\W b}^{(0)}\right)\nn
& = &\sum_{b\in S\setminus S_1} 2(z_{b}^{(1)}-z_{S_1}){\cal E}_{b\in S\setminus S_1; \eps^{-1}}  =0~~~ ~~~~ \label{dolan-gen1-4-2}
\eea
Combining \eref{dolan-gen1-4-1} and \eref{dolan-gen1-4-2} we have
\bea (k^{(0)}_{S})^{2} & = & \sum_{\substack{a,b\in S_1\\a<b}}2 k_a^{(0)}\cdot k_b^{(0)}+
\sum_{a\in S_1, b\in S\setminus S_1} 2k_a^{(0)} \cdot k_{b}^{(0)}+ \sum_{b,\W b\in S\setminus S_1,b<\W b} 2k_b^{(0)} \cdot k_{\W b}^{(0)}=0~~~ \label{dolan-gen1-5}\eea
In other words, under the limit $\eps\to 0$, two poles go to zero at the same time.

\begin{figure}
\begin{center}
\begin{tikzpicture}

\draw (-0.5,0)--(0.5,0);
\draw [fill] (0,1) circle [radius=0.05];
\node [] at (0,0.6) {$ \vdots $};
\draw [fill] (0,0) circle [radius=0.05];
\draw (-0.5,1)--(0.5,1);
\draw [decorate,decoration={calligraphic brace,mirror,amplitude=2mm},thick] (0.8,0) -- (0.8,1);
\node [right] at (1.0,0.5) {$\O S$};

\draw [fill] (0,2) circle [radius=0.05];
\node [left] at (0,2) {$z_S$};
\draw (0,2)--(1.5,2);
\node [] at (1.5,1.8) {$ \vdots $};
\draw (0,2)--(1.5,1.4);
\draw [decorate,decoration={calligraphic brace,mirror,amplitude=1mm},thick] (1.7,1.4) -- (1.7,2);
\node [left] at (3.1,1.7) {$ S\setminus S_1$};

\draw (0,2)--(1.5,3);
\node [above left] at (1.5,3) {$ z_{S_1}$};
\draw [fill] (1.5,3) circle [radius=0.05];
\draw (1.5,3)--(3.0,3);
\draw (1.5,3)--(3.0,2.4);
\node [] at (3,2.8) {$ \vdots $};
\draw [decorate,decoration={calligraphic brace,mirror,amplitude=1mm},thick] (3.2,2.4) -- (3.2,3);
\node [left] at (4.0,2.7) {$ S_1$};

\node [above] at (0,4) {$ \eps^0$};
\node [above] at (1.5,4) {$ \eps^1$};
\node [above] at (3,4) {$ \eps^2$};

\end{tikzpicture}
\caption{
  The pictorial representation of conditions A and B exhibited in \eref{Cond-A} and \eref{Cond-B}. Since $z_{i\in \O S}^{(0)}$'s are different values, we represent them by a node under the column of $\eps^0$. For the subset $S$, they have the same $z_{i\in  S}^{(0)}=z_S$,  which is represented by a node with several lines under the column of $\eps^0$. Similarly, for the subset $S_1$, they have the same $z_{i\in  S_1}^{(1)}=z_{S_1}$, which is represented by a node with several lines under the column of $\eps^1$.
  The ending of each line has meaning too. Different positions in a given column mean the different $z_i$'s at the given order of $\eps$. For the subset $\O S$, the ending under the column of $\eps^0$ means that all $z_{i\in\O S}^{(0)}$ and $z_{S}$ are different, so the leading scattering equation is at the order of $\eps^0$ as given in \eref{dolan-1-2}. Similarly, for the subset $ S\setminus S_1$, the ending under the column of $\eps^1$ means the leading scattering equation is at the order of $\eps^1$ as given in \eref{dolan-gen1-3-3}, while for the subset $  S_1$, the ending under the column of $\eps^2$ means the leading scattering equation is at the order of $\eps^2$ as given in \eref{dolan-gen1-3-1}.
} 
\label{Type-A-1}
\end{center}
\end{figure}
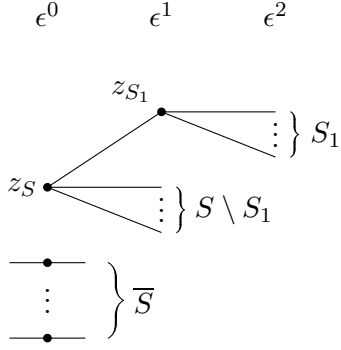

\begin{figure}
\begin{center}
\begin{tikzpicture}

\node [above] at (0,-1) {$ \eps^0$};
\node [above] at (1,-1) {$ \eps^1$};
\node [above] at (2,-1) {$ \eps^2$};
\node [above] at (3,-1) {$ \dots $};
\node [above] at (4,-1) {$ \eps^{k-2}$};
\node [above] at (5,-1) {$ \eps^{k-1}$};
\node [above] at (6,-1) {$ \eps^{k}$};

\node [] at (0,-7.1) {$ \vdots $};
\node [right] at (0.2,-7.2) {$\O S$};
\draw [fill] (0,-6.4) circle [radius=0.05];
\draw (0,-6.4)--(1,-6.4);
\draw (0,-6.4)--(1,-6.6);
\draw (0,-6.4)--(1,-6.8);
\draw (0,-6.4)--(1,-5.4);
\draw [fill] (1,-5.4) circle [radius=0.05];
\node [right] at (1,-6.6) {$S_1$};
\node [above left] at (0,-6.4) {$z_{S_1}$};

\draw (1,-5.4)--(2,-5.4);
\draw (1,-5.4)--(2,-5.6);
\draw (1,-5.4)--(2,-5.8);
\draw (1,-5.4)--(2,-4.4);
\draw [fill] (2,-4.4) circle [radius=0.05];
\node [right] at (2,-5.6) {$S_2$};
\node [above left] at (1,-5.4) {$z_{S_2}$};

\draw (2,-4.4)--(3,-4.4);
\draw (2,-4.4)--(3,-4.6);
\draw (2,-4.4)--(3,-4.8);
\draw [dashed, ultra thick] (2,-4.4)--(4,-3);
\draw [fill] (4,-3) circle [radius=0.05];
\node [right] at (3,-4.6) {$S_3$};
\node [above left] at (2,-4.4) {$z_{S_3}$};

\draw (4,-3)--(5,-3);
\draw (4,-3)--(5,-3.2);
\draw (4,-3)--(5,-3.4);
\draw (4,-3)--(5,-2);
\draw [fill] (5,-2) circle [radius=0.05];
\node [right] at (5,-3.2) {$S_{k-1}$};
\node [above left] at (4,-3) {$z_{S_{k-1}}$};

\draw (5,-2)--(6,-1.8);
\draw (5,-2)--(6,-2.2);
\draw (5,-2)--(6,-2.6);
\node [right] at (6,-2.2) {$S_{k}$};
\node [above left] at (5,-2) {$z_{S_k}$};
\end{tikzpicture}
\caption{The pictorial representation of the Type A generalization. For this type, where each node has several branches, but at most one branch
has substructure. We have also used the $z$ to denote the common position at each order of $\eps$.    } \label{Type-A}
\end{center}
\end{figure}
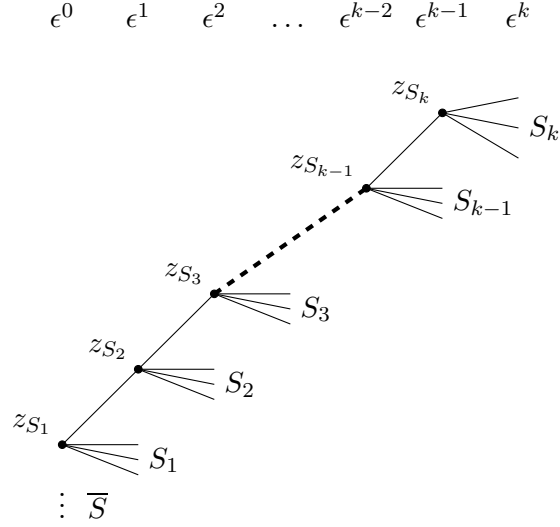

The above case can be generalized to more configurations. To see that, let us first rephrase the condition A and B in \eref{Cond-A} and \eref{Cond-B} in Figure \ref{Type-A-1}. With the pictorial representation, we consider the more general one given in Figure \ref{Type-A}. We will call it the "Type A" structure, where each node has several branches, but at most, one branch has a substructure. From Figure \ref{Type-A}, we can read out the following kinematic information. First, using the leading order part (at the order $\eps^{-k}$) of scattering equations of the subset $S_k$.
\bea {\cal E}_{a\in S_k; \eps^{-k}} =\sum_{t\neq a, t\in S_k} { k_{a}^{(0)}\cdot k_{t}^{(0)}\over z_{a}^{(k)}-z_{t}^{(k)}}=0~~~~\label{A-1-1}\eea
we can derive
\bea (k^{(0)}_{S_k})^{2}&\equiv&   \sum_{\substack{a,b\in S_k\\a<b}}2 k_a^{(0)}\cdot k_b^{(0)}
       =\sum_{\substack{a,b\in S_k\\a\neq b}}\frac{z_a^{(k)}}{z_a^{(k)}-z_b^{(k)}}2 k_a^{(0)}\cdot k_b^{(0)}=2\sum_{\substack{a\in S_k}}z_a^{(k)} {\cal E}_{a\in S_1;\eps^{-k}}=0~~~~~~\label{A-1-2}\eea
Next we consider the leading order part (at the order $\eps^{-(k-1)}$) of  scattering equations  of the subset $S_{k-1}$
\bea {\cal E}_{a\in S_{k-1}; \eps^{-(k-1)}} &= &\sum_{t\in S_k}{ k_{a}^{(0)}\cdot k_{t}^{(0)}\over z_{a}^{(k-1)}-z_{S_k}}+  \sum_{t\neq a, t\in S_{k-1}} { k_{a}^{(0)}\cdot k_{t}^{(0)}\over z_{a}^{(k-1)}-z_{t}^{(k-1)}}\nn
& = &{ k_{a}^{(0)}\cdot K_{S_k}^{(0)}\over z_{a}^{(k-1)}-z_{S_k}}+  \sum_{t\neq a, t\in S_{k-1}} { k_{a}^{(0)}\cdot k_{t}^{(0)}\over z_{a}^{(k-1)}-z_{t}^{(k-1)}}=0 ~~~~\label{A-2-1}\eea
Using it, we can show
\bea & & 2 K_{S_k}\cdot K_{S_{k-1}}+ K^2_{S_{k-1}}  \nn
& = & 2 \sum_{a\in S_{k-1}} { z_{a}^{(k-1)}-z_{S_k}\over z_{a}^{(k-1)}-z_{S_k}}
k_{a}^{(0)}\cdot K_{S_k}^{(0)}+2\sum_{t\neq a, t\in S_{k-1}} { (z_{a}^{(k-1)}-z_{S_k})k_{a}^{(0)}\cdot k_{t}^{(0)}\over z_{a}^{(k-1)}-z_{t}^{(k-1)}}\nn
& = & 2 \sum_{a\in S_{k-1}} (z_{a}^{(k-1)}-z_{S_k}){\cal E}_{a\in S_{k-1}; \eps^{-(k-1)}}=0 ~~~~\label{A-2-2}\eea
When combining \eref{A-1-2} and  \eref{A-2-2}, we reach $K^2_{S_k\bigcup S_{k-1}}=0$. Similarly
from the leading order part (at the order $\eps^{-(k-2)}$) of  scattering equations  of the subset $S_{k-2}$
\bea {\cal E}_{a\in S_{k-2}; \eps^{-(k-2)}}
& = &{ k_{a}^{(0)}\cdot K_{S_k\bigcup S_{k-1}}^{(0)}\over z_{a}^{(k-2)}-z_{S_{k-1}}}+  \sum_{t\neq a, t\in S_{k-2}} { k_{a}^{(0)}\cdot k_{t}^{(0)}\over z_{a}^{(k-2)}-z_{t}^{(k-2)}}=0, ~~~~\label{A-3-1}\eea
we can derive $2 K_{S_k\bigcup S_{k-1}}\cdot K_{S_{k-2}}+ K^2_{S_{k-2}}=0$ and further
$K^2_{S_k\bigcup S_{k-1}\bigcup S_{k-2}}=0$. Iterating the procedure, one can easily see that
 from the Figure \ref{Type-A}, we can derive
\bea K^2_{\bigcup_{t=j}^k S_{t}}=0,~~~~~j=1,...,k~~~~\label{A-4-1}\eea
An important point of above iterating procedure is that when going to the lower orders of $\eps$,
see for example \eref{A-3-1}, the detail structure of the node $z_{S_{k-1}}$ does not appear. In other words, we can treat  the subset $S_k\bigcup S_{k-1}$ as an effective single leg when considering the subset $S_{k-2}$.

\subsection{The second case}

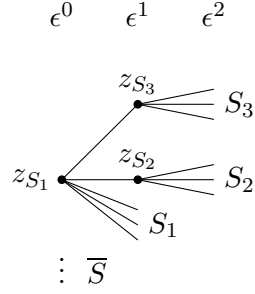
\begin{figure}
\begin{center}
\begin{tikzpicture}

\node [above] at (0,-4.5) {$ \eps^0$};
\node [above] at (1,-4.5) {$ \eps^1$};
\node [above] at (2,-4.5) {$ \eps^2$};

\node [] at (0,-7.5) {$ \vdots $};
\node [right] at (0.2,-7.6) {$\O S$};
\draw [fill] (0,-6.4) circle [radius=0.05];
\draw (0,-6.4)--(1,-5.4);
\draw (0,-6.4)--(1,-6.4);
\draw (0,-6.4)--(1,-6.8);
\draw (0,-6.4)--(1,-7.0);
\draw (0,-6.4)--(1,-7.2);
\draw [fill] (1,-5.4) circle [radius=0.05];
\draw [fill] (1,-6.4) circle [radius=0.05];
\node [right] at (1,-7.0) {$S_1$};
\node [left] at (0,-6.4) {$z_{S_1}$};

\draw (1,-5.4)--(2,-5.2);
\draw (1,-5.4)--(2,-5.4);
\draw (1,-5.4)--(2,-5.6);
\node [right] at (2,-5.4) {$S_3$};
\node [above] at (1,-5.4) {$z_{S_3}$};

\draw (1,-6.4)--(2,-6.2);
\draw (1,-6.4)--(2,-6.4);
\draw (1,-6.4)--(2,-6.6);
\node [right] at (2,-6.4) {$S_2$};
\node [above] at (1,-6.4) {$z_{S_2}$};
\end{tikzpicture}
\caption{The pictorial representation of one simple case of the Type B,  where a node has two branches
having substructure.   } \label{Type-B-1}
\end{center}
\end{figure}

In the first case, we have required that in the pictorial representation of solution $z_i$'s, although each node has several branches, there is at most one branch having a substructure. Now we consider the case that there are at least two branches having substructures. For simplicity, let consider the situation given in Figure \ref{Type-B-1}.
For $a\in S_3$, we have the scattering equations as
\bea & & {\cal E}_{a\in S_3}  =  {1\over \eps^2} \left\{ \sum_{t\neq a, t\in S_3}
{ k_a^{(0)}\cdot k_t^{(0)}\over z_a^{(2)}- z_t^{(2)}}\right\}+{1\over \eps}
\left\{ \sum_{t\neq a, t\in S_3}
{ k_a^{(0)}\cdot k_t^{(1)}+k_a^{(1)}\cdot k_t^{(0)}\over z_a^{(2)}- z_t^{(2)}}
\right. \nn & & \left. +\sum_{t\neq a, t\in S_3}
{- (z_a^{(3)}- z_t^{(3)})(k_a^{(0)}\cdot k_t^{(0)})\over (z_a^{(2)}- z_t^{(2)})^2}
+\sum_{t\in S_1}{ k_a^{(0)}\cdot k_t^{(0)}\over z_{S_3}- z_t^{(1)}}
+ \sum_{t\in S_2}{ k_a^{(0)}\cdot k_t^{(0)}\over z_{S_3}- z_{S_2}}\right\}+{\cal O}(\eps^0)~~~~~~\label{B-I-1}\eea
thus we get
\bea {\cal E}_{a\in S_3; \eps^{-2}} & = & \sum_{t\neq a, t\in S_3}
{ k_a^{(0)}\cdot k_t^{(0)}\over z_a^{(2)}- z_t^{(2)}}=0 ~~~\label{B-I-1-1}\\
{\cal E}_{a\in S_3; \eps^{-1}} & = &\sum_{t\neq a, t\in S_3}
{ k_a^{(0)}\cdot k_t^{(1)}+k_a^{(1)}\cdot k_t^{(0)}\over z_a^{(2)}- z_t^{(2)}}
+\sum_{t\neq a, t\in S_3}
{- (z_a^{(3)}- z_t^{(3)})(k_a^{(0)}\cdot k_t^{(0)})\over (z_a^{(2)}- z_t^{(2)})^2}
\nn & & +\sum_{t\in S_1}{ k_a^{(0)}\cdot k_t^{(0)}\over z_{S_3}- z_t^{(1)}}
+ { k_a^{(0)}\cdot K_{S_2}^{(0)}\over z_{S_3}- z_{S_2}}=0~~~\label{B-I-1-2}\eea
Similarly, for $a\in S_2$ we get
\bea {\cal E}_{a\in S_2; \eps^{-2}} & = & \sum_{t\neq a, t\in S_2}
{ k_a^{(0)}\cdot k_t^{(0)}\over z_a^{(2)}- z_t^{(2)}}=0 ~~~\label{B-I-2-1}\\
{\cal E}_{a\in S_2; \eps^{-1}} & = &\sum_{t\neq a, t\in S_2}
{ k_a^{(0)}\cdot k_t^{(1)}+k_a^{(1)}\cdot k_t^{(0)}\over z_a^{(2)}- z_t^{(2)}}
+\sum_{t\neq a, t\in S_2}
{- (z_a^{(3)}- z_t^{(3)})(k_a^{(0)}\cdot k_t^{(0)})\over (z_a^{(2)}- z_t^{(2)})^2}
\nn & & +\sum_{t\in S_1}{ k_a^{(0)}\cdot k_t^{(0)}\over z_{S_2}- z_t^{(1)}}
+ { k_a^{(0)}\cdot K_{S_3}^{(0)}\over z_{S_2}- z_{S_3}}=0~~~\label{B-I-2-2}\eea
For $a\in S_1$ we get
\bea {\cal E}_{a\in S_1; \eps^{-1}} & = & \sum_{t\neq a, t\in S_1}
{ k_a^{(0)}\cdot k_t^{(0)}\over z_a^{(1)}- z_t^{(1)}}+\sum_{t\in S_2}
{ k_a^{(0)}\cdot k_t^{(0)}\over z_a^{(1)}- z_{S_2}}
+\sum_{t\in S_3}
{ k_a^{(0)}\cdot k_t^{(0)}\over z_a^{(1)}- z_{S_3}}\nn
& = &\sum_{t\neq a, t\in S_1}
{ k_a^{(0)}\cdot k_t^{(0)}\over z_a^{(1)}- z_t^{(1)}}+
{ k_a^{(0)}\cdot k_{S_2}^{(0)}\over z_a^{(1)}- z_{S_2}}
+
{ k_a^{(0)}\cdot k_{S_3}^{(0)}\over z_a^{(1)}- z_{S_3}} =0 ~~~\label{B-I-3-1}\eea
%

Now let us see which kinematic information can be derived using above identities. First,
using \eref{B-I-1-1} and \eref{B-I-2-1} we get
\bea K^2_{S_2}=0,~~~~K^2_{S_3}=0~~~~\label{B-I-4-1}\eea
Secondly using \eref{B-I-1-2} and \eref{B-I-2-2} we get
\bea 0 & = & \sum_{a\in S_3} {\cal E}_{a\in S_3; \eps^{-1}}  = \sum_{t\in S_1}{ K_{S_3}^{(0)}\cdot k_t^{(0)}\over z_{S_3}- z_t^{(1)}}
+ { K_{S_3}^{(0)}\cdot K_{S_2}^{(0)}\over z_{S_3}- z_{S_2}}~~~~~\label{B-I-5-1}\\
0 & = & \sum_{a\in S_2} {\cal E}_{a\in S_2; \eps^{-1}}  = \sum_{t\in S_1}{ K_{S_2}^{(0)}\cdot k_t^{(0)}\over z_{S_2}- z_t^{(1)}}
+ { K_{S_2}^{(0)}\cdot K_{S_3}^{(0)}\over z_{S_2}- z_{S_3}}~~~~~\label{B-I-5-2}\eea
Now we consider
\bea & & 2K_{S_2}^{(0)}\cdot K_{S_3}^{(0)}+ 2\sum_{a\in S_1} (K_{S_3}^{(0)}\cdot k_a^{(0)}+
K_{S_2}^{(0)}\cdot k_a^{(0)})+2 \sum_{t< a, t,a\in S_1}k_a^{(0)}\cdot k_t^{(0)}\nn
& = & 2 \sum_{t< a, t,a\in S_1}{(z_a^{(1)}-z_{S_2})- (z_t^{(1)}-z_{S_2}) \over z_a^{(1)}- z_t^{(1)}} k_a^{(0)}\cdot k_t^{(0)}+2\sum_{a\in S_1} {(z_a^{(1)}-z_{S_2})\over (z_a^{(1)}-z_{S_2})}
K_{S_2}^{(0)}\cdot k_a^{(0)}\nn & & +2\sum_{a\in S_1} {(z_a^{(1)}-z_{S_2})+(z_{S_2}-z_{S_3})\over (z_a^{(1)}-z_{S_3})}K_{S_3}^{(0)}\cdot k_a^{(0)}+ 2K_{S_2}^{(0)}\cdot K_{S_3}^{(0)} \nn
& = &  2\sum_{a\in S_1} (z_a^{(1)}-z_{S_2})\sum_{t\neq  a, t\in S_1} { k_a^{(0)}\cdot k_t^{(0)}\over z_a^{(1)}- z_t^{(1)}} +2\sum_{a\in S_1} (z_a^{(1)}-z_{S_2}){K_{S_2}^{(0)}\cdot k_a^{(0)} \over (z_a^{(1)}-z_{S_2})}\nn
& & +2\sum_{a\in S_1} (z_a^{(1)}-z_{S_2}){K_{S_3}^{(0)}\cdot k_a^{(0)} \over (z_a^{(1)}-z_{S_3})}+ 2(z_{S_2}-z_{S_3})\sum_{a\in S_1} {K_{S_3}^{(0)}\cdot k_a^{(0)}\over (z_a^{(1)}-z_{S_3})}+ 2K_{S_2}^{(0)}\cdot K_{S_3}^{(0)}\nn
& = & 2\sum_{a\in S_1}(z_a^{(1)}-z_{S_2}){\cal E}_{a\in S_1; \eps^{-1}}=0~~~~~\label{B-I-6-1}\eea
where \eref{B-I-5-1} has been used to reach the last line. When combining \eref{B-I-6-1} with
\eref{B-I-4-1} we arrive
\bea K^2_{S_1\bigcup S_2\bigcup S_3}=0~~~~~\label{B-I-7-1}\eea

We can generalize above argument for more than two branches having substructure. Let us assume
that (1) when $i\in S$, $z_{i}^{(0)}= z_0$; (2) when $i\in S_j\subset S$, $z_{i}^{(1)}= z_{S_j}$
with $j=1,...,m$; (3) $S= S_0\bigcup S_1\bigcup ...\bigcup S_m$. For each $a\in S_{j},j=1,...,m$, we have following two identities
\bea {\cal E}_{a\in S_j; \eps^{-2}} & = & \sum_{t\neq a, t\in S_j}
{ k_a^{(0)}\cdot k_t^{(0)}\over z_a^{(2)}- z_t^{(2)}}=0 ~~~\label{B-I-8-1}\\
{\cal E}_{a\in S_j; \eps^{-1}} & = &\sum_{t\neq a, t\in S_j}
{ k_a^{(0)}\cdot k_t^{(1)}+k_a^{(1)}\cdot k_t^{(0)}\over z_a^{(2)}- z_t^{(2)}}
+\sum_{t\neq a, t\in S_j}
{- (z_a^{(3)}- z_t^{(3)})(k_a^{(0)}\cdot k_t^{(0)})\over (z_a^{(2)}- z_t^{(2)})^2}
\nn & & +\sum_{t\in S_0}{ k_a^{(0)}\cdot k_t^{(0)}\over z_{S_j}- z_t^{(1)}}
+ \sum_{p=1,p\neq j}^m { k_a^{(0)}\cdot K_{S_p}^{(0)}\over z_{S_j}- z_{S_p}}=0~~~\label{B-I-8-2}\eea
and using them we get
\bea 0 & = & K_{S_j}^2,~~~~~~0= \sum_{a\in S_j}{\cal E}_{a\in S_j; \eps^{-1}}=\sum_{t\in S_0}{ K_{S_j}^{(0)}\cdot k_t^{(0)}\over z_{S_j}- z_t^{(1)}}
+ \sum_{p=1,p\neq j}^m { K_{S_j}^{(0)}\cdot K_{S_p}^{(0)}\over z_{S_j}- z_{S_p}}~~~\label{B-I-9-2} \eea
For $a\in S_0$, the leading part of scattering equations is
\bea {\cal E}_{a\in S_0; \eps^{-1}} & = & \sum_{t\neq a, t\in S_0}
{ k_a^{(0)}\cdot k_t^{(0)}\over z_a^{(1)}- z_t^{(1)}}+\sum_{j=1}^m \sum_{t\in S_j}
{ k_a^{(0)}\cdot k_t^{(0)}\over z_a^{(1)}- z_{S_j}}\nn &= & \sum_{t\neq a, t\in S_0}
{ k_a^{(0)}\cdot k_t^{(0)}\over z_a^{(1)}- z_t^{(1)}}+\sum_{j=1}^m
{ k_a^{(0)}\cdot K_{S_j}^{(0)}\over z_a^{(1)}- z_{S_j}}=0~~~\label{B-I-10-1}
\eea
One important point of expressions \eref{B-I-9-2} and \eref{B-I-10-1} is that the subset $S_j$ has been effectively treated as a single leg. This fact is very useful to see in an arbitrary tree graph, how the computation done explicitly here is generalized.
Using above results, we can see
\bea \Sigma&\equiv & \sum_{1\leq i<j\leq m} K_{S_i}^{(0)}\cdot K_{S_j}^{(0)}
+\sum_{j=1}^m \sum_{t\in S_0} K_{S_j}^{(0)} \cdot k_t^{(0)}+ \sum_{t< a| a,t\in S_0}
 k_a^{(0)}\cdot k_t^{(0)}\nn
 & = &  \sum_{t< a| a,t\in S_0} {z_a^{(1)}-z_t^{(1)}\over
 z_a^{(1)}- z_t^{(1)}}  k_a^{(0)}\cdot k_t^{(0)}+\sum_{1\leq i<j\leq m}{z_{S_i}-z_{S_j}\over z_{S_i}-z_{S_j}} K_{S_i}^{(0)}\cdot K_{S_j}^{(0)}+  \sum_{j=1}^m \sum_{a\in S_0}{z_a^{(1)}- z_{S_j}\over z_a^{(1)}- z_{S_j}} K_{S_j}^{(0)} \cdot k_a^{(0)}\nn
& = & \sum_{t\neq  a; a,t\in S_0} {z_a^{(1)}\over
 z_a^{(1)}- z_t^{(1)}}  k_a^{(0)}\cdot k_t^{(0)}+\sum_{i\neq j}{z_{S_i}\over z_{S_i}-z_{S_j}} K_{S_i}^{(0)}\cdot K_{S_j}^{(0)}+  \sum_{j=1}^m \sum_{a\in S_0}{z_a^{(1)}- z_{S_j}\over z_a^{(1)}- z_{S_j}} K_{S_j}^{(0)} \cdot k_a^{(0)}\nn
& = & \sum_{a\in S_0}z_a^{(1)}
 \left\{\sum_{t\neq  a, t\in S_0} {k_a^{(0)}\cdot k_t^{(0)}\over
 z_a^{(1)}- z_t^{(1)}}+\sum_{j=1}^m {K_{S_j}^{(0)} \cdot k_a^{(0)}\over z_a^{(1)}- z_{S_j}}\right\} +\sum_{j=1}^m z_{S_j}\left\{ \sum_{a\in S_0}{ K_{S_j}^{(0)} \cdot k_a^{(0)}\over z_{S_j}-z_a^{(1)}}+ \sum_{i=1,i\neq j}^m { K_{S_j}^{(0)}\cdot K_{S_i}^{(0)}\over z_{S_j}- z_{S_i}} \right\}\nn
 & = & 0~~~\label{B-I-11-1}\eea
where \eref{B-I-10-1} and \eref{B-I-9-2} have been used to reach the last line. When combining
\eref{B-I-11-1} and \eref{B-I-9-2} we get
\bea K_S^2=0~~~\label{B-I-11-2}\eea

{\bf Summary:}
Although we have considered only a tree graph with two levels for Type B, it is easy to see combining results for Type A and B. The above observation can be easily generalized to a tree graph with an arbitrary structure. The basic conclusion is that for each node in the tree graph, we have $(\sum_{i\in {\rm node}} k_i)^2=0$. One important remark is that for all subsets $S_i$ with $K^2_{S_i}=0$, they are compatible to each others, i.e., either $S_i\bigcap S_j=\emptyset$ or $S_i\subset S_j$.

With results in this section, one can also show that {\sl if $k_S^2\neq 0$ for all subsets $S$ with between $2$ and $(N-2)$ elements, the values of the $z_a, a\in A,$ are distinct, i.e., there are no singular solutions.} This claim is proved in \cite{Dolan:2014ega} by contradiction. Assuming there is a singular solution, by analysis in this section, we will reach at least one $S_A\to 0$ no matter which configuration of the singular solution is.

\section{From singular kinematics to solutions I: the factorization limit}

Starting from this section, we discuss what we can learn about the solutions of scattering equations when there are one or more poles going to on-shell.

\subsection{With only one $S_A\to 0$}

This case has been well studied in \cite{Cachazo:2013iea,Dolan:2014ega}. Under the limit $\eps\to 0$, assuming there is one and only one $S_A\to 0$, what we can say about the solutions of scattering equations? Under the limit, we could have two possibilities: (R1) There is still no singular solution;(R2) There are some singular solutions.

Let us argue that the possibility (R1) could not be true first. We could assign a particular CHY-integrand, such as the multiplication of two PT-factors, which produces amplitude containing one and only one cubic Feynman diagram having the pole $S_A$. Since $S_A\to 0$, the amplitude will be divergent. Now let us check the expression \eref{gen-A-2-1} for the amplitude with gauge choice $(ijk)=(rst)$ and three fixed points at the finite locations.  We see that the numerator $(z_{ij} z_{jk} z_{ki})^2$ is not divergent, as well as  the factor ${1\over  |\Phi|_{ijk}^{rst}}$ and the particular CHY-integrand if the possibility (R1) is true.  Thus by the contradiction, we see that the possibility (R1) could not be true.

Now the possibility (R2) is realized, and we discuss the properties of singular solutions.
For a given singular solution, from the careful analysis in section two, we see that (a) the singular set can be and only be the subset $A$\footnote{Again, we use $A$ to denote the one with at most one gauge fixed point among the subsets $A, \O A$}; (b) there is no substructure in $A$, i.e., every point in $A$ going to the same location with the same speed.
If (a) is not true, we will have $S_B\to 0$ with $B\neq A$. If (b) is not true, we will have
other $S_B\to 0$ with $B\subset A$. In fact more information is given in \cite{Cachazo:2013iea}:  the number of singular solution is $(n_A+1-3)!\times (n_{\O A}+1-3)!$ where $n_A, n_{\O A}$ are the number of elements in the subsets $A, \O A$. The counting of the number of singular solutions can be easily seen from the left picture given in Figure \ref{Factorize}.

In fact, we can get more accurate information, i.e., the speed of every point in $A$ going to the same location. Assuming $|z_i-z_j|\sim t^N$  $\forall i,j\in A$, let us study the $\epsilon$ order of various part in the formula\footnote{For the regular solutions,
both measure and integrand will give $\eps^0$ contributions.} \eref{gen-A-2-1}. Without loss of generality, we choose $S_{12...k}\to 0$  and fix $\{r,s,t\}={1,2,3}$ and $\{i,j,k\}=(n,n-1,n-2)$. And choose $S_{12...k}\to 0$.
For the measure part
\begin{equation}
    J=\frac{(z_r-z_s)(z_s-z_t)(z_t-z_r)(z_i-z_j)(z_j-z_k)(z_k-z_i)}{|\Phi|^{rst}_{ijk}}
    \label{eq:Jacobi}
\end{equation}
where matrix $\Phi$ is defined in \eref{Phi-Jacobi}, one can check that
\bea & & (z_1-z_2)(z_2-z_3)(z_3-z_1)(z_n-z_{n-1})(z_{n-1}-z_{n-2})(z_{n-2}-z_n)\sim \left\{ \begin{array}{ll}\epsilon^{N},  & ~~~|A|=2 \\
\epsilon^{3N},  & ~~~|A|\geq 3 \end{array}\right. \nn
& & { |\Phi|^{rst}_{ijk}}\sim \left\{ \begin{array}{ll}\epsilon^{0},  & ~~~|A|=2 \\
\epsilon^{-2(|A|-3)N},  & ~~~|A|\geq 3 \end{array}\right.\nn
& & J\sim \epsilon^{(2|A|-3)N}~~~~\label{measure-BCFW-1}\eea
For the integrand part, the leading behaviour is $\eps^{-N\mathbb{L}[A]}$ where the $\mathbb{L}[A]$ is the linking number. Adding all together, we get the behaviour
$\eps^{ -N (\mathbb{L}[A]-2(|A|-1)+1)}=\eps^{-N(\chi[A]+1)}$. From the integration rule
\cite{Baadsgaard:2015voa,Baadsgaard:2015ifa,Cardona:2016gon}, the behaviour
is $S_A^{-(\chi[A]+1)}\sim \eps^{-(\chi[A]+1)}$, thus the match up of arbitrary choice of
$\chi[A]$ requires $N=1$.
{\sl One interesting point of above analysis of contributions of singular solutions is that
when $\chi[A]=-1$, we will have $\eps^0$.} In other words, to get the right numerical
result is the $\eps\to 0$ limit, we must include the contributions of singular solutions although analytic expression does not contain the pole $S_A$.

\subsubsection{The numerical checking}

Now we present some numerical computations to verify the above analysis. In numerical computation, one key is the construction of the kinematics having the wanted limit behaviour. The detailed discussion has been given in Appendix \ref{kine}.
For current case, i.e., with one and only one $S_A\to 0$, a straightforward way is to use the  BCFW-deformation (see \eref{One-deform-1-1}, \eref{One-deform-1-2} and \eref{One-deform-1-4}).

~\\{\bf Example I: six points with $S_{123}\to 0$} \\ The kinematics is given by
\begin{equation}
    \begin{aligned}
       k_1\to & \{-0.152-0.593 i,-1.019-0.981 i,0.106 +2.464i,-0.560+1.268 i,-2.642+0.175 i\} \\
       +& \{0.254,0.4205,-0.238-0.484 i,-0.1242-0.248 i,0.461-0.317 i\}\epsilon \\
       k_2\to& \{1.220,-0.852,-0.262,-0.825,0.115\} \\
       k_3\to & \{1.586,-0.999,-0.459,-0.664,0.931\} \\
       k_4\to& \{1.203,-0.240,0.419,0.968,0.527\} \\
       k_5\to& \{-1.416,-0.0899,-0.750,-0.688,0.980\} \\
       k_6\to &\{-2.441 +0.5935 i,+3.201 +0.981 i,+0.946 -2.464i, +1.770-1.267 i,0.089 -0.175i\} \\
       &-\{0.254,0.4205,-0.238-0.484 i,-0.1242-0.248 i,0.461-0.317 i\}\epsilon
    \end{aligned}
    \label{eq:numer k 001}
\end{equation}
We fix the gauge of $z_4=1,z_5=0,z_6=1$. There are $6$ solutions with one singular and $5$ regular solutions. The solutions are shown below as the series of $\eps$ up to $\epsilon^1$ terms:
\begin{equation}
        \begin{aligned}
        & Singular (1) \quad \left\{
            \begin{array}{l}
                z_1\to -(0.701112\, -0.144145 i)+(-3.14275-0.280886 i) \epsilon \\
                z_2\to -(0.701112\, -0.144145 i)+(-3.20442-0.255032 i) \epsilon \\
                z_3\to -(0.701112\, -0.144145 i)+(-3.18786-0.260227 i) \epsilon \\
            \end{array}
        \right.
        \end{aligned}
        \label{eq:singular sol 01}
    \end{equation}
\bea
        & Regular (1) \quad \left\{
            \begin{array}{l}
                z_1\to (0.0452115\, +0.636781 i)-(0.0913283\, +0.0933938 i) \epsilon \\
                z_2\to -(2.63638\, +0.252226 i)+(1.32192\, +0.0640334 i) \epsilon \\
                z_3\to -(8.15521\, -3.66499 i)+(7.11037\, -9.61685 i) \epsilon \\
            \end{array}
            \right.
\eea
\bea
        & Regular (2) \quad \left\{
            \begin{array}{l}
                z_1\to (0.0308196\, +0.774219 i)-(0.0791472\, +0.145376 i) \epsilon \\
                z_2\to -(0.71447\, +6.1477 i) +(-6.30372+0.80879 i) \epsilon\\
                z_3\to -(1.67537\, -0.0415368 i)+(0.253797\, -0.132973 i) \epsilon \\
            \end{array}
            \right.\eea
\bea
        & Regular (3) \quad \left\{
            \begin{array}{l}
                z_1\to -(0.28604\, -0.376317 i)+ (-0.0471353-0.097799 i) \epsilon\\
                z_2\to -(4.08226\, +0.917673 i)+(2.66599\, +1.09849 i) \epsilon \\
                z_3\to -(0.377007\, -0.515852 i)+(-0.0729845-0.0909321 i) \epsilon \\
               \end{array}
            \right.\eea
\bea
        & Regular (4) \quad \left\{
            \begin{array}{l}
                z_1\to -(0.435329\, -0.28238 i)+(0.0150451\, -0.0647125 i) \epsilon \\
                z_2\to -(0.297746\, -0.197856 i)+ (-0.0197837-0.00948061 i) \epsilon\\
                z_3\to -(3.40278\, -1.35514 i)+(0.562885\, -0.896193 i) \epsilon \\
            \end{array}
            \right.\eea
\bea
        & Regular (5) \quad \left\{
            \begin{array}{l}
                z_1\to -(0.72372\, -0.14116 i)+(0.0430294\, -0.0225452 i) \epsilon \\
                z_2\to -(0.558638\, -0.130488 i)+(-0.00452728-0.0035687 i) \epsilon \\
                z_3\to -(0.855428\, -0.12072 i)+(-0.0054946-0.00384179 i) \epsilon \\
            \end{array}
            \right.
        \label{eq:regular sol 01}\eea
The measure part of each solution is shown in table\eqref{tab:Jacobi001}.
\begin{table}[htbp]
    \centering
    \begin{tabular}{|c|c|}
    \hline
    Solution   &    Measure \\
    \hline
    Singular (1) &$\left(1.08633 \times 10^{-7}-3.24352 \times 10^{-8} i\right) \epsilon^{3}$   \\
    Regular (1)  &   $-(516.832+220.062 i)$   \\
    Regular (2)  &   $-(6.23504+50.3245 i)$   \\
    Regular (3)  &   $(0.0453058-0.141481 i)$   \\
    Regular (4)  &   $(0.0188278+0.0202573 i)$   \\
    Regular (5)  &   $-\left(3.4399 \times 10^{-6}+8.80267 \times 10^{-6} i\right)$   \\
    \hline
    \end{tabular}
    \caption{The measure part of each solution}
    \label{tab:Jacobi001}
\end{table}
which is indeed the behaviour $\epsilon^{(2|A|-3)}=\epsilon^{2\times 3-3}=\epsilon^3$. Now we choose three different CHY-integrands with pole index of $\chi[S_{123}]=0,-1,-2$ respectively to calculate the contribution of each solution.

\begin{itemize}
    \item $I_1=PT(123456)PT(124563)$ with $\chi[S_{123}]=0$.
The contribution of each solution is shown in table\eqref{tab:Integrands01}.
    \begin{table}[htbp]
    \centering
    \begin{tabular}{|c|c|}
    \hline
    Solution   &   Integration \\
    \hline
    Singular (1) &$\frac{1}{\epsilon}(0.0801588+0.0445719 i)$   \\
    Regular (1)  &   $-(0.00189857+0.00277649 i)$   \\
    Regular (2)  &   $(0.00417627+0.00436192 i)$   \\
    Regular (3)  &   $-(0.00185648+0.00211634 i)$   \\
    Regular (4)  &   $-(0.00963083-0.000139444 i)$  \\
    Regular (5)  &   $(0.0509707+0.00619341 i)$   \\
    \hline
    \end{tabular}
    \caption{The contribution of each solution for the CHY-Integrand $PT(123456)PT(124563)$}
    \label{tab:Integrands01}
\end{table}
Summing up all six solutions, we have
\begin{equation}
    \frac{0.0801588+0.0445719 i}{\epsilon}+O(\epsilon^0)
    \label{eq:sum 001}
\end{equation}
If we substitute the numerical kinematics \eqref{eq:numer k 001} to the analytic expression
\bea
       \frac{1}{S_{12} S_{45} S_{123}}+\frac{1}{S_{12} S_{56} S_{123}}
    =\frac{0.0802693+0.0448299 i}{\epsilon}+O(\epsilon^0)
\eea
This result is essentially in agreement with \eqref{eq:sum 001}. For this case, the leading contribution comes from the singular solution only.

\item $I_2=PT(123456)PT(124536)$ with $\chi[S_{123}]=-1$. The result is shown in table\eqref{tab:Integrands2}
\begin{table}[htbp]
    \centering
    \begin{tabular}{|c|c|}
    \hline
    Solution   &   Integration \\
    \hline
    Singular (1) & $-(0.0188869-0.00321464 i)$   \\
    Regular (1)  & $(0.00236768+0.00127038 i)$  \\
    Regular (2)  & $-(0.00430291+0.00290199 i)$   \\
    Regular (3)  & $(0.000865392+0.000279345 i)$  \\
    Regular (4)  & $(0.011857-0.00567046 i)$  \\
    Regular (5)  & $-(0.0254912-0.00146168 i)$   \\
    \hline
    \end{tabular}
    \caption{The contribution of each solution of CHY-Integrand $PT(123456)PT(124536)$}
    \label{tab:Integrands2}
\end{table}
The summation of six solutions is
\begin{equation}
     -(0.033591+0.0023464 i)+O\epsilon)
\end{equation}
which is identical with the numerical result by substituting \eqref{eq:numer k 001}
to analytic expression
\bea
     \frac{1}{S_{12} S_{45} S_{126}}+\frac{1}{S_{16} S_{45} S_{126}}
    =-(0.033591+0.0023464 i)+O(\eps),
\eea
This shown the correctness of \eqref{eq:singular sol 01} and \eqref{eq:regular sol 01}. From table \eqref{tab:Integrands2} we see all solutions (including the singular solution) gives $\epsilon^0$ contribution.

\item $I_3=PT(123456)PT(143526)$ with $\chi[S_{123}]=-2$.
The result is shown in\eqref{tab:Integrands3}.
\begin{table}[htbp]
    \centering
    \begin{tabular}{|c|c|}
    \hline
    Solution   &   Integration \\
    \hline
    Singular (1) &$(0.00100813-0.000287969 i) \epsilon$   \\
    Regular (1)  &  $-(0.001916+0.00405212 i)$   \\
    Regular (2)  &  $(0.000756028+0.000975639 i)$   \\
    Regular (3)  &  $-(0.0000417965-0.000573095 i)$  \\
    Regular (4)  &   $-(0.00286251-0.00330538 i)$  \\
    Regular (5)  &  $-(0.0012266+0.00080199 i)$   \\
    \hline
    \end{tabular}
    \caption{The contribution of each solution of CHY-Intergrand $PT(123456)PT(143526)$}
    \label{tab:Integrands3}
\end{table}
The summation of all six solutions is
\begin{equation}
    -0.00529087+O(\epsilon)
\end{equation}
which is identical to the analytic expression  with the  numerical kinematic \eqref{eq:numer k 001}
\bea
       -\frac{1}{S_{16}  S_{34} S_{126}}
    =-0.00529087+O(\epsilon)
\eea
where singular solution will give no contribution when $\epsilon\to 0$.

\end{itemize}

~\\{\bf Example II: six points with $S_{12}\to 0$}.
The choice of numerical kinematics is
\begin{equation}
    \begin{aligned}
        k_1\to &\{0.111 +0.384 i,-1.887
            -0.493 i, 0.221 -2.605 i,-0.751 -1.544 i, 2.344 -0.627 i\} \\
       +&\{0.361,0.755+0.428 i,-0.363+0.731 i,-0.123+0.466 i,-0.589\}\epsilon\\
       k_2\to& \{0.894,-0.453,-0.411,0.647,-0.0763\}\\
       k_3\to& \{1.065,-0.183,0.780,0.557,0.427\} \\
       k_4\to& \{1.387,-0.0137,-0.981,0.840,0.504\} \\
       k_5\to& \{-1.7085,1.1005,-1.083,0.2038,-0.7025\} \\
       k_6\to &\{-1.748-0.384 i,1.436+0.493 i,1.475+2.605 i,-1.497+1.544 i,-2.497+0.627 i\} \\
       -&\{0.361,0.755+0.428 i,-0.363+0.731 i,-0.123+0.466 i,-0.589\}\epsilon.
    \end{aligned}
    \label{eq:numer k 002}
\end{equation}
We fix the gauge choice $z_4=1,z_5=0,z_6=1$, there are two  singular and four regular solutions:

\begin{equation}
        \begin{aligned}
        & Singular (1) \quad \left\{
            \begin{array}{l}
                z_1\to (0.762074\, +0.364077 i)-(0.0646452\, +0.838813 i) \epsilon \\
                z_2\to (0.762074\, +0.364077 i)+(0.515056\, -0.754896 i) \epsilon \\
                z_3\to -(0.0322539\, -0.0615108 i)+(0.01392\, +0.00452414 i) \epsilon \\
            \end{array}
        \right.\\
        & Singular (2) \quad \left\{
            \begin{array}{l}
                z_1\to (0.112028\, -0.54179 i)-(0.047236\, -0.318322 i) \epsilon \\
                z_2\to (0.112028\, -0.54179 i)+ (0.225107\, +0.143289 i) \epsilon\\
                z_3\to (0.396924\, +0.0453154 i)-(0.0887612\, -0.0250629 i) \epsilon \\
            \end{array}
        \right.
        \end{aligned}
        \label{eq:singular sol 02}
    \end{equation}

    \begin{equation}
        \begin{aligned}
        & Regular (1) \quad \left\{
            \begin{array}{l}
                z_1\to (1.05565\, +0.418985 i)-(0.51277\, +0.296097 i) \epsilon \\
                z_2\to (4.03218\, +1.32966 i)+ (6.55886\, +28.4338 i) \epsilon\\
                z_3\to -(0.0320861\, -0.0733717 i)+(-0.0125772+0.00408331 i) \epsilon \\
            \end{array}
        \right.\\
        & Regular (2) \quad \left\{
            \begin{array}{l}
                z_1\to (0.305812\, -0.703593 i)-(0.205823\, -0.327967 i) \epsilon \\
                z_2\to -(4.30772\, +0.129581 i)+(15.332\, +1.98828 i) \epsilon \\
                z_3\to (0.452278\, -0.0114679 i)-(0.109894\, -0.0917084 i) \epsilon \\
            \end{array}
        \right.\\
        & Regular (3) \quad \left\{
            \begin{array}{l}
                z_1\to (0.932462\, +0.508345 i)-(0.0889137\, +0.704417 i) \epsilon \\
                z_2\to -(0.147739\, -0.337423 i)+(-0.0381437+0.0380429 i) \epsilon \\
                z_3\to -(0.0558585\, -0.162427 i)+(-0.0398685-0.0252863 i) \epsilon \\
            \end{array}
        \right.\\
        & Regular (4) \quad \left\{
            \begin{array}{l}
                z_1\to (0.290776\, -0.394965 i)-(0.208807\, -0.274043 i) \epsilon \\
                z_2\to (0.548109\, +0.0301496 i)-(0.132396\, -0.052737 i) \epsilon\\
                z_3\to (0.296957\, +0.0219136 i)-(0.142326\, -0.0932441 i) \epsilon \\
            \end{array}
        \right.
        \end{aligned}
        \label{eq:regular sol 02}
    \end{equation}
The measure part of each solution is shown in table \eqref{tab:Jacobi02}.

\begin{table}[htbp]
    \centering
    \begin{tabular}{|c|c|}
    \hline
    Solution   &   Measure \\
    \hline
    Singular (1) &$-(0.000529048+0.0005798 i) \epsilon$   \\
    Singular (2) &$(0.000892278+0.000357788 i) \epsilon$   \\
    Regular (1)  &$-(0.186684+0.928303 i)$   \\
    Regular (2)  &$-(1.95152-1.51335 i)$   \\
    Regular (3)  &$-(0.0000405857-0.000616897 i)$   \\
    Regular (4)  &$-(0.000169266-0.0000142748 i)$   \\
    \hline
    \end{tabular}
    \caption{The measure part of each solution for $S_{12}\to 0$.}
    \label{tab:Jacobi02}
\end{table}
which agrees with our analysis $\eps^0$ for regular solutions and $\eps^{2|A|-3}$ for
singular solutions.
Again we choose three different CHY-Intergrands with pole index $\chi[S_{12}]=0,-1,-2$ respectively.
\begin{itemize}
    \item $I_1=PT(123456)PT(124563)$ with $\chi[S_{12}]=0$.  The result is shown in \eqref{tab:Integrands4}.
\begin{table}[htbp]
    \centering
    \begin{tabular}{|c|c|}
    \hline
    Solution   &   Integration \\
    \hline
    Singular (1) &$\frac{1}{\epsilon}(0.00333026+0.00227831 i)$   \\
    Singular (2) &$-\frac{1}{\epsilon}(0.0142701-0.0137308 i)$  \\
    Regular (1)  &$-(0.00260062-0.0012496 i)$   \\
    Regular (2)  &$(0.00122598-0.00527255 i)$   \\
    Regular (3)  &$-(0.000501431-0.000889243 i)$   \\
    Regular (4)  &$(0.0115214-0.00241242 i)$   \\
    \hline
    \end{tabular}
    \caption{The contribution of each solution of CHY-Intergrand $PT(123456)PT(124563)$}
    \label{tab:Integrands4}
\end{table}
The summation of six solutions is
    \begin{equation}
    -\frac{0.0109398-0.0160091 i}{\epsilon}+O(\epsilon)
\end{equation}
which   is identical with the analytic expression  with the  numerical kinematic \eqref{eq:numer k 002}
\bea\frac{1}{S_{12} S_{45} S_{123}}+\frac{1}{S_{12} S_{56} S_{123}} =
      -\frac{0.0109398-0.0160091 i}{\epsilon}+O(\epsilon)
\eea
where the leading contribution  comes from the singular solutions only.

\item $I_2=PT(123456)PT(134526)$ with $\chi[S_{12}]=-1$.
The result is shown in \eqref{tab:Integrands5}.
\begin{table}[htbp]
    \centering
    \begin{tabular}{|c|c|}
    \hline
    Solution   &   Integration \\
    \hline
    Singular (1) &$-(0.000196838-0.000292468 i)$   \\
    Singular (2) &$-(0.000377502-0.0182017 i)$  \\
    Regular (1)  &$-(0.000561198-0.000217058 i)$   \\
    Regular (2)  &$(0.00839347-0.0144827 i)$   \\
    Regular (3)  &$(0.00167404-0.00059116 i)$   \\
    Regular (4)  &$-(0.00221397+0.00363733 i)$   \\
    \hline
    \end{tabular}
    \caption{The contribution of each solutions of CHY-Intergrand $PT(123456)PT(134526)$}
    \label{tab:Integrands5}
\end{table}
The fact that summation of six solutions
\begin{equation}
    0.006718+O(\epsilon)
\end{equation}
is identical with analytic result
\bea\frac{1}{S_{16} S_{34} S_{126}}+\frac{1}{S_{16} S_{45} S_{126}}
        =0.006718+O(\epsilon)
\eea
shows that when $\epsilon\to 0$, both regular and singular solutions give nonzero contributions.

\item $I_3=PT(132456)PT(134526)$ with pole index $\chi[S_{12}]=-2$.
The result is shown in \eqref{tab:Integrands6}.
\begin{table}[htbp]
    \centering
    \begin{tabular}{|c|c|}
    \hline
    Solution   &   Integration \\
    \hline
    Singular (1) &$-(0.000557274-0.000151919 i) \epsilon$   \\
    Singular (2) &$-(0.00394239-0.00346853 i) \epsilon$  \\
    Regular (1)  &$(0.00033303-0.00039 i)$   \\
    Regular (2)  &$-(0.00932611+0.00647014 i)$   \\
    Regular (3)  &$-(0.00153402-0.00062326 i)$   \\
    Regular (4)  &$(0.00680103+0.00398464$ i)   \\
    \hline
    \end{tabular}
    \caption{The contribution of each solutions of CHY-Intergrand $PT(132456)PT(134526)$}
    \label{tab:Integrands6}
\end{table}
The summation of six solutions is
\begin{equation}
    -(0.00372607+0.00225224 i)+O(\epsilon)
\end{equation}
which agrees with the analytic result
\bea \frac{1}{S_{13}  S_{45} S_{136}}+\frac{1}{S_{16}  S_{45} S_{136}}=-(0.00372607+0.00225224 i)+O(\epsilon)
\eea
For this case, when $\epsilon\to 0$, the regular solutions give whole contribution and singular solutions give no contributions.

\end{itemize}

\subsection{With $S_A, S_B\to 0$}

Now we consider the factorization limit  where two poles $S_A, S_B$ go to zero
at the same time. As discussed in the Appendix,
a way to reach the kinematic configuration is to use two pairs of BCFW-deformations,
for example,
\bea p_1(w_1)&= & p_1+w_{1} q_{1n},~~~p_n(w_1)=p_n-w_{1} q_{1n},~~~q_{1n}^2=q_{1n}\cdot p_1=q_{1n}\cdot
p_n=0\nn
 p_2(w_2) &= &p_2+w_{2} q_{2m},~~~p_m(w_2)=p_m-w_{2} q_{2m},~~~q_{2m}^2=q_{2m}\cdot p_1=q_{2m}\cdot
p_m=0~~~\label{Two-BCFW}\eea
Using $S_A=S_B=0$ we can solve $w_{1}^*, w_{2}^*$, then we write $w_1=w_1^*+\eps$ and
$w_2=w_2^*+\eps$ and $\eps\to 0$ will give $S_A, S_B\to 0$ at the same time.

A new feature of two poles going to zero at the same time is that
there are two different cases we should consider. The first case is that two poles are compatible, i.e., two poles appear in a single Feynman diagram. The second case is that two poles are not compatible, so there is no Feynman diagram containing them simultaneously. While the second case is that they are not compatible, let us discuss them one by one.

\subsubsection{ $S_A, S_B$ are not compatible}

For this case, we will have $A\not\subset B$ and $A\not\subset \O B$, so at least one intersection $A\cap B$ or $A\cap \O B$ will be the true nonempty subset. Therefore, the first claim we can give is that there are singular solutions for $S_A$ and $S_B$. The second question is that are there  "common singular solutions"? If there is a common singular solution, $z_i\to z_A, \forall i\in A$ and $z_j\to z_B,\forall j\in B$ will lead $z_A=z_B$, thus by the analysis in section two. We will get $S_{A\cup B}\to 0$, which is a conflict with our assumption of kinematic configuration. Therefore, we conclude that there is no common singular solution when $S_A, S_B$ are not compatible. This result should be easy to guess since there is no Feynman diagram containing them simultaneously.

Thirdly, we discuss the collapse speed of singular solutions, i.e., $|z_i-z_j|\sim \eps^N, \forall i,j \in A$. The analysis will be exact same as in previous subsection and we conclude that $N=1$, thus contributions of singular solutions of $a_A$ will be the order $ \epsilon^{-1+\chi(A)}$ (similar for $S_B$). Finally the number of singular solutions for $S_A$ and $S_B$ will be $(n_A-2)!(n-n_A-2)!$ and  $(n_B-2)!(n-n_B-2)!$ respectively.

\subsubsection{ $S_A, S_B$ are  compatible}

For this case, we can make $A\cap B=\emptyset$. Again the first result is that there are singular solutions for $S_A$ and $S_B$. To see if there is any common singular solution, we can consider the CHY-integrands, which give  Feynman diagrams containing both poles $S_A, S_B$. If there is no common singular solution, by the same analysis (see \eref{measure-BCFW-1}), we can see that the measure gives either $\eps^{(2|A|-3) N_A}$ or $\eps^{(2|A|-3) N_B}$, thus the leading behavior will be either $\eps^{-N_A(\chi[A]+1)}$ or $\eps^{-N_B(\chi[A]+1)}$, but from Feynman diagrams, we see the leading behaviour $\eps^{-(\chi[A]+1)} \eps^{-(\chi[B]+1)}$. We cannot get the match result since $N_A, N_B$ are fixed while choosing different $\chi[A],\chi[B]$. Thus the contradiction tells us that there are common singular solutions.

\begin{figure}
  \centering
  \begin{minipage}[thtbp]{0.8\linewidth}
  \centering
  \begin{tikzpicture}
    \draw [fill] (0,0) circle [radius=0.05];
    \draw (0,0)--(-1,1);
    \draw (0,0)--(-1,0);
    \draw (0,0)--(-1,-1);

    \draw [decorate,decoration={calligraphic brace,mirror,amplitude=2mm},thick] (-1.5,1) -- (-1.5,-1);
    \node [left] at (-1.7,0) {$A$};

    \draw (0,0)--(2,0);
    \node [above] at (1,0.1) {$S_A$};

    \draw [fill] (2,0) circle [radius=0.05];
    \draw (2,0)--(3,1);
    \draw (2,0)--(3,0);
    \draw (2,0)--(3,-1);

    \draw [decorate,decoration={calligraphic brace,amplitude=2mm},thick] (3.5,1) -- (3.5,-1);
    \node [right] at (3.7,0) {$\O A$};
    \end{tikzpicture}
  \end{minipage}

  \begin{minipage}[thtbp]{0.8\linewidth}
  \centering

  \begin{tikzpicture}
    \draw [fill] (5,0) circle [radius=0.05];
    \draw (5,0)--(4,1);
    \draw (5,0)--(4,0);
    \draw (5,0)--(4,-1);

    \draw [decorate,decoration={calligraphic brace,mirror,amplitude=2mm},thick] (3.5,1) -- (3.5,-1);
    \node [left] at (3.3,0) {$A$};

    \draw (5,0)--(9,0);
    \node [below] at (6,-0) {$S_A$};

    \draw [fill] (7,0) circle [radius=0.05];
    \draw (7,0)--(6,1);
    \draw (7,0)--(7,1);
    \draw (7,0)--(8,1);
    \draw [decorate,decoration={calligraphic brace,amplitude=2mm},thick] (6,1.5) -- (8,1.5);
    \node [above] at (7,1.7) {$\O{A\bigcup B}$};
    \draw [fill] (7,3.1) circle [radius=0.0001];

    \node [below] at (8,-0) {$S_B$};

    \draw [fill] (9,0) circle [radius=0.05];
    \draw (9,0)--(10,1);
    \draw (9,0)--(10,0);
    \draw (9,0)--(10,-1);
    \draw [decorate,decoration={calligraphic brace,amplitude=2mm},thick] (10.5,1) -- (10.5,-1);
    \node [right] at (10.7,0.0) {$B$};
    \end{tikzpicture}
  \end{minipage}
  \caption{The effective Feynman diagrams: the above is for only one singular poles,while the below is for two compatible singular poles.}
 \label{Factorize}
\end{figure}
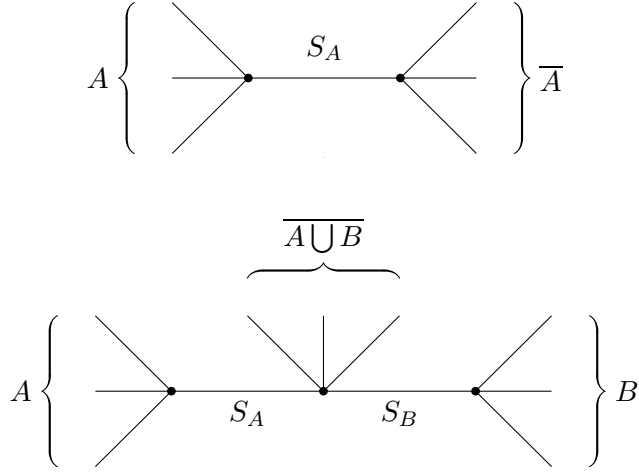

One intuitive way to see the number of common singular solution is to draw the effective Feynman diagram (see Figure \ref{Factorize}). For only one $S_A\to 0$, the left effective Feynman diagram gives the right accounting. For two compatible poles, the effective Feynman diagrams are given at the right hand side. Thus all legs have been divided into three part: the left part contains $n_A+1$ external legs, the middle part contains $n_{\overline{A\cup B}}+2$ external legs, and the right part includes $n_{B}+1$ external legs. With the picture, the whole $n$-point will contains $(n_A+1-3)!\times (n_{\overline{A\cup B}}+2-3)!\times (n_{B}+1-3)!=(n_A-2)!(n_{\overline{A\cup B}}-1)!(n_{B}-2)!$ common singular solutions.

Now we discuss the collapse speed of singular solutions, i.e., $|z_i-z_j|\sim \eps^N,\forall i,j \in A(B)$. For the three types of singular solutions, by matching up the singular behaviour of Feynman diagrams, one can see that for the singular solution of $S_A$ only, we will have $|z_i-z_j|\sim \eps, \forall i,j \in A$ and $|z_i-z_j|\sim \eps^0, \forall i,j \in B$\footnote{Noticing that we have assumed $A\cap B=\emptyset$. Otherwise, we use $\O B$ to replace $B$.} Similar result holds for singular solution of $S_B$ only, i.e., $|z_i-z_j|\sim \eps,
\forall i,j \in B$ and $|z_i-z_j|\sim \eps^0, \forall i,j \in A$. For the common singular solutions, the analysis is tricky. Because the asymptotic behaviour depends on the choice of gauge fixing $z_i, z_j, z_k$, we can have two types of asymptotic behaviour. The first one is
\bea type-I:& & z_{i\in A}\to z_A+ a_i\eps^{N_1},~~~|a_i-a_j|\neq 0,~~~i,j\in A \nn
& & z_{t\in B}\to z_B+ a_t\eps^{N_2},~~~|a_t-a_s|\neq 0,~~~t,s\in B,~~~~z_A\neq z_B
\label{type-I}
\eea
The second type is
\bea type-II:& & z_{i\in A}\to z_B+ z_A \eps^{N_1}+ a_i\eps^{N_2},~~~|a_i-a_j|\neq 0,~~~i,j\in A,~~~N_2>N_1 \nn
& & z_{t\in B\backslash A}\to z_B+ a_t\eps^{N_1},~~~|a_t-z_A|\neq 0, |a_t-a_s|\neq 0,~~~t,s\in B\backslash A
\label{type-II}
\eea
The type-I can be realised that if we take one gauge fixed point in  $ A$, one gauge fixed point in $B$ and the third point in $\O{ A\bigcup B}$ with the choice $A\bigcap B=\emptyset$. The type-II can be realised by taking one gauge fixed point in $B\backslash A$ and two gauge-fixed points in $\O B$ with the choice $A\subset B$. Suppose the gauge choice makes the type-I behaviour by analysing the leading behaviour of measure and integrand in the previous subsection. In that case, one can check that we should have $N_1=N_2=1$ to match up the result from Feynman diagrams. If the gauge choice makes the type-II behaviour, a similar argument leads $N_1=1, N_2=2$.

Knowing the value of $N_1,N_2$, we see the measure part of the common singular solutions gives $\epsilon^{2|A|-3}\cdot \epsilon^{2|B|-3}$, while the leading contribution gives $\epsilon^{-(\chi[A]+1)}\cdot \epsilon^{-(\chi[B]+1)}$.\footnote{Although the asymptotic behaviour of type-$I$ and type-$II$ are different, their leading behaviour of measure and contribution have the same $\epsilon-$order. Which can be calculated by the same procedure of the previous subsection staring by staring with (\ref{type-I}) and (\ref{type-II}) respectively.}

\subsubsection{Numerical Check}

In this part, we will present several numerical examples to demonstrate above analysis.

~\\
{\bf Example I: $n=7$, $S_{12}=S_{45}=0$}

We first consider the case of $S_A$ is compatible with $S_B$. We will choice $n=7$, $A=\{1,2\},B=\{4,5\}$. The kinematics configuration is
\begin{equation}
    \begin{array}{ll}
    k_1 \rightarrow &\{0.502+0.165 i,-0.303+0.197 i,-0.326-0.315 i,-0.109-0.0949 i,-0.409-0.0717 i\}
    +\\ &\{0.616 \epsilon ,(0.336-0.195 i) \epsilon ,(-0.119+0.516 i) \epsilon ,(-0.719-0.177 i) \epsilon ,-0.269 \epsilon \}\\
    k_2 \rightarrow &\{1.2,-0.256,-0.731,0.424,-0.818\}\\
    k_3 \rightarrow &\{1.19-0.165 i,0.728-0.197 i,0.11+0.315 i,-0.99+0.0949 i,0.102+0.0717 i\}
    +\\ &\{-0.616 \epsilon ,(-0.336+0.195 i) \epsilon ,(0.119-0.516 i) \epsilon ,(0.719+0.177 i) \epsilon ,0.269 \epsilon \}\\
    k_4 \rightarrow &\{0.361+0.689 i,-0.646-0.902 i,0.705-0.421 i,0.0459-0.54 i,-0.162+0.0756 i\}
    +\\ &\{0.96 \epsilon ,(-0.779+0.414 i) \epsilon ,(-0.99-0.349 i) \epsilon ,(-0.0376+0.619 i) \epsilon ,0.105 \epsilon \}\\
    k_5 \rightarrow &\{1.,-0.394,0.288,-0.796,0.366\}\\
    k_6 \rightarrow &\{-1.76,0.214,-1.52,0.246,-0.82\}\\
    k_7 \rightarrow &\{-2.49-0.689 i,0.657+0.902 i,1.48+0.421 i,1.18+0.54 i,1.74-0.0756 i\}
    +\\ &\{-0.96 \epsilon ,(0.779-0.414 i) \epsilon ,(0.99+0.349 i) \epsilon ,(0.0376-0.619 i) \epsilon ,-0.105 \epsilon \}\\
    \end{array}
\end{equation}
We fix $z_3=1,z_6=5,z_7=-7$, thus the type-I behaviour is realized. There are 10 singular solutions in total:
\bea
 \text{Singular }S_{12}S_{45} \text{ (1)}  & &\left\{
\begin{array}{l}
z_1\to (-0.19+7.49 i)-(19.4-1.93 i) \epsilon -(21.3+32.7 i) \epsilon ^2 \\
z_2\to (-0.19+7.49 i)-(4.74+3.58 i) \epsilon +(3.28-6.28 i) \epsilon ^2 \\
z_4\to (11.4+0.699 i)+(7.43-8.4 i) \epsilon +(2.68-4.86 i) \epsilon ^2 \\
z_5\to (11.4+0.699 i)+(3.62+1.88 i) \epsilon -(4.62-7.69 i) \epsilon ^2 \\
\end{array}
\right.\nn
 \text{Singular }S_{12}S_{45} \text{ (2)}  & &\left\{
\begin{array}{l}
z_1\to (-8.16-2.93 i)+(4.47+4.09 i) \epsilon -(2.34+4.47 i) \epsilon ^2 \\
z_2\to (-8.16-2.93 i)+(2.59-1.07 i) \epsilon +(2.6-5.33 i) \epsilon ^2 \\
z_4\to (2.41+2.42 i)+(1.69+6.53 i) \epsilon -(1.56+0.251 i) \epsilon ^2 \\
z_5\to (2.41+2.42 i)+(0.484+0.193 i) \epsilon -(1.83-1.95 i) \epsilon ^2 \\
\end{array}
\right. \nonumber
\eea

\bea
 \text{Singular }S_{12} \text{ (1)}  & &\left\{
\begin{array}{l}
z_1\to (6.09+26.5 i)-(147.-111. i) \epsilon -(850.+129. i) \epsilon ^2 \\
z_2\to (6.09+26.5 i)-(36.3-28.2 i) \epsilon -(129.-2.6 i) \epsilon ^2 \\
z_4\to (23.2-27. i)-(42.3+79.6 i) \epsilon +(173.-208. i) \epsilon ^2 \\
z_5\to (2.47+1.81 i)-(1.14-0.729 i) \epsilon -(5.27+1.87 i) \epsilon ^2 \\
\end{array}
\right.\nn
 \text{Singular }S_{12} \text{ (2)}  & &\left\{
\begin{array}{l}
z_1\to (-0.326+7.62 i)-(21.1-5.41 i) \epsilon -(45.6+41.3 i) \epsilon ^2 \\
z_2\to (-0.326+7.62 i)-(4.44+0.898 i) \epsilon +(1.-3.95 i) \epsilon ^2 \\
z_4\to (6.96+16.3 i)-(11.6-14. i) \epsilon -(17.4+30.4 i) \epsilon ^2 \\
z_5\to (4.61-4.29 i)-(2.76+1.31 i) \epsilon +(1.47-1.54 i) \epsilon ^2 \\
\end{array}
\right.\nn
\text{Singular }S_{12} \text{ (3)}  & &\left\{
\begin{array}{l}
z_1\to (-8.34+0.879 i)-(3.38+4.78 i) \epsilon +(7.54+1.7 i) \epsilon ^2 \\
z_2\to (-8.34+0.879 i)-(5.22+2.61 i) \epsilon -(2.47+2.18 i) \epsilon ^2 \\
z_4\to (1.54+1.85 i)-(0.644-3.99 i) \epsilon -(4.87+1.35 i) \epsilon ^2 \\
z_5\to (-4.29-5.26 i)+(18.2+3.21 i) \epsilon -(5.86-9.66 i) \epsilon ^2 \\
\end{array}
\right.\nn
\text{Singular }S_{12} \text{ (4)}  & &\left\{
\begin{array}{l}
z_1\to (-7.12-3.13 i)+(6.76+4. i) \epsilon -(10.1+10.1 i) \epsilon ^2 \\
z_2\to (-7.12-3.13 i)+(6.87-1.01 i) \epsilon -(0.647+10.5 i) \epsilon ^2 \\
z_4\to (1.67+1.96 i)-(0.965-3.22 i) \epsilon -(2.25-6.98 i) \epsilon ^2 \\
z_5\to (2.82+6.01 i)+(5.02+19.3 i) \epsilon +(30.2-13.2 i) \epsilon ^2 \\
\end{array}
\right.\nonumber \eea
\bea
\text{Singular }S_{45} \text{ (1)}  & &\left\{
\begin{array}{l}
z_1\to (-2.33+29.4 i)-(85.6+2.5 i) \epsilon +(24.1-22.3 i) \epsilon ^2 \\
z_2\to (0.596+6.55 i)-(6.3+2.18 i) \epsilon -(1.86+8.77 i) \epsilon ^2 \\
z_4\to (14.4+2.23 i)+(15.6-21.3 i) \epsilon +(21.6-13.2 i) \epsilon ^2 \\
z_5\to (14.4+2.23 i)+(6.56-4.88 i) \epsilon +(11.8+0.324 i) \epsilon ^2 \\
\end{array}
\right.\nn
 \text{Singular }S_{45} \text{ (2)}  & &\left\{
\begin{array}{l}
z_1\to (-0.959+1.13 i)-(1.06+0.33 i) \epsilon -(1.02+0.891 i) \epsilon ^2 \\
z_2\to (4.6+6.79 i)-(7.89-6.2 i) \epsilon -(9.72+0.0344 i) \epsilon ^2 \\
z_4\to (8.42-0.152 i)+(5.02-4.83 i) \epsilon +(3.21-5.33 i) \epsilon ^2 \\
z_5\to (8.42-0.152 i)+(4.05-0.0335 i) \epsilon +(3.11+3.2 i) \epsilon ^2 \\
\end{array}
\right.\nn
 \text{Singular }S_{45} \text{ (3)}  & &\left\{
\begin{array}{l}
z_1\to (-2.69+1.77 i)-(2.51+0.646 i) \epsilon -(1.06+0.197 i) \epsilon ^2 \\
z_2\to (-5.6-7.92 i)-(2.68-2.24 i) \epsilon +(1.2+5.65 i) \epsilon ^2 \\
z_4\to (2.4+2.42 i)+(1.74+6.79 i) \epsilon -(0.271-3.09 i) \epsilon ^2 \\
z_5\to (2.4+2.42 i)+(0.631+0.616 i) \epsilon -(2.43-3.81 i) \epsilon ^2 \\
\end{array}
\right.\nn
 \text{Singular }S_{45} \text{ (4)}  & &\left\{
\begin{array}{l}
z_1\to (1.06+2.21 i)-(3.67-1.25 i) \epsilon -(5.58+1.63 i) \epsilon ^2 \\
z_2\to (-4.1-6.76 i)-(0.276-0.326 i) \epsilon -(1.36-0.429 i) \epsilon ^2 \\
z_4\to (3.47+3.83 i)+(3.84+5.81 i) \epsilon +(0.612+0.62 i) \epsilon ^2 \\
z_5\to (3.47+3.83 i)+(0.535+0.828 i) \epsilon -(2.6-0.521 i) \epsilon ^2 \\
\end{array}
\right.\nonumber
\eea
Among them there are two common singular solutions, six solutions for $S_{12}$, six solutions for $S_{45}$. The quantity and the asymptotic behaviours confirm our claim.
The measure part for common singular solutions are given by
\begin{table}[htbp]
\centering
\begin{tabular}{|c|c|}
\hline
Solution   &   measure \\
\hline
Singular $S_{12}S_{45}$ (1) & $\left(3.7\times 10^{11}+\left(6.26\times 10^{11}\right) i\right) \epsilon ^2+O\left(\epsilon ^3\right)$   \\
Singular $S_{12}S_{45}$ (2) & $-\left(1.83\times 10^8+\left(1.23\times 10^9\right) i\right) \epsilon ^2+O\left(\epsilon ^3\right)$   \\
Singular $S_{12}$ (1) & $\left(5.41\times 10^{15}+\left(1.2\times 10^{16}\right) i\right) \epsilon +O\left(\epsilon ^2\right)$   \\
Singular $S_{12}$ (2) & $\left(3.74\times 10^{11}+\left(1.57\times 10^{12}\right) i\right) \epsilon +O\left(\epsilon ^2\right)$   \\
Singular $S_{12}$ (3) & $-\left(1.22\times 10^8-\left(6.38\times 10^7\right) i\right) \epsilon +O\left(\epsilon ^2\right)$   \\
Singular $S_{12}$ (4) & $\left(1.53\times 10^9-\left(1.66\times 10^8\right) i\right) \epsilon +O\left(\epsilon ^2\right)$   \\
Singular $S_{45}$ (1) & $-\left(9.4\times 10^{13}-\left(5.12\times 10^{12}\right) i\right) \epsilon +O\left(\epsilon ^2\right)$   \\
Singular $S_{45}$ (2) & $-\left(3.92\times 10^7+\left(1.17\times 10^9\right) i\right) \epsilon +O\left(\epsilon ^2\right)$   \\
Singular $S_{45}$ (3) & $-\left(9.05\times 10^8+\left(2.25\times 10^9\right) i\right) \epsilon +O\left(\epsilon ^2\right)$   \\
Singular $S_{45}$ (4) & $\left(6.1\times 10^8-\left(4.74\times 10^8\right) i\right) \epsilon +O\left(\epsilon ^2\right)$   \\

\hline
\end{tabular}
\caption{Measure part of each singular solution of $S_{12}=S_{45}=0$}
\label{tab:2 poles measure part 01}
\end{table}
where for common singular solutions it is $\epsilon^{2|A|+2|B|-6}=\epsilon^2$, and for singular solutions of only one pole it is  $\epsilon^{2|A|-3}=\epsilon^{2|B|-3}=\epsilon^1$.

We choose a few CHY-integrands to see the effects of singular solutions. For the first Integrand
  \begin{equation}
    I_1=\frac{1}{z_{12}^2 z_{17}^2 z_{23} z_{24} z_{34} z_{35} z_{37} z_{45} z_{46} z_{56}^2 z_{67}}
  \end{equation}
with $\chi(12)=0$ and $\chi(45)=-1$. The contribution of each singular solution is

\begin{table}[htbp]
\centering
\begin{tabular}{|c|c|}
\hline
Solution   &   Contribution \\
\hline
Singular $S_{12}S_{45}$ (1) & $-\frac{0.00877-0.00326 i}{\epsilon }+O\left(\epsilon ^0\right)$   \\
Singular $S_{12}S_{45}$ (2) & $\frac{0.107+0.13 i}{\epsilon }+O\left(\epsilon ^0\right)$   \\
Singular $S_{12}$ (1) & $\frac{0.00589+0.000659 i}{\epsilon }+O\left(\epsilon ^0\right)$   \\
Singular $S_{12}$ (2) & $\frac{0.00559+0.00778 i}{\epsilon }+O\left(\epsilon ^0\right)$   \\
Singular $S_{12}$ (3) & $-\frac{0.00599-0.0111 i}{\epsilon }+O\left(\epsilon ^0\right)$   \\
Singular $S_{12}$ (4) & $-\frac{0.0197+0.0828 i}{\epsilon }+O\left(\epsilon ^0\right)$   \\
Singular $S_{45}$ (5) & $(0.00198-0.00667 i)+O\left(\epsilon ^1\right)$   \\
Singular $S_{45}$ (6) & $(0.00348+0.00699 i)+O\left(\epsilon ^1\right)$   \\
Singular $S_{45}$ (7) & $(0.0172+0.0341 i)+O\left(\epsilon ^1\right)$   \\
Singular $S_{45}$ (8) & $(0.000144-0.00111 i)+O\left(\epsilon ^1\right)$   \\
All Regulars & $-(0.0285+0.0362 i)-(0.0596+0.314 i) \epsilon +O\left(\epsilon ^2\right)$   \\
\hline
\end{tabular}
\caption{Contribution of each singular solution of $I_1$}
\label{tab:2 poles contribution 01}
\end{table}
The summation is
  \begin{equation}
    \frac{0.0839+0.0704 i}{\epsilon }+O\left(\epsilon ^0\right)
  \end{equation}
while the corresponding amplitude result by substitute numerical kinematics is
  \begin{equation}
  \begin{aligned}
  &\frac{1}{S_{12} S_{56} S_{127} S_{456}}+\frac{1}{S_{17} S_{56} S_{127} S_{456}}\\
  =&\frac{0.0839+0.0704 i}{\epsilon }+(0.0809+0.209 i)-(0.128-0.302 i) \epsilon +O\left(\epsilon ^2\right)
  \end{aligned}
  \end{equation}

The second Integrand as
  \begin{equation}
    I_2=-\frac{1}{z_{12}^2 z_{13} z_{17} z_{23} z_{26} z_{34} z_{35} z_{45}^2 z_{47} z_{56} z_{67}^2}
  \end{equation}
with $\chi(12)=\chi(45)=0$. The contribution of each singular solution is
\begin{table}[htbp]
\centering
\begin{tabular}{|c|c|}
\hline
Solution   &   Contribution \\
\hline
Singular $S_{12}S_{45}$ (1) & $\frac{0.0021+0.00137 i}{\epsilon ^2}+O\left(\frac{1}{\epsilon }\right)$   \\
Singular $S_{12}S_{45}$ (2) & $-\frac{0.00282-0.00584 i}{\epsilon ^2}+O\left(\frac{1}{\epsilon }\right)$   \\
Singular $S_{12}$ (1) & $\frac{0.000623+0.000173 i}{\epsilon }+O\left(\epsilon ^0\right)$   \\
Singular $S_{12}$ (2) & $-\frac{0.0013-0.000968 i}{\epsilon }+O\left(\epsilon ^0\right)$   \\
Singular $S_{12}$ (3) & $-\frac{0.000436+0.000344 i}{\epsilon }+O\left(\epsilon ^0\right)$   \\
Singular $S_{12}$ (4) & $-\frac{0.00963-0.00522 i}{\epsilon }+O\left(\epsilon ^0\right)$   \\
Singular $S_{45}$ (1) & $-\frac{0.0017+0.00104 i}{\epsilon }+O\left(\epsilon ^0\right)$   \\
Singular $S_{45}$ (2) & $-\frac{0.00234+0.00102 i}{\epsilon }+O\left(\epsilon ^0\right)$   \\
Singular $S_{45}$ (3) & $\frac{0.00355-0.00477 i}{\epsilon }+O\left(\epsilon ^0\right)$   \\
Singular $S_{45}$ (4) & $-\frac{0.000814+0.000133 i}{\epsilon }+O\left(\epsilon ^0\right)$   \\
All Regulars & $(0.0112-0.00641 i)+(0.00411+0.0247 i) \epsilon +O\left(\epsilon ^2\right)$   \\
\hline
\end{tabular}
\caption{Contribution of each singular solution of $I_2$}
\label{tab:tab:2 poles contribution 02}
\end{table}
The summation is
  \begin{equation}
    -\frac{0.000722-0.00721 i}{\epsilon ^2}+O\left(\frac{1}{\epsilon }\right)
  \end{equation}
while the amplitude result by substitute numerical kinematics is
  \begin{equation}
  \begin{aligned}
  &-\frac{1}{S_{12} S_{45} S_{67} S_{123}}-\frac{1}{S_{12} S_{45} S_{67} S_{345}}\\
  =&-\frac{0.000722-0.00721 i}{\epsilon ^2}+\frac{0.00438-0.00799 i}{\epsilon }-(0.0122-0.00712 i) +O(\epsilon)
  \end{aligned}
  \end{equation}

From the result of these two Integrands, we can verify the behaviour of the singular solutions. The contribution of the common singular solution is exact $\epsilon^{-2-\chi(12)-\chi(45)}$, $\epsilon^{-1-\chi(12)}$ for the singular solutions of $S_{12}$ and $\epsilon^{-1-\chi(45)}$ for the singular solutions of $S_{45}$, which agree with our analysis.

~\\
{\bf Example II: $n=7,S_{123}=S_{12}=0$}

Next we consider another compatible case $A\subset B$ with $A=\{12\},B=\{123\}$. Our gauge choice is that  $z_i,z_j,z_k\notin B$, thus the type-II behaviour is realized. The kinematics of 7-point construction numerically we choose as
\begin{equation}
    \begin{array}{ll}
    k_1 \rightarrow &\{5.78+1.74 i,21.2-7.05 i,0.742+0.329 i,-3.03+14.8 i,-11.8-17.3 i\}
    +\\ &\{0.638 \epsilon ,(2.15-1.77 i) \epsilon ,0.12 \epsilon ,(0.2+1.95 i) \epsilon ,(-2.23-1.53 i) \epsilon \}\\
    k_2 \rightarrow &\{1.27+0.0208 i,-0.315-1.17 i,0.27+0.0766 i,-0.115+0.124 i,-1.69+0.206 i\}
    +\\ &\{0.106 \epsilon ,(0.0134+1.15 i) \epsilon ,0.389 \epsilon ,(0.817+0.036 i) \epsilon ,(0.726-0.0618 i) \epsilon \}\\
    k_3 \rightarrow &\{0.875,0.379,-0.717,0.329,-0.00131\}\\
    k_4 \rightarrow &\{1.7,-0.972,0.513,-0.912,0.918\}\\
    k_5 \rightarrow &\{1.05-0.0208 i,0.283+1.17 i,-0.234-0.0766 i,0.655-0.124 i,1.41-0.206 i\}
    +\\ &\{-0.106 \epsilon ,(-0.0134-1.15 i) \epsilon ,-0.389 \epsilon ,(-0.817-0.036 i) \epsilon ,(-0.726+0.0618 i) \epsilon \}\\
    k_6 \rightarrow &\{-7.17-1.74 i,-20.1+7.05 i,0.435-0.329 i,5.4-14.8 i,13.6+17.3 i\}
    +\\ &\{-0.638 \epsilon ,(-2.15+1.77 i) \epsilon ,-0.12 \epsilon ,(-0.2-1.95 i) \epsilon ,(2.23+1.53 i) \epsilon \}\\
    k_7 \rightarrow &\{-3.51,-0.482,-1.01,-2.33,-2.37\}\\
    \end{array}
\end{equation}
We fix $z_5=1,z_6=5,z_7=-7$, the singular solutions are
\bea \text{Singular }S_{123}S_{12} \text{ (1)}  & &\left\{
\begin{array}{l}
z_1\to (4.96-0.203 i)+(0.0866+0.743 i) \epsilon +(17.9+39.6 i) \epsilon ^2 \\
z_2\to (4.96-0.203 i)+(0.0866+0.743 i) \epsilon -(94.6-51.5 i) \epsilon ^2 \\
z_3\to (4.96-0.203 i)+(0.502-14. i) \epsilon +(224.-904. i) \epsilon ^2 \\
z_4\to (3.51+1.78 i)+(3.4+1.59 i) \epsilon -(2.95-7.46 i) \epsilon ^2 \\
\end{array}
\right.\nn
 \text{Singular }S_{123}S_{12} \text{ (2)}  & &\left\{
\begin{array}{l}
z_1\to (4.89-0.552 i)+(0.158+2.21 i) \epsilon +(49.+119. i) \epsilon ^2 \\
z_2\to (4.89-0.552 i)+(0.158+2.21 i) \epsilon -(254.-148. i) \epsilon ^2 \\
z_3\to (4.89-0.552 i)+(1.68-37.7 i) \epsilon +(539.-2460. i) \epsilon ^2 \\
z_4\to (1.88+3.74 i)+(3.94+3.07 i) \epsilon +(59.4-0.813 i) \epsilon ^2 \\
\end{array}
\right.\nonumber \eea
\bea
 \text{Singular }S_{12} \text{ (1)}  & &\left\{
\begin{array}{l}
z_1\to (4.9+0.163 i)-(0.148-1.15 i) \epsilon +(0.803+3.69 i) \epsilon ^2 \\
z_2\to (4.9+0.163 i)-(2.89+0.276 i) \epsilon -(18.6-1.77 i) \epsilon ^2 \\
z_3\to (1.46-0.781 i)-(0.66+0.418 i) \epsilon -(3.94+1.17 i) \epsilon ^2 \\
z_4\to (8.4-0.412 i)+(20.3-13.4 i) \epsilon +(55.5-60.9 i) \epsilon ^2 \\
\end{array}
\right.\nn
 \text{Singular }S_{12} \text{ (2)}  & &\left\{
\begin{array}{l}
z_1\to (4.99-0.209 i)-(0.974-1.16 i) \epsilon -(17.+23.2 i) \epsilon ^2 \\
z_2\to (4.99-0.209 i)+(1.95+2.3 i) \epsilon -(72.1+13. i) \epsilon ^2 \\
z_3\to (4.56-0.333 i)+(13.8-7.47 i) \epsilon +(44.9+397. i) \epsilon ^2 \\
z_4\to (3.47+1.82 i)+(4.5-2.73 i) \epsilon +(107.+67.9 i) \epsilon ^2 \\
\end{array}
\right.\nn
 \text{Singular }S_{12} \text{ (3)}  & &\left\{
\begin{array}{l}
z_1\to (4.94-0.54 i)-(3.24-0.525 i) \epsilon +(11.7-45.7 i) \epsilon ^2 \\
z_2\to (4.94-0.54 i)+(3.78+3.17 i) \epsilon -(77.5+23.5 i) \epsilon ^2 \\
z_3\to (3.88-0.729 i)+(28.6-16.3 i) \epsilon +(208.+725. i) \epsilon ^2 \\
z_4\to (1.99+3.72 i)-(1.77-7.54 i) \epsilon -(39.6+131. i) \epsilon ^2 \\
\end{array}
\right.\nn
\text{Singular }S_{12} \text{ (4)}  & &\left\{
\begin{array}{l}
z_1\to (4.81-0.807 i)-(2.27+1.54 i) \epsilon -(13.8+5.69 i) \epsilon ^2 \\
z_2\to (4.81-0.807 i)+(7.88-1.43 i) \epsilon +(28.9-7.46 i) \epsilon ^2 \\
z_3\to (2.35+0.69 i)-(3.98-0.141 i) \epsilon -(15.2-5.13 i) \epsilon ^2 \\
z_4\to (1.46+5.15 i)-(10.9-10. i) \epsilon -(44.2-60.1 i) \epsilon ^2 \\
\end{array}
\right.\nonumber \eea
There are two common singular solutions and  six solutions for $S_{12}$ only. The quantity and asymptotic behaviour are same as our analysis. The measure part is shown in table \eqref{tab:2 poles measure part 02}
\begin{table}[htbp]
\centering
\begin{tabular}{|c|c|}
\hline
Solution   &   Measure \\
\hline
Singular $S_{123}S_{12}$ (1) & $\left(691000.-\left(5.97\times 10^6\right) i\right) \epsilon ^4+O\left(\epsilon ^5\right)$   \\
Singular $S_{123}S_{12}$ (2) & $-\left(3.53\times 10^9-\left(2.51\times 10^9\right) i\right) \epsilon ^4+O\left(\epsilon ^5\right)$   \\
Singular $S_{12}$ (1) & $-(50.8+792. i) \epsilon +O\left(\epsilon ^2\right)$   \\
Singular $S_{12}$ (2) & $(158.+104. i) \epsilon +O\left(\epsilon ^2\right)$   \\
Singular $S_{12}$ (3) & $-(27300.+65900. i) \epsilon +O\left(\epsilon ^2\right)$   \\
Singular $S_{12}$ (4) & $(153000.+66900. i) \epsilon +O\left(\epsilon ^2\right)$   \\
All Regulars & $-\left(84300.+\left(3.87\times 10^6\right) i\right)+\left(3.42\times 10^7-\left(1.04\times 10^7\right) i\right) \epsilon +O\left(\epsilon ^2\right)$   \\
\hline
\end{tabular}
\caption{Measure part of each singular solution of $S_{123}=S_{12}=0$}
\label{tab:2 poles measure part 02}
\end{table}
The measure part of common singular solutions is exact $\epsilon^{2|A|+2|B|-6}=\epsilon^{2\times 2+2\times 3-6}=\epsilon^4$, and $\epsilon^{2|A|-3}=\epsilon^{2\times 2-3}=\epsilon^1$ for the solutions only for pole $S_{12}$.

The first Integrand we have chosen is
  \begin{equation}
    I_1=-\frac{1}{z_{12} z_{13} z_{14} z_{17} z_{23}^2 z_{27} z_{34} z_{45}^2 z_{56}^2 z_{67}^2}
  \end{equation}
with pole index $\chi(123)=0,\chi(12)=-1$. The contribution of each solution is in table\eqref{tab:2 poles contribution 03}
\begin{table}[htbp]
\centering
\begin{tabular}{|c|c|}
\hline
Solution   &   Contribution \\
\hline
Singular $S_{123}S_{12}$ (1) & $-\frac{5.93\times 10^{-7}+\left(6.32\times 10^{-7}\right) i}{\epsilon }+O\left(\epsilon ^0\right)$   \\
Singular $S_{123}S_{12}$ (2) & $\frac{1.62\times 10^{-6}+\left(4.45\times 10^{-7}\right) i}{\epsilon }+O\left(\epsilon ^0\right)$   \\
Singular $S_{12}$ (1) & $1.17\times 10^{-8}+\left(5.19\times 10^{-9}\right) i$   \\
Singular $S_{12}$ (2) & $0.0000351+\left(6.06\times 10^{-6}\right) i$   \\
Singular $S_{12}$ (3) & $-0.0000613+\left(6.88\times 10^{-6}\right) i$   \\
Singular $S_{12}$ (4) & $-2.07\times 10^{-7}+\left(2.57\times 10^{-6}\right) i$   \\
All Regulars & $-(0.000127+0.0000511 i)-(0.0426+0.0299 i) \epsilon +O\left(\epsilon ^2\right)$   \\
\hline
\end{tabular}
\caption{Contribution of each singular solution of $I_1$}
\label{tab:2 poles contribution 03}
\end{table}
The summation is
  \begin{equation}
    \frac{1.03\times 10^{-6}-\left(1.88\times 10^{-7}\right) i}{\epsilon }+O\left(\epsilon ^0\right)
  \end{equation}
We verify the summation with numerical kinematics
  \begin{equation}
  \begin{aligned}
   &-\frac{1}{S_{23} S_{45} S_{67} S_{123}}-\frac{1}{S_{23} S_{45} S_{123} S_{456}}-\frac{1}{S_{23} S_{56} S_{123} S_{456}}-\frac{1}{S_{23} S_{56} S_{123} S_{567}}-\frac{1}{S_{23} S_{67} S_{123} S_{567}}\\
   =&\frac{1.03\times 10^{-6}-\left(1.88\times 10^{-7}\right) i}{\epsilon }+\left(7.19\times 10^{-7}-\left(5.16\times 10^{-6}\right) i\right) +O(\epsilon )
  \end{aligned}
  \end{equation}

The second Integrand is
\begin{equation}
    -\frac{1}{z_{14}^2 z_{16}^2 z_{23}^2 z_{25}^2 z_{34} z_{37} z_{47} z_{56} z_{57} z_{67}}
  \end{equation}
with pole index $\chi(123)=-2,\chi(12)=-2$. The contribution of each solution is in table \eqref{tab:2 poles contribution 04}
\begin{table}[htbp]
\centering
\begin{tabular}{|c|c|}
\hline
Solution   &   Contribution \\
\hline
Singular $S_{123}S_{12}$ (1) & $-(0.00292-0.0564 i) \epsilon ^2+O\left(\epsilon ^3\right)$   \\
Singular $S_{123}S_{12}$ (2) & $(0.0703-0.0516 i) \epsilon ^2+O\left(\epsilon ^3\right)$   \\
Singular $S_{12}$ (1) & $-\left(0.0000246+\left(3.35\times 10^{-6}\right) i\right) \epsilon +O\left(\epsilon ^2\right)$   \\
Singular $S_{12}$ (2) & $(0.00186-0.000129 i) \epsilon +O\left(\epsilon ^2\right)$   \\
Singular $S_{12}$ (3) & $-(0.00148+0.00198 i) \epsilon +O\left(\epsilon ^2\right)$   \\
Singular $S_{12}$ (4) & $-(0.000134-0.000217 i) \epsilon +O\left(\epsilon ^2\right)$   \\
All Regulars & $-(0.000476-0.0000642 i)-(0.000257-0.00181 i) \epsilon +O\left(\epsilon ^2\right)$   \\
\hline
\end{tabular}
\caption{Contribution of each singular solution of $I_2$}
\label{tab:2 poles contribution 04}
\end{table}
The summation is
  \begin{equation}
    -(0.000476-0.0000642 i)-(0.0000339+0.0000789 i) \epsilon +O\left(\epsilon ^2\right)
  \end{equation}
We verify the summation with numerical kinematics:
  \begin{equation}
  \begin{aligned}
   &-\frac{1}{S_{14} S_{23} S_{146} S_{235}}-\frac{1}{S_{16} S_{23} S_{146} S_{235}}-\frac{1}{S_{14} S_{25} S_{146} S_{235}}-\frac{1}{S_{16} S_{25} S_{146} S_{235}}\\
   =& -(0.000476-0.0000642 i)-(0.0000339+0.0000789 i) \epsilon +O\left(\epsilon ^2\right)
  \end{aligned}
  \end{equation}

From the result of these two Integrand, we verify the contribution of common singular solution is $\epsilon^{-2-\chi(12)-\chi(123)}$, and $\epsilon^{-1-\chi(12)}$ for singular solutions of $S_{12}$.

~\\
{\bf Example III: $n=7,S_{123}=S_{234}=0$}

Now we consider the case two poles are not compatible with each other, we choose $A=\{123\}$ and $B=\{234\}$. The numerical kinematics are
\begin{equation}
    \begin{array}{ll}
    k_1 \rightarrow &\{-0.198-0.128 i,-0.0155-4.35 i,-3.65+0.112 i,-0.829-0.0698 i,-2.23-0.138 i\}
    +\\ &\{0.16 \epsilon ,(0.0591+0.504 i) \epsilon ,(0.499-0.06 i) \epsilon ,(0.0301+0.00534 i) \epsilon ,0.172 \epsilon \}\\
    k_2 \rightarrow &\{1.54,0.667,0.963,0.368,-0.931\}\\
    k_3 \rightarrow &\{1.17,-0.525,-0.27,0.459,-0.902\}\\
    k_4 \rightarrow &\{0.524-0.123 i,0.1+0.999 i,0.406-0.0327 i,0.343-0.73 i,-1.23-0.0812 i\}
    +\\ &\{0.839 \epsilon ,(-0.951-0.84 i) \epsilon ,(0.187+0.00511 i) \epsilon ,(-0.942+0.85 i) \epsilon ,0.552 \epsilon \}\\
    k_5 \rightarrow &\{0.951,-0.376,-0.594,0.164,-0.618\}\\
    k_6 \rightarrow &\{-1.17+0.123 i,0.325-0.999 i,1.23+0.0327 i,0.0673+0.73 i,1.12+0.0812 i\}
    +\\ &\{-0.839 \epsilon ,(0.951+0.84 i) \epsilon ,(-0.187-0.00511 i) \epsilon ,(0.942-0.85 i) \epsilon ,-0.552 \epsilon \}\\
    k_7 \rightarrow &\{-2.82+0.128 i,-0.176+4.35 i,1.92-0.112 i,-0.573+0.0698 i,4.78+0.138 i\}
    +\\ &\{-0.16 \epsilon ,(-0.0591-0.504 i) \epsilon ,(-0.499+0.06 i) \epsilon ,(-0.0301-0.00534 i) \epsilon ,-0.172 \epsilon \}\\
    \end{array}
\end{equation}
We fix $z_5=1,z_6=5,z_7=-7$, the singular solutions are
\begin{align*}
& \text{Singular }S_{123} \text{ (1)}  & &\left\{
\begin{array}{l}
z_1\to (-6.59+5.61 i)+(0.988-7.57 i) \epsilon +(13.1+2.66 i) \epsilon ^2 \\
z_2\to (-6.59+5.61 i)+(0.622-8.5 i) \epsilon +(16.5+10.3 i) \epsilon ^2 \\
z_3\to (-6.59+5.61 i)+(0.143-8.57 i) \epsilon +(21.1+11. i) \epsilon ^2 \\
z_4\to (5.71+5.41 i)-(2.95-11.1 i) \epsilon -(12.+9.39 i) \epsilon ^2 \\
\end{array}
\right.\\
& \text{Singular }S_{123} \text{ (2)}  & &\left\{
\begin{array}{l}
z_1\to (-5.2-0.937 i)-(0.505+0.588 i) \epsilon -(1.03+0.394 i) \epsilon ^2 \\
z_2\to (-5.2-0.937 i)-(0.839+0.422 i) \epsilon -(0.412+0.42 i) \epsilon ^2 \\
z_3\to (-5.2-0.937 i)-(0.852+0.241 i) \epsilon -(0.226+0.714 i) \epsilon ^2 \\
z_4\to (1.42-1.1 i)-(3.76+0.476 i) \epsilon -(2.6+0.484 i) \epsilon ^2 \\
\end{array}
\right.\\
& \text{Singular }S_{234} \text{ (1)}  & &\left\{
\begin{array}{l}
z_1\to (0.948+9. i)-(14.1+3.67 i) \epsilon -(40.7+39.5 i) \epsilon ^2 \\
z_2\to (6.25+3.19 i)-(0.79-13.5 i) \epsilon -(31.8-47.6 i) \epsilon ^2 \\
z_3\to (6.25+3.19 i)-(3.64-7.5 i) \epsilon -(32.8-22.1 i) \epsilon ^2 \\
z_4\to (6.25+3.19 i)+(0.0579+8.31 i) \epsilon -(20.6-16.6 i) \epsilon ^2 \\
\end{array}
\right.\\
& \text{Singular }S_{234} \text{ (2)}  & &\left\{
\begin{array}{l}
z_1\to (-6.25-0.445 i)+(7.43+7.67 i) \epsilon +(11.3-45.2 i) \epsilon ^2 \\
z_2\to (-2.7-1.43 i)-(4.19-3.88 i) \epsilon +(37.1+13.9 i) \epsilon ^2 \\
z_3\to (-2.7-1.43 i)+(28.6+19.4 i) \epsilon -(9.49+134. i) \epsilon ^2 \\
z_4\to (-2.7-1.43 i)+(10.2+28.8 i) \epsilon +(49.1-127. i) \epsilon ^2 \\
\end{array}
\right.
\end{align*}
The quantity and asymptotic behaviour is conform with our analysis. The measure part given in the table
\begin{table}[htbp]
\centering
\begin{tabular}{|c|c|}
\hline
Solution   &   Measure  \\
\hline
Singular $S_{123}$ (1) & $\left(6.05\times 10^6-\left(7.15\times 10^6\right) i\right) \epsilon ^3+O\left(\epsilon ^4\right)$   \\
Singular $S_{123}$ (2) & $(1020.+829. i) \epsilon ^3+O\left(\epsilon ^4\right)$   \\
Singular $S_{234}$ (1) & $\left(4.46\times 10^8+\left(2.5\times 10^8\right) i\right) \epsilon ^3+O\left(\epsilon ^4\right)$   \\
Singular $S_{234}$ (2) & $\left(2.8\times 10^9+\left(6.32\times 10^9\right) i\right) \epsilon ^3+O\left(\epsilon ^4\right)$   \\
All Regulars & $-\left(1.62\times 10^{11}+\left(2.12\times 10^{11}\right) i\right)+\left(5.05\times 10^{11}+\left(8.03\times 10^{12}\right) i\right) \epsilon +O\left(\epsilon ^2\right)$   \\
\hline
\end{tabular}
\caption{Measure part of each singular solution of $S_{123}=S_{234}=0$}
\label{tab:2 poles measure 03}
\end{table}
shows $\epsilon^{2|A|-3}=\epsilon^{2|B|-3}=\epsilon^{2\times 3-3}=\epsilon^3$, which fit our analysis.

We choose the first Integrand as
  \begin{equation}
    I_1=-\frac{1}{z_{12} z_{13} z_{14} z_{17} z_{23}^2 z_{27} z_{34} z_{45}^2 z_{56}^2 z_{67}^2}
  \end{equation}
with pole index $\chi(123)=0,\chi(234)=-1$. The contribution of each singular solution is in table\eqref{tab:2 poles contribution 05}.
\begin{table}[htbp]
  \centering
  \begin{tabular}{|c|c|}
  \hline
  Solution   &   Integration \\
  \hline
  Singular $S_{123}$ (1)  & $\frac{0.00281+0.054 i}{\epsilon }+O\left(\epsilon ^0\right)$   \\
  Singular $S_{123}$ (2) & $\frac{0.185-0.331 i}{\epsilon }+O\left(\epsilon ^0\right)$   \\
  Singular $S_{234}$ (1) & $0.0000327+0.000439 i$   \\
  Singular $S_{234}$ (2) & $-0.00337+0.0353 i$   \\
  All Regulars & $-(0.0143+0.0791 i)+(1.62-0.868 i) \epsilon +O\left(\epsilon ^2\right)$   \\
  \hline
  \end{tabular}
  \caption{The contribution of each singular solution of $I_1$}
  \label{tab:2 poles contribution 05}
  \end{table}
The summation
  \begin{equation}
    \frac{0.188-0.277 i}{\epsilon }+O\left(\epsilon ^0\right)
  \end{equation}
and be  verified with the amplitude after putting the  kinematics
  \begin{equation}
  \begin{aligned}
   &-\frac{1}{S_{23} S_{45} S_{67} S_{123}}-\frac{1}{S_{23} S_{45} S_{123} S_{456}}-\frac{1}{S_{23} S_{56} S_{123} S_{456}}-\frac{1}{S_{23} S_{56} S_{123} S_{567}}-\frac{1}{S_{23} S_{67} S_{123} S_{567}}\\
   =&\frac{0.188-0.277 i}{\epsilon }+(1.78-0.469 i)+(6.24+3.76 i) \epsilon +O\left(\epsilon ^2\right)
  \end{aligned}
  \end{equation}

We choose the second Integrand as
  \begin{equation}
   I_2= -\frac{1}{z_{12} z_{15} z_{17}^2 z_{23}^2 z_{25} z_{34}^2 z_{45} z_{46} z_{56} z_{67}^2}
  \end{equation}
with pole index $\chi(123)=-1,\chi(234)=0$. The contribution of each singular solution is
\begin{table}[htbp]
\centering
\begin{tabular}{|c|c|}
\hline
Solution   &   Contribution \\
\hline
Singular $S_{123}$ (1) & $(0.00415-0.000281 i)+O\left(\epsilon ^1\right)$   \\
Singular $S_{123}$ (2) & $-(0.00517-0.0029 i)+O\left(\epsilon ^1\right)$   \\
Singular $S_{234}$ (1) & $-\frac{0.000857-0.00062 i}{\epsilon }+O\left(\epsilon ^0\right)$   \\
Singular $S_{234}$ (2) & $-\frac{0.0000737+0.00856 i}{\epsilon }+O\left(\epsilon ^0\right)$   \\
All Regulars & $-(0.241-0.0685 i)-(8.06+9.33 i) \epsilon +O\left(\epsilon ^2\right)$   \\
\hline
\end{tabular}
\caption{The contribution of each singular solution of $I_2$}
\label{tab:2 poles contribution 06}
\end{table}
The summation is
  \begin{equation}
    -\frac{0.000931+0.00794 i}{\epsilon }+O\left(\epsilon ^0\right)
  \end{equation}
which are same as the amplitude after putting the kinematics
  \begin{equation}
  \begin{aligned}
     & -\frac{1}{S_{17} S_{23} S_{167} S_{234}}-\frac{1}{S_{17} S_{34} S_{167} S_{234}}-\frac{1}{S_{23} S_{67} S_{167} S_{234}}-\frac{1}{S_{34} S_{67} S_{167} S_{234}}\\
     =&-\frac{0.000931+0.00794 i}{\epsilon }+(0.00343-0.0000467 i)-(0.0108+0.0421 i) \epsilon +O\left(\epsilon ^2\right)
  \end{aligned}
  \end{equation}

We choose the third Integrand as
  \begin{equation}
    I_3= \frac{1}{z_{12}^2 z_{15} z_{17} z_{23}^2 z_{34}^2 z_{45} z_{47} z_{56}^2 z_{67}^2}
  \end{equation}
with pole index $\chi(123)=\chi(234)=0$. The contribution of each singular solution is in table \eqref{tab:2 poles contribution 07}
\begin{table}[htbp]
\centering
\begin{tabular}{|c|c|}
\hline
Solution   &   Contribution \\
\hline
  Singular $S_{123}$ (1) & $-\frac{0.0177+0.0131 i}{\epsilon }+O\left(\epsilon ^0\right)$   \\
  Singular $S_{123}$ (2) & $-\frac{0.0223+0.00101 i}{\epsilon }+O\left(\epsilon ^0\right)$   \\
  Singular $S_{234}$ (1) & $-\frac{0.000521-0.000342 i}{\epsilon }+O\left(\epsilon ^0\right)$   \\
  Singular $S_{234}$ (2) & $-\frac{0.00196-0.00286 i}{\epsilon }+O\left(\epsilon ^0\right)$   \\
  All Regulars & $(0.1+0.0492 i)+(0.903+5.43 i) \epsilon +O\left(\epsilon ^2\right)$   \\
\hline
\end{tabular}
\caption{The contribution of each singular solution of $I_3$}
\label{tab:2 poles contribution 07}
\end{table}
The summation is
  \begin{equation}
    -\frac{0.0424+0.0109 i}{\epsilon }+O\left(\epsilon ^0\right)
  \end{equation}
and the amplitude after putting the kinematics is
  \begin{equation}
  \begin{aligned}
     &\frac{1}{S_{12} S_{34} S_{56} S_{567}}+\frac{1}{S_{12} S_{34} S_{67} S_{567}}+\frac{1}{S_{12} S_{56} S_{123} S_{567}}+\frac{1}{S_{23} S_{56} S_{123} S_{567}}+\frac{1}{S_{12} S_{67} S_{123} S_{567}}+\\
     &\frac{1}{S_{23} S_{67} S_{123} S_{567}}+\frac{1}{S_{23} S_{56} S_{234} S_{567}}+\frac{1}{S_{34} S_{56} S_{234} S_{567}}+\frac{1}{S_{23} S_{67} S_{234} S_{567}}+\frac{1}{S_{34} S_{67} S_{234} S_{567}}\\
     =& -\frac{0.0424+0.0109 i}{\epsilon }-(0.154+0.0133 i)-(0.397-0.0194 i) \
     \epsilon +O\left(\epsilon ^2\right)
  \end{aligned}
  \end{equation}
We found that the contribution of each singular solution is exact $\epsilon^{-1-\chi(A)}$ and $\epsilon^{-1-\chi(B)}$ respectively.

\section{From singular kinematics to solutions II: the soft  limit}

Now we consider the second type of singularity, the soft limit, for example, the $k_1\to 0$. Different from preview case, when $k_1\to 0$, we have $S_{1i}\to 0$ for all $i=2,...,n$ automatically. If we naively use the picture from the previous section, we will conclude that there will be some singular solutions with $z_2\to z_1$ and some other solutions with $z_3\to z_1$, etc. Since poles $S_{1i}$ are not compatible, the singular solutions will be the union of all these $z_i\to z_1$. However, as will be shown in this section, the above naive picture is wrong. In fact, for the soft limit, there is no singular solution.

With the behaviour  $k_1=k_1^{(1)}\epsilon$ and $k^{(0)}_{a\neq 1}\neq 0$ when $\eps\to 0$, the
scattering equation ${\cal E}_1$ is
\begin{equation}
    \mathcal{E}_1=\epsilon[\sum_{\substack{a\in\ \{2,...,N\}}}\frac{k_1^{(1)}\cdot k_a^{(0)}}{z_1-z_a}+\mathcal{O}(\epsilon)]=0.
    \label{eq:E1ofsoft}
\end{equation}
Assuming there is at most one location, for example, $z_2\to z_1$ while other $z_i$'s are separated from each other, the term $\frac{k_1^{(1)}\cdot k_2^{(0)}}{z_1-z_2}$ in \eref{eq:E1ofsoft} will be
singular comparing to other terms, since in general $k_a^{(0)}\neq 0, k_1^{(1)}\cdot k_a^{(0)}\neq 0 $ and all numerators in \eqref{eq:E1ofsoft} are nonzero. Thus \eref{eq:E1ofsoft} could not be satisfied under above assumption.

There are two possibilities to make \eref{eq:E1ofsoft} true. The first possibility is that there are at least two locations, for example, $z_i,z_j\to z_1$. Then by our careful analysis given in section two, we will have $S_A\to 0$ with $3\leq |A|\leq n-3$, which conflicts with our kinematic condition that under $k_1\to 0$, only  $S_{1i}\to 0$.
Thus there is left with only another possibility, i.e., all $z_i$'s are separated from each other and $z_1$, so all terms in \eqref{eq:E1ofsoft} will have equivalent weight to make it zero.

Having proved that there is no singular solution for scattering equations, we need to understand how soft singularities are produced in the amplitude. The key is the measure part. Unlike the one $J\sim \eps^{2|A|-3}\to 0$ given in \eref{measure-BCFW-1}, for soft limit
\bea & & (z_1-z_2)(z_2-z_3)(z_3-z_1)(z_n-z_{n-1})(z_{n-1}-z_{n-2})(z_{n-2}-z_n)\sim \epsilon^0\nn
& & |\Phi|^{rst}_{ijk}\to \eps^1,~~~~~~J\sim \eps^{-1}\eea
Thus it is the measure providing the singularity.

Now we present an example to demonstrate the behaviour of soft limit.
First we construct the kinematics for the soft limit. According to the discussion given in the Appendix \ref{kine}, we construct $\W p_i, i=2,...,n$ such that $\sum_{i=2}^{n} \widetilde{p}_{i}=0$, then we choose $u,v$ such that
\bea p_{1}=\epsilon(u+v),~~~p_{2}=\widetilde{p}_{2}-\epsilon u \quad
    p_{3}=\widetilde{p}_{3}-\epsilon v,~~~~p_{a}=\W p_a,~~a=4,...,n\eea
The $u$ and $v$ are not arbitrary. To make all momenta on-shell, they need to satisfy
\bea u \cdot v=0 \quad u^{2}=v^{2}=0,~~~
            u \cdot \widetilde{p}_{2}=0,~~~~
            v \cdot \widetilde{p}_{3}=0 \eea
For space-time dimension $D\geq 4$, there are solutions for $u,v$.

Using above frame, let us present one example, i.e., the six point case with following choice of
kinematics
\begin{equation}
 \begin{aligned}
 k_1\to &\{0.774 ,-2.039+1.645 i ,-2.358-2.948 i ,2.216\, -1.623 i ,-0.0675\}\epsilon,\\
 k_2\to &\{1.125,0.595,-0.105,0.802,0.507\}\\
 +&\{-0.162,-0.876,-0.059-0.969 i,0.454-0.127 i,-0.0628i\}\epsilon,\\
 k_3\to &\{0.950,-0.683,-0.0312,-0.572,-0.329\}\\
 +&\{-0.612,2.915-1.645 i,2.417+3.917 i,-2.668+1.750 i,0.130\}\epsilon\\
 k_4\to &\{1.17,-0.617,-0.284,-0.951,0.0575\},\\
 k_5\to &\{-1.247,-0.299,0.288,-0.936,-0.71\},\\
 k_6\to &\{-1.999,1.003,0.132,1.66,0.475\}
 \end{aligned}
 \label{eq:numer k 004}
 \end{equation}
We fix the gauge choice $z_4=1,z_5=0,z_6=-1$ and find following six solutions

 \begin{equation}
      \begin{aligned}
         & \left\{
         \begin{aligned}
             & z_1\to (-0.236755-0.255543 i)-(1.47022\, -0.800113 i) \epsilon+O\left(\epsilon^2\right)\\
             & z_2\to (-0.0108264+0.011729 i)+(0.0134425\, +0.0210137 i) \epsilon+O\left(\epsilon^2\right)\\
             & z_3\to (-0.28117-0.303133 i)-(1.54701\, -0.967519 i) \epsilon+O\left(\epsilon^2\right)
         \end{aligned}
         \right.
         \\
         & \left\{
         \begin{aligned}
             & z_1\to (-0.00979118+0.011149 I)-(0.0327251 +0.0713408 I) \epsilon+O\left(\epsilon^2\right)\\
             & z_2\to (-0.0108264+0.011729 i)+(0.0134425\, +0.0210137 i)\epsilon+ O\left(\epsilon^2\right)\\
             & z_3\to (-0.28117-0.303133 i)-(1.54701\, -0.967519 i) \epsilon+O\left(\epsilon^2\right)
         \end{aligned}
         \right.
         \\
         & \left\{
         \begin{aligned}
             & z_1\to (0.261008\, +0.153645 i)-(0.587951\, -0.0543757 i) \epsilon+O\left(\epsilon^2\right)\\
             & z_2\to (-0.0108264+0.011729 i)+(0.0134425\, +0.0210137 i) \epsilon+O\left(\epsilon^2\right)\\
             & z_3\to (-0.28117-0.303133 i)-(1.54701\, -0.967519 i) \epsilon+O\left(\epsilon^2\right)
         \end{aligned}
         \right.
         \end{aligned}
         \label{eq:soft limit sol 001}
  \end{equation}

 \begin{equation}
             \begin{aligned}
             & \left\{
             \begin{aligned}
                 &z_1\to (-0.19289+0.33945 i)-(0.36536\, +0.659837 i) \epsilon+O\left(\epsilon^2\right)\\
                 &z_2\to (-0.0108264-0.011729 i)+(0.017517\, +0.0426134 i) \epsilon+O\left(\epsilon^2\right)\\
                 &z_3\to (-0.28117+0.303133 i)-(0.225438\, +0.604153 i) \epsilon+O\left(\epsilon^2\right)
             \end{aligned}
             \right.
             \\
             & \left\{
             \begin{aligned}
                 &z_1\to (-0.010308-0.0107239 i)+(0.130698 +0.227897 i) \epsilon+O\left(\epsilon^2\right)\\
                 &z_2\to (-0.0108264-0.011729 i)+(0.017517\, +0.0426134 i) \epsilon+O\left(\epsilon^2\right)\\
                 &z_3\to (-0.28117+0.303133 i)-(0.225438\, +0.604153 i) \epsilon+O\left(\epsilon^2\right)
             \end{aligned}
             \right.
             \\
             & \left\{
             \begin{aligned}
                 &z_1\to (0.214106 +0.185642 i)-(0.276318 +0.500216 i) \epsilon+O\left(\epsilon^2\right)\\
                 &z_2\to (-0.0108264-0.011729 i)+(0.017517\, +0.0426134 i) \epsilon+O\left(\epsilon^2\right)\\
                 &z_3\to (-0.28117+0.303133 i)-(0.225438\, +0.604153 i) \epsilon+O\left(\epsilon^2\right)
             \end{aligned}
             \right.
             \end{aligned}\label{eq:soft limit sol 002}
  \end{equation}

We have presented these six solutions into two groups. When checking carefully, one can see that the values of $z_2,z_3$ are the same in each group.  In fact, as shown in \cite{Cachazo:2013hca}, for the general case of $n$ points with soft limit $k_1\to 0$, all $(n-3)!$ solutions can divided into $(n-4)!$ groups: in each group  the values of $z_i,i=2,3,...,n$ are same while the $(n-3)$ values of $z_1$ are different. More explicitly, at the leading order, the scattering equations become
\begin{equation}
    \mathcal{E}_a=\sum_{b=2, b\neq a}^n \frac{k_a\cdot k_b}{z_a-z_b}~~~~~\label{eq:E1ofsoft-1}
\end{equation}
for $a=2,3,...,n$ and
\begin{equation}
    \mathcal{E}_1=\epsilon[\sum_{\substack{a\in\ \{2,...,N\}}}\frac{k_1^{(1)}\cdot k_a^{(0)}}{z_1-z_a}+\mathcal{O}(\epsilon)]=0.
    ~~~~~\label{eq:E1ofsoft-2}
\end{equation}
Using \eref{eq:E1ofsoft-1} we get $(n-4)!$ different solutions for $z_i,i=2,3,...,n$. Putting them to \eref{eq:E1ofsoft-2} we get $(n-3)$ solutions for $z_1$.

Using \eref{eq:soft limit sol 001} and \eref{eq:soft limit sol 002}  the measure $
     J=\frac{z_{12}z_{23}z_{31}z_{45}z_{56}z_{64}}{|\Phi|^{123}_{456}}$ of
 each solution is
\begin{small}
         \begin{equation}
             \begin{aligned}
                 & J_1=-\frac{8.28657 \times 10^{-6}+5.97512 \times 10^{-6} i}{\epsilon}-(0.00010047-0.000143733 i)+\mathcal{O}(\epsilon)\\
                 & J_2=\frac{9.27966 \times 10^{-9}+1.71473 \times 10^{-8} i}{\epsilon}+\left(6.26664 \times 10^{-7}-5.07552 \times 10^{-9} i\right)+\mathcal{O}(\epsilon)\\
                 & J_3=\-\frac{0.0000888246+0.0000219678 i}{\epsilon}-(0.00413616-0.00188736 i)+\mathcal{O}(\epsilon)\\
                 & J_4=-\frac{7.12674 \times 10^{-6}+0.0000270679 i}{\epsilon}+(0.000134457+0.000191776 i)+\mathcal{O}(\epsilon)\\
                 & J_5=\frac{1.73122 \times 10^{-8}-4.0953 \times 10^{-9} i}{\epsilon}-\left(5.04811 \times 10^{-7}+1.2758 \times 10^{-7} i\right)+\mathcal{O}(\epsilon)\\
                 & J_6=\frac{0.0000631231-0.0000764316 i}{\epsilon}-(0.0000541201+0.00243992 i)+\mathcal{O}(\epsilon)
             \end{aligned}
         \end{equation}
     \end{small}
where each one has the leading behaviour ${1\over \eps}$ indeed.
Now we choose two different CHY-Intergrands:  one contains pole $S_{1a}$ and  another does not.
\begin{itemize}
    \item $I_{CHY}=\frac{1}{z_{12}^2 z_{23} z_{25} z_{34}^2 z_{41} z_{45} z_{56}^2 z_{61} z_{63}}$ with pole $S_{12}$.  The analytic result gives
     \begin{equation}
     \frac{1}{S_{12}S_{34}S_{56}}=\frac{1.83042-0.257211 i}{\epsilon}+(6.21704-21.8651 i)+O\left(\epsilon\right)
     \label{eq:amp soft 01}
 \end{equation}
after substituting \eqref{eq:numer k 004}. The contributions of each solution are

    \begin{small}
     \begin{equation}
         \begin{aligned}
             & S_1: ~~~~~\frac{0.00419196-0.00814611 i}{\epsilon}-(0.00751988-0.0171901 i)+O\left(\epsilon\right)\\
             & S_2:~~~~~\frac{0.635358-1.40928 i}{\epsilon}-(18.0379+34.6097 i)+O\left(\epsilon\right)\\
             & S_3: ~~~~~\frac{0.0838495-0.0761965 i}{\epsilon}-(0.293397+2.66308 i)+O\left(\epsilon^2\right)\\
             & S_4: ~~~~~-\frac{0.0173534+0.0060881 i}{\epsilon}+(0.0108633+0.0298797 i)+O\left(\epsilon\right)\\
             & S_5:~~~~~ \frac{1.05731+1.13749 i}{\epsilon}+(23.2755+14.1021 i)+O\left(\epsilon\right)\\
             & S_6:~~~~~-\frac{0.0670696+0.105014 i}{\epsilon}+(1.26947+1.25845 i)+O\left(\epsilon\right)\\
             & \sum_{i=1}^6 S_i:~~~~~\frac{1.83042-0.257211 i}{\epsilon}+(6.21704-21.8651 i)+O\left(\epsilon\right)
         \end{aligned}
     \end{equation}
 \end{small}

\item $I_{CHY}=\frac{1}{z_{12} z_{13} z_{23} z_{25} z_{32} z_{34} z_{41} z_{45} z_{56}^2 z_{61} z_{64}}$ withour the pole $S_{1a}$. The analytic expression gives
 \begin{equation}
     \frac{1}{S_{23}S_{56}S_{123}}=-0.00611146-(0.00176926 +0.000337693 i) \epsilon+O\left(\epsilon^2\right)
 \end{equation}
 which is not divergence  when $\epsilon\to 0$. The contributions of each solution are
 \begin{small}
     \begin{equation}
         \begin{aligned}
             & S_1: \frac{0.0294162+0.0533801 i}{\epsilon}-(0.15429+0.0531782 i)-(0.399708-0.492001 i) \epsilon+O\left(\epsilon^2\right)\\
             & S_2:-\frac{0.00502886-0.00214344 i}{\epsilon}-(0.0153742+0.0794493 i)+(3.43002+0.885407 i) \epsilon+O\left(\epsilon^2\right)\\
             & S_3: -\frac{0.0243873+0.0555235 i}{\epsilon}+(0.166609+0.697732 i)-(37.8097+16.3499 i) \epsilon+O\left(\epsilon^2\right)\\
             & S_4: \frac{0.0936548-0.0110378 i}{\epsilon}-(0.0469898+0.113523 i)+(0.563917-0.0892481 i) \epsilon+O\left(\epsilon^2\right)\\
             & S_5: -\frac{0.00413562+0.00322131 i}{\epsilon}-(0.0000649134-0.0756482 i)-(0.337054-0.169771 i) \epsilon+O\left(\epsilon^2\right)\\
             & S_6:-\frac{0.0895192-0.0142591 i}{\epsilon}+(0.043999-0.527229 i)+(2.47751+42.7249 i) \epsilon+O\left(\epsilon^2\right)\\
             & \sum_{i=1}^6 S_i:~~-0.00611146-(0.00176926 +0.000337693 i) \epsilon+O\left(\epsilon^2\right)
         \end{aligned}
     \end{equation}
 \end{small}

For this CHY-integrand, the critical point is how the cancellation of ${1\over \eps}$ part happens. When checking carefully, one can see the cancellation happens inside each group, i.e., for $S_1+S_2+S_3$ and $S_4+S_5+S_6$, the  ${1\over \eps}$ part disappears. This is a general phenomenon, and we can easily prove using the trick given in \cite{Cachazo:2013hca}:
\begin{equation}
\begin{aligned}
    &A_{n}(1,2,...,n)\\ =&\int\prod_{\substack{i=1}}^{n-3}d z_{i}\  z_{n-2, n-1}^{2} z_{n-1, n}^{2} z_{n, n-2}^{2} \prod_{a \neq n-2, n-1, n}  \delta\left(\mathcal{E}_{a}\right) PT\left(1,\alpha_{2}, \cdots,\alpha_{n}\right)PT\left(1,\beta_{2}, \cdots,\beta_{n}\right) \\
    = &\int\prod_{\substack{i=1}}^{n-3}d z_{i} \delta\left(E_{1}\right) \frac{z_{\alpha_n\alpha_{2}} }{z_{1\alpha_2} z_{1\alpha_{n}}} \frac{z_{\beta_{n\beta_{2}}} }{z_{1\beta_{2}}  z_{1\beta_{n}} }\\
    &\times \prod_{a\neq1, n-2,\atop n-1, n} \delta\left(\mathcal{E}_{a}\right) PT\left(\alpha_{2}, \cdots,\alpha_{n}\right)PT\left(\beta_{2}, \cdots,\beta_{n}\right)z_{n-2, n-1}^{2} z_{n-1, n}^{2} z_{n, n-2}^{2}\\
    = &\int d z_{1} \delta\left(E_{1}\right) \frac{z_{\alpha_n\alpha_{2}} }{z_{1\alpha_2} z_{1\alpha_{n}}} \frac{z_{\beta_{n\beta_{2}}} }{z_{1\beta_{2}}  z_{1\beta_{n}} }\\ &\times\int\prod_{\substack{i=2}}^{n-3}d z_{i}\prod_{a\neq1, n-2,\atop n-1, n} \delta\left(\mathcal{E}_{a}\right) PT\left(\alpha_{2}, \cdots,\alpha_{n}\right)PT\left(\beta_{2}, \cdots,\beta_{n}\right)z_{n-2, n-1}^{2} z_{n-1, n}^{2} z_{n, n-2}^{2}\\
    = &\int d z_{1} \delta\left(E_{1}\right) \frac{z_{\alpha_n\alpha_{2}} }{z_{1\alpha_2} z_{1\alpha_{n}}} \frac{z_{\beta_{n\beta_{2}}} }{z_{1\beta_{2}}  z_{1\beta_{n}} } A_{n-1}(2,3,...,n)
\end{aligned}
\label{eq:proof of chy integral}
\end{equation}
where at the leading part of $\eps$, the $\prod_{a\neq1, n-2,\atop n-1, n} \delta\left(\mathcal{E}_{a}\right)$ takes the form \eref{eq:E1ofsoft-1} so
$A_{n-1}(2,3,...,n)$ can be treated as a constant.
To carry out the  contour integration of $z_1$, using \eref{eq:E1ofsoft-2} we write $\mathcal{E}_1=\eps \sum_{i=2}^n\frac{k_1^{(1)}\cdot k_i^{(0)}}{z_1-z_{i}}=\eps\frac{f(z_1;z_2,...,z_n)}{z_{12}z_{13}\cdots z_{1n}}$, thus
\begin{equation}
\begin{aligned}
    & \int d z_{1} \  \delta\left(\mathcal{E}_{1}\right) \frac{z_{\alpha_n\alpha_{2}} }{z_{1\alpha_2} z_{1\alpha_{n}}} \frac{z_{\beta_{n\beta_{2}}} }{z_{1\beta_{2}}  z_{1\beta_{n}} }\\
    &=\oint _{\Gamma_f} d z_{1} \frac{1}{\mathcal{E}_{1}} \frac{z_{\alpha_n\alpha_{2}} }{z_{1\alpha_2} z_{1\alpha_{n}}} \frac{z_{\beta_{n\beta_{2}}} }{z_{1\beta_{2}}  z_{1\beta_{n}} }\\
    &=\oint _{\Gamma_f} d z_{1} \frac{z_{12} z_{13} \cdots z_{1 n}}{\eps f\left(z_{1} ; z_{2}, \cdots, z_{n}\right)}
    \frac{z_{\alpha_n\alpha_{2}} }{z_{1\alpha_2} z_{1\alpha_{n}}} \frac{z_{\beta_{n\beta_{2}}} }{z_{1\beta_{2}}  z_{1\beta_{n}}}
\end{aligned}
\label{soft-z1}
\end{equation}
Where the contour $\Gamma_f$ means taking the sum only over poles coming from $f$. One can see that because the numerator $z_{12} z_{13} \cdots z_{1 n}$, as long as all $\a_1,\a_n,\b_2,\b_n$ are different, there is no extra pole in the denominator besides these coming from $f$, thus by the Cauchy residue theorem\footnote{The large $z_1$ behaviour is ${z_1^{n-1}\over z_1^{(n-3)+4}}\sim {1\over z_1^2}$, so there is no boundary contribution at $z_1=\infty$.}, the integral is zero, which is exact  the case of the CHY-Integrand
 containing no poles of $S_{1a}$.
On the other hand, if there are at least two labels of $\alpha_2$,$\alpha_n$,$\beta_2$,$\beta_n$  same, the integral is nonzero by the extra poles in the denominator,  which corresponding to the CHY-Integrand with pole $S_{1a}$.
One further point is that by \eref{soft-z1}, no matter if the $S_{1a}$ is a single-pole or poles of higher degrees, the leading behaviour is always ${1\over \eps}$.

\end{itemize}
%

\section{From singular kinematics to solutions III: the forward  limit}

The last singular kinematics we will discussed in the paper is the forward limit
\begin{equation}
    k_1+k_2=\epsilon q\to 0.~~~~\label{f-limit}
\end{equation}
As we have mentioned in the section on motivation, the forward limit is very 
useful when discussing the loop-level CHY-integrands. In this section, we will 
give a general disucssion of forward limit first, and then apply it to the one-loop
CHY-integrals. 

\subsection{The general discussion}

Similarly to the soft limit, under the limit \eref{f-limit} we will have $S_{12i}\to 0$ for all $i=3,...,n$ automatically under the forward limit \eref{f-limit}.
As we will see, the forward limit can be considered as the mixture of the factorization limit and the soft limit. While all poles $S_{12i}$ are not compatible, all of them contain the common pole $S_{12}\to 0$. Now we ask what the implication of the singular kinematics to the is
the behaviour of solutions of scattering equations?

Let us start with the singular kinematics $S_{12}\to 0$. Unlike the factorization limit
where $S_A\to \eps^1$, the \eref{f-limit} implies $S_{12}\to \eps^2$. The difference will lead to the different behaviour of singular solutions,
As given in \cite{He:2015yua}, He and Yuan have shown that there are two kinds of singular solutions
\begin{equation}
    \begin{aligned}
        Singular-I:\ \ |z_1-z_2|\sim \epsilon^1,\\
        Singular-II:\ \ |z_1-z_2|\sim \epsilon^2.
    \end{aligned}
    \label{f-limit-2}
\end{equation}
where each kind has $(n-4)!$ solutions. In other words, like the factorization limit, the singular kinematics $S_{12}\to 0$ implies the singular solution. The counting of $(n-4)!$ of each type matches the counting of factorization limit of pole $S_{12}$ too. However, unlike the
factorization limit with only one type of singular solutions, there are two types of singular solutions, especially type II with the limit behaviour $|z_1-z_2|\sim \epsilon^2$~~\footnote{ A naive understanding is that \eref{f-limit} implies there are two kinematic singularities simultaneously. The first one is the
$(k_1+k_2)\to \eps q$, which is the soft limit type. The second one is the $(k_1-k_2)\to 2k_1$, which is
more like the factorization limit.}.

Now we move to the automatic singular kinematics $S_{12i}\to 0$. Do they imply the singular solution of $z_1,z_2,z_i\to z$? With the experience from the soft limit, we need to do more careful analysis. Assuming there is a singular solution where, for example, $z_3$ tend to $z_{1,2}$ while all other $z_i$ are separated from each other.
In the scattering equation
\begin{equation}
    \mathcal{E}_1=\frac{k_1\cdot k_2}{z_1-z_2}+\frac{k_1\cdot k_3}{z_1-z_3}+\frac{k_1\cdot k_4}{z_1-z_4}+\cdots +\frac{k_1\cdot k_n}{z_1-z_n}=0.
\end{equation}
since $k_1\cdot k_2=\frac{\eps^2}{2}q$, term $\frac{k_1\cdot k_2}{z_1-z_2}\sim \eps^1/\eps^0$ depending if it is singular I or singular II solutions, while other terms $\frac{k_1\cdot k_i}{z_1-z_i}\sim \eps^0, i\neq 2,3$ because $k_1\cdot k_i\sim \eps^0$ and $|z_1-z_i|\sim \eps^0$. But the term $\frac{k_1\cdot k_3}{z_1-z_3}$ will be singular because $z_3\to z_1$, unless $S_{13}=2k_1\cdot k_3\to 0 $, which is conflict with our
kinematic configuration. Therefore, we learned that there is no other $z_i(i\neq 1,2)$ tend to $z_{1,2}$. The only possible asymptotic behaviour of singular solutions must be \eqref{eq:forward limit BH}.
\begin{figure}
  \centering
  \begin{minipage}[thtbp]{0.4\linewidth}
  \centering
  \begin{tikzpicture}
      \draw (-0.5,0)--(0.5,0);
      \draw [fill] (0,1) circle [radius=0.05];
      \node [] at (0,0.6) {$ \vdots $};
      \draw [fill] (0,0) circle [radius=0.05];
      \draw (-0.5,1)--(0.5,1);
      \draw [decorate,decoration={calligraphic brace,mirror,amplitude=2mm},thick] (0.8,0) -- (0.8,1);
      \node [right] at (1.0,0.5) {$z_{i, i \neq 1,2}$};

      \draw [fill] (0,2.5) circle [radius=0.05];
      \draw (0,2.5)--(1.5,3.5);
      \node [right] at (1.5,3.5) {$ z_1$};
      \draw (0,2.5)--(1.5,1.5);
      \node [right] at (1.5,1.5) {$ z_2$};

      \node [above] at (0,4) {$ \eps^0$};
      \node [above] at (1.5,4) {$ \eps^1$};
      \end{tikzpicture}
      \caption{Singular-I}
  \end{minipage}%
  \begin{minipage}[thtbp]{0.4\linewidth}
  \centering

  \begin{tikzpicture}
      \draw (-0.5,0)--(0.5,0);
      \draw [fill] (0,1) circle [radius=0.05];
      \node [] at (0,0.6) {$ \vdots $};
      \draw [fill] (0,0) circle [radius=0.05];
      \draw (-0.5,1)--(0.5,1);
      \draw [decorate,decoration={calligraphic brace,mirror,amplitude=2mm},thick] (0.8,0) -- (0.8,1);
      \node [right] at (1.0,0.5) {$z_{i, i \neq 1,2}$};

      \draw [fill] (0,2.5) circle [radius=0.05];
      \draw (0,2.5)--(1.5,2.5);
      \draw [fill] (1.5,2.5) circle [radius=0.05];
      \draw (1.5,2.5)--(3,3.5);
      \node [right] at (3,3.5) {$ z_1$};
      \draw (1.5,2.5)--(3,1.5);
      \node [right] at (3,1.5) {$ z_2$};

      \node [above] at (0,4) {$ \eps^0$};
      \node [above] at (1.5,4) {$ \eps^1$};
      \node [above] at (3,4) {$ \eps^2$};
    \end{tikzpicture}
    \caption{Singular-II}
  \end{minipage}%
  \label{eq:forward limit BH}
\end{figure}
Now let us consider the contributions of singular solutions to amplitudes. We chose $(rst)=(123)$ and $(ijk)=(n-2,n-1,n)$. For the measure part,
\bea
& & (z_1-z_2)(z_2-z_3)(z_3-z_1)(z_n-z_{n-1})(z_{n-1}-z_{n-2})(z_{n-2}-z_n)\sim \left\{ \begin{array}{ll}\epsilon^{1},  & {\rm Singuar~I} \\
\epsilon^{2},  & {\rm Singuar~II} \end{array}\right. \nn
& & { |\Phi|^{rst}_{ijk}}\sim \eps^1 \nn
& & J\sim \left\{ \begin{array}{ll}\epsilon^{0},  & {\rm Singuar~I} \\
\epsilon^{1},  & {\rm Singuar~II} \end{array}\right.~~~~\label{measure-BCFW-1}
\eea
With above knowledge we summary the contributions of each type of solutions as follows
\begin{table}[h]
    \centering
    \begin{tabular}{cm{1.1cm}<{\centering}|c|c|c|c|}
      \hline
      \multicolumn{1}{|m{2.3cm}|}{ }& $z_1-z_2$ &measure & $I_{_{CHY}}\! \sim \! \frac{1}{(z_1-z_2)^2}$ & $I_{_{CHY}}\! \sim \! \frac{1}{(z_1-z_2)^1}$ & $I_{_{CHY}} \! \sim \! \frac{1}{(z_1-z_2)^0}$\\
     \hline
      \multicolumn{1}{|m{2.3cm}|}{Singular-I} & $\epsilon$ & 1 & $\epsilon^{-2}$ & $\epsilon^{-1}$  & 1 \\
        \hline
      \multicolumn{1}{|m{2.3cm}|}{Singular-II} & $\epsilon^2$ & $\epsilon$ & $\epsilon^{-4}$ & $\epsilon^{-2}$ & 1\\
        \hline
        \multicolumn{1}{|m{2.3cm}|}{Regular} & 1 & 1 & 1 & 1 & 1 \\
        \hline
        \multicolumn{1}{|m{2.3cm}|}{Contribution of Singular-I} & \  & \  & $\epsilon^{-2}$ & $\epsilon^{-1}$ & 1 \\
        \hline
        \multicolumn{1}{|m{2.3cm}|}{Contribution of Singular-II} & \  & \  & $\epsilon^{-3}$ & $\epsilon^{-1}$ & $\epsilon$ \\
        \hline
        \multicolumn{1}{|m{2.3cm}|}{Contribution of Regular} & 1 & 1 & 1 & 1 & 1 \\
        \hline
    \end{tabular}
    \caption{Contribution of two Singular solutions and regular respectively}
    \label{tab:Summaryofforward}
\end{table}
One interesting point is that the singular I solutions will contribute $I_{CHY}\sim \frac{1}{(z_1-z_2)^0}$ at the order $\eps^0$, just like the regular solutions. Thus we can not naively remove their contributions\footnote{In \cite{Cachazo:2015aol} it has been shown that  although
singular I solutions give nonzero contributions at the integrand level, it can be removed for some theories since the one-loop CHY-integrand is defined up to terms integrated to zero.}

Having the above discussion, we present one numerical example to show the above conclusions. The choice of kinematics has been given in Appendix \ref{kine}. We choose the six-point example since it is the simplest nontrivial case under the forward limit. The numerical kinematics is given by
\begin{equation}
    \begin{aligned}
        k_1\to& \{0.824,-0.421,0.306,0.283,-0.572,0\} \\
        k_2\to & \{1.169,0.705,0.291,0.341,-0.817,0\} \\
        k_3\to &\{-0.677,0.284,0.136
        ,-0.316,+0.509,0\} \\
        +&\{-0.323,-0.387+0.282 i,-0.185-0.591 i,-0.320,0.496,0\}\epsilon\\
        k_4\to &\{-1.315,-0.568,-0.733,-0.309,0.880,0\} \\
        +&\{-0.714,0.417,0.0996,-0.443-1.142 i,1.262\, -0.401 i,0\}\epsilon\\
        k_+\to &\{0.410,-0.158,0.186 +0.0370 i,-0.0794
        +0.0870 i,-0.333,0\} \\
        +&\{0.518,-0.0148-0.141 i,0.043+0.295 i,0.382+0.571 i,-0.879+0.201 i,0.128-0.442 i\}\epsilon\\
        k_-\to &-\{0.410,-0.158,0.186 +0.0370 i,-0.0794
        +0.0870 i,-0.333,0\}\\
        +&\{0.518,-0.0148-0.141 i,0.043+0.295 i,0.382+0.571 i,-0.879+0.201 i,-0.128+0.442 i\}\epsilon
    \end{aligned}
    \label{eq:numer k 003}
\end{equation}
We fix the gauge $z_1=1,z_2=2,z_3=3$. There are two  singular-I  solutions, two singular-II solutions and two regular solutions:
\begin{equation}
        \begin{aligned}
        & Regular (1) \quad \left\{
            \begin{array}{l}
                z_4\to (1.925\, -0.124 i)+(0.0527\, +0.0364 i) \epsilon+(0.330\, -0.043 i) \epsilon^2 \\
                z_+\to (2.079\, -0.567 i)+(0.654\, +0.288 i) \epsilon +(1.460\, +1.530 i) \epsilon^2 \\
                z_-\to (1.549\, +0.129 i)-(0.590\, -0.787 i) \epsilon+(0.296\, +1.238 i) \epsilon^2 \\
            \end{array}
        \right.\\
        & Regular (2) \quad \left\{
            \begin{array}{l}
                z_4\to (1.996\, +0.0401 i)+(0.0904\, +0.0436 i) \epsilon+(0.141\, -0.512 i) \epsilon^2 \\
                z_+\to (2.049\, +0.132 i)+(0.0966\, -0.0383 i) \epsilon+(-0.346-1.292 i) \epsilon^2 \\
                z_-\to (1.590\, -0.286 i)-(0.123\, -2.415 i) \epsilon+(9.723\, -7.520 i) \epsilon^2 \\
            \end{array}
        \right.
        \end{aligned}
        \label{eq:forward regular 01}
    \end{equation}
\begin{equation}
        \begin{aligned}
        & SingularI (1) \quad \left\{
            \begin{array}{l}
                z_4\to 1.911-(0.267\, -0.265 i) \epsilon+(-0.599+1.144 i) \epsilon^2 \\
                z_+\to (1.726\, -0.479 i)+(0.112-1.692 i) \epsilon+(-0.604-4.715 i) \epsilon^2 \\
                z_-\to (1.726 -0.479 i)-(0.112\, -1.692 i) \epsilon+(0.604\, +4.715 i) \epsilon^2 \\
            \end{array}
        \right.\\
        & SingularI (2) \quad \left\{
            \begin{array}{l}
                z_4\to 1.911+(0.152\, +0.0200 i) \epsilon+(0.0931\, -0.514 i) \epsilon^2 \\
                z_+\to (1.785\, +0.232 i)+(0.158\, -0.509 i) \epsilon+(-1.055-0.466 i) \epsilon^2 \\
                z_-\to (1.785\, +0.232 i)-(0.158\, -0.509 i) \epsilon+(1.055\, +0.466 i) \epsilon^2 \\
            \end{array}
        \right.
        \end{aligned}
        \label{eq:forward singular-I 01}
    \end{equation}
    \begin{equation}
        \begin{aligned}
        & SingularII (1) \quad \left\{
            \begin{array}{l}
                z_4\to 1.911-(0.130\, -0.336 i) \epsilon +(0.359\, +1.232 i) \epsilon^2\\
                z_+\to (1.958-0.123 i)+(0.285-0.0967 i)\epsilon-(0.0694\, +0.129 i) \epsilon^2 \\
                z_-\to (1.958\, -0.123 i)+(0.285-0.0967 i)\epsilon +(0.0694+0.129 i) \epsilon^2\\
            \end{array}
        \right.\\
        & SingularII (2) \quad \left\{
            \begin{array}{l}
                z_4\to 1.911+(0.0151\, -0.0503 i) \epsilon+(0.231\, -0.189 i) \epsilon^2 \\
                z_+\to (2.064\, +0.345 i)+(0.0666+1.0254 i)\epsilon-(0.223\, -0.747 i) \epsilon^2 \\
                z_-\to (2.064\, +0.345 i)+(0.0666+1.0254 i)\epsilon+(0.223 -0.747 i) \epsilon^2 \\
            \end{array}
        \right.
        \end{aligned}
        \label{eq:forward singular-II 01}
    \end{equation}
The measure part of each solution is in table \eqref{tab:jacobi001}.

\begin{table}[htbp]
    \centering
    \begin{tabular}{|c|c|}
    \hline
    Solution   &   Measure  \\
    \hline
    Regular (1)     &  $0.201709+0.315957 i+O(\epsilon)$   \\
    Regular (2)     &  $0.00828429+0.000758566 i+O(\epsilon)$   \\
    SingularI (1)   &  $0.09663-0.37947 i +O(\epsilon)$  \\
    SingularI (2)   &   $-0.0222706-0.0106388 i+O(\epsilon)$  \\
    SingularII (1)   &  $(-0.000290196+0.0000925484i)\epsilon+O(\epsilon^2)$  \\
    SingularII (2)   &   $(0.606991+0.0403871 i)\epsilon+O(\epsilon^2)$  \\

    \hline
    \end{tabular}
    \caption{The measure part of each solution in Forward limit}
    \label{tab:jacobi001}
\end{table}
which agree with our computation (the third column of table \eqref{tab:Summaryofforward}).
Next we choose three different CHY-Integrands with pole index $\chi[S_{+-}]=0,-1,-2$ respectively to show the behaviour of table\eqref{tab:Summaryofforward}:
\begin{itemize}
    \item $I_1=PT(1234+-)PT(1342-+)$ with $\chi[S_{+-}]=0$:
    \begin{table}[htbp]
    \centering
    \begin{tabular}{|c|c|}
    \hline
    Solution   &   Integration \\
    \hline
    Regular (1)     &  $-(9.3103+1.98347 \dot{i})$   \\
    Regular (2)     &  $(2.81938-6.61778 i)$   \\
    SingularI (1)   &  $-\frac{1}{\epsilon^2}(0.622372+0.426007 i)$  \\
    SingularI (2)   &   $-\frac{1}{\epsilon^2}(0.726396+1.70513 i)$  \\
    SingularII (1)   &  $\frac{1}{\epsilon^3}(0.657355+0.819951 i)$  \\
    SingularII (2)   &   $-\frac{1}{\epsilon^3}(0.862234-0.0145693 i)$  \\
    Summation & $-\frac{1}{\epsilon^3}(0.204879-0.83452 \mathrm{i})$ \\
    Amplitude & $-\frac{1}{S_{34} S_{+-} S_{1+-}}$\\
    Numerical amplitude & $-\frac{1}{\epsilon^3}(0.204879-0.83452 \mathrm{i})$\\

    \hline
    \end{tabular}
    \caption{The contribution of each solution of CHY-Integrand $PT(1234+-)PT(1342-+)$}

    \label{tab:Integrand001}
\end{table}
For this example (see table
\ref{tab:Integrand001}), there are poles $S_{+-}\sim \eps^{-2}$ and $S_{1+-}\sim \eps^{-1}$,
so the leading contribution comes only from Singular II solutions.

\item $I_2=PT(1234+-)PT(1+234-)$ with $\chi[S_{+-}]=-1$: For this example (see table
\ref{tab:Integrand002}), there is  pole $S_{1+-}\sim \eps^{-1}$,
so the leading contribution comes  from both Singular I and Singular II solutions.
\begin{table}[htbp]
    \centering
    \begin{tabular}{|c|c|}
    \hline
    Solution   &   Integration \\
    \hline
    Regular (1)     &  $-(2.00263-7.49773 i)$   \\
    Regular (2)     &  $(3.43278+1.47441 i)$   \\
    SingularI (1)   &  $-\frac{1}{\epsilon}(0.0475165-1.0134i)$  \\
    SingularI (2)   &   $-\frac{1}{\epsilon}(1.25809+1.02357i)$  \\
    SingularII (1)   &  $-\frac{1}{\epsilon}(0.423532+0.0715691i)$  \\
    SingularII (2)   &   $-\frac{1}{\epsilon}(0.524593+0.208376i)$  \\
    Summation & $\frac{1}{\epsilon}(1.20454-0.126638 i)$ \\
    Amplitude & $\frac{1}{S_{1-}S_{23}S_{1+-}}+\frac{1}{S_{1-}S_{34}S_{1+-}}$\\
    Numerical amplitude & $\frac{1}{\epsilon}(1.20454-0.126638 i)$\\
    \hline
    \end{tabular}
    \caption{The contribution of each solution of CHY-Integrand $PT(1234+-)PT(1+234-)$}
    \label{tab:Integrand002}
\end{table}

\item $I_3=PT(123+4-)PT(1+423-)$ with $\chi[S_{+-}]=-2$: For this example (see table
\ref{tab:Integrand003}), there is  no singular pole,
so the leading contribution comes  from regular solutions as well as  Singular I solutions.

\begin{table}[htbp]
    \centering
    \begin{tabular}{|c|c|}
    \hline
    Solution   &   Integration \\
    \hline
    Regular (1)     &  $(36.3435-21.9682 i)$   \\
    Regular (2)     &  $-(49.0302-26.6536 i)$   \\
    SingularI (1)   &  $(19.3907-18.34 i)$  \\
    SingularI (2)   &   $-(18.0838+0.689813i)$  \\
    SingularII (1)   &  $(1.43509+0.475868 i)\epsilon$  \\
    SingularII (2)   &   $-(10.9922+0.404811 i)\epsilon$  \\
    Summation & $-(11.3798+14.3444i)$ \\
    Amplitude & $\frac{1}{S_{1-}S_{23}S_{4+}}$\\
    Numerical amplitude & $-(11.3798+14.3444i)$\\
    \hline
    \end{tabular}
    \caption{The contribution of each solution of CHY-Integrand$PT(123+4-)PT(1+423-)$}
    \label{tab:Integrand003}
\end{table}
\end{itemize}


\subsection{Applications in one-loop CHY-Integrands}

One main motivation of our current study is to understand that under the forward limit, what is the contributions of singular solutions for one-loop CHY-integrands constructed using the forward limit method.  In this subsection, we will use two types of one-loop CHY-integrands of the bi-adjoint scalar theory to provide some general picture, especially the different patterns of cancellation of singularities. 

\subsubsection{The first type of integrands}
In \cite{Cachazo:2015aol} the  one-loop CHY-integrand is given by
\begin{equation}
  m_{n}^{1-\operatorname{loop}}[\pi \mid \rho]=\int \frac{d^{D} l}{(2 \pi)^{D}} \frac{1}{l^{2}} \lim _{k_{\pm} \rightarrow \pm l} \sum_{\alpha \in c y c(\pi) \atop \beta \in c y c(\rho)} m_{n+2}^{\text {tree }}[-\alpha+\mid-\beta+]
  \label{loop-tree}
\end{equation}
where $\alpha$ and $\beta$ are two PT factors and
\begin{equation}
  m_{n+2}^{\text {tree }}[-\alpha+\mid-\beta+]=\int d \Omega_{C H Y} P T(-, \alpha(1), \cdots, \alpha(n),+) P T(-, \beta(1), \cdots, \beta(n),+)
\end{equation}
is the tree-level amplitudes of $(n+2)$ particles defined by the CHY-integrand of two PT factors.

To have a concrete picture, let us consider a simple example, i.e., one loop  integrands $m_{4}^{1-\text { loop }}[1234 \mid 1243]$. From \eref{loop-tree}, this loop integrand is given by the sum of  all possible combinations of PT-factor sets $\{\mathrm{PT}(1+-234), \mathrm{PT}(12+-34), \mathrm{PT}(123+-4), \mathrm{PT}(1234+-)\}$ and $\{\mathrm{PT}(1+-243), \mathrm{PT}(12+-43), \mathrm{PT}(124+-3), \mathrm{PT}(1243+-)\}$. As in the previous section, a set of momentum with forward limit parameter $\eps$ has been taken as following:

\begin{equation}
  \begin{array}{ll}
  k_1 \rightarrow &\{1.04,0.113,-0.156,0.775,0.667,0\}\\
  k_2 \rightarrow &\{1.43,-0.913,-0.771,-0.727,-0.295,0\}\\
  k_3 \rightarrow &\{-0.931,0.383,-0.508,-0.344,-0.587,0\}
  +\\ &\{-0.699 \epsilon ,(0.4+0.304 i) \epsilon ,(-0.529+0.23 i) \epsilon ,0.0745 \epsilon ,-0.434 \epsilon ,0\}\\
  k_4 \rightarrow &\{-1.54,0.417,1.43,0.296,0.215,0\}
  +\\ &\{-1.15 \epsilon ,(0.354-0.641 i) \epsilon ,(1.22+0.186 i) \epsilon ,-0.191 \epsilon ,-0.342 \epsilon ,0\}\\
  k_+ \rightarrow &\{0.45,-1.02-0.34 i,-0.548,0.287-1.21 i,-0.597,0\}
  +\\ &\{0.923 \epsilon ,(-0.377+0.168 i) \epsilon ,(-0.344-0.208 i) \epsilon ,0.0584 \epsilon ,0.388 \epsilon ,(0.714-0.0115 i) \epsilon \}\\
  k_- \rightarrow &\{-0.45,1.02+0.34 i,0.548,-0.287+1.21 i,0.597,0\}
  +\\ &\{0.923 \epsilon ,(-0.377+0.168 i) \epsilon ,(-0.344-0.208 i) \epsilon ,0.0584 \epsilon ,0.388 \epsilon ,(-0.714+0.0115 i) \epsilon \}\\
  \end{array}
\end{equation}

For current example, there are  $(n-3)!=6$  solutions. Under the forward limit, they are divided into three classes according to the asymptotic behaviour, i.e.,  regular solutions, singular-I and singular-II solutions, and each of them contains two solutions:
\begin{align*}
  & \text{Regular } \text{ (1)} \quad & &\left\{
  \begin{array}{l}
  z_4\to (1.99+0.074 i)-(0.21-0.0669 i) \epsilon+(0.0316-0.313 i) \epsilon^2 \\
  z_+\to (2.3+0.135 i)-(0.283-0.543 i) \epsilon-(1.36+1.09 i) \epsilon^2 \\
  z_-\to (2.01+0.175 i)-(0.0946-0.132 i) \epsilon+(0.0981-0.0337 i) \epsilon^2 \\
  \end{array}
  \right.\\
  & \text{Regular }\text{ (2)} \quad & &\left\{
  \begin{array}{l}
  z_4\to (2.04-0.483 i)-(0.496+0.172 i) \epsilon-(0.182-1.33 i) \epsilon^2 \\
  z_+\to (1.78-0.594 i)-(0.234-0.198 i) \epsilon+(0.678+1.02 i) \epsilon^2 \\
  z_-\to (2.52+0.273 i)+(0.516+0.15 i) \epsilon+(0.354-0.104 i) \epsilon^2 \\
  \end{array}
  \right.
\end{align*}

\begin{align*}
  & \text{SingularI } \text{ (1)} \quad & &\left\{
  \begin{array}{l}
  z_4\to 1.89-(0.0953+0.0598 i) \epsilon-(0.469+0.562 i) \epsilon^2 \\
  z_+\to (1.82+0.0335 i)+(0.00702-0.103 i) \epsilon+(0.207-0.0868 i) \epsilon^2 \\
  z_-\to (1.82+0.0335 i)-(0.00702-0.103 i) \epsilon-(0.207-0.0868 i) \epsilon^2 \\
  \end{array}
  \right.\\
  & \text{SingularI } \text{ (2)} \quad & &\left\{
  \begin{array}{l}
  z_4\to 1.89-(0.0845+0.0676 i) \epsilon-(0.0545+0.0212 i) \epsilon^2 \\
  z_+\to (2.39+0.184 i)-(0.344-0.0399 i) \epsilon+(0.176-0.0308 i) \epsilon^2 \\
  z_-\to (2.39+0.184 i)+(0.344-0.0399 i) \epsilon-(0.176-0.0308 i) \epsilon^2 \\
  \end{array}
  \right.
\end{align*}

\begin{align*}
  & \text{SingularII } \text{ (1)} \quad & &\left\{
  \begin{array}{l}
  z_4\to 1.89-(0.0923+0.236 i) \epsilon+(0.317+0.0928 i) \epsilon^2 \\
  z_+\to (1.89-0.376 i)+(0.103+0.191 i) \epsilon-(6.61-1.44 i) \epsilon^2 \\
  z_-\to (1.89-0.376 i)+(0.103+0.191 i) \epsilon-(6.37-1.27 i) \epsilon^2 \\
  \end{array}
  \right.\\
  & \text{SingularII } \text{ (2)} \quad & &\left\{
  \begin{array}{l}
  z_4\to 1.89-(0.0874-0.109 i) \epsilon-(0.00897-0.265 i) \epsilon^2 \\
  z_+\to (1.97+0.196 i)+(0.0645+0.0918 i) \epsilon-(2.57-1.42 i) \epsilon^2 \\
  z_-\to (1.97+0.196 i)+(0.0645+0.0918 i) \epsilon-(2.64-1.18 i) \epsilon^2 \\
  \end{array}
  \right.
\end{align*}

For the numerical evaluation, we will take two different approaches. In the first approach, for each term in the $16$ combinations of CHY-integrand \eref{loop-tree}, we sum over $6$ solutions first, and the result is given in the table \eref{tab:he1tree6}.
 
In the table \eref{tab:he1tree6}, although most of the terms will have leading divergence ${1\over \eps^3}$, when summing over $16$ terms, all divergences will cancel each other, and we are left with finite results under the forward limit as expected.

\begin{table}[htbp]
\begin{tabular}{|c|c|c|}
  \hline Integrand &  Asymptotic level of integrand & Leading Order \\
  \hline $PT(1+-234)PT(1+-243)$ & $\frac{1}{z_{+-}^2}$ & $\frac{0.074+0.006 i}{ \epsilon^3}$ \\
  \hline $PT(1+-234)PT(12+-43)$ & $\frac{1}{z_{+-}^2}$ & $-\frac{0.032+0.0013 i}{ \epsilon^3}$ \\
  \hline $PT(1+-234)PT(124+-3)$ & $\frac{1}{z_{+-}^2}$ & 0 \\
  \hline $PT(1+-234)PT(1243+-)$ & $\frac{1}{z_{+-}^2}$ & $-\frac{0.042+0.0047 i}{ \epsilon^3}$ \\
  \hline $PT(12+-34)PT(1+-243)$ & $\frac{1}{z_{+-}^2}$ & $-\frac{0.032+0.0013 i}{ \epsilon^3}$ \\
  \hline $PT(12+-34)PT(12+-43)$ & $\frac{1}{z_{+-}^2}$ & $\frac{0.032+0.0013 i}{ \epsilon^3}$ \\
  \hline $PT(12+-34)PT(124+-3)$ & $\frac{1}{z_{+-}^2}$ & $-\frac{0.04-0.0093 i}{ \epsilon^3}$ \\
  \hline $PT(12+-34)PT(1243+-)$ & $\frac{1}{z_{+-}^2}$ & $\frac{0.04-0.0093 i}{ \epsilon^3}$ \\
  \hline $PT(123+-4)PT(1+-243)$ & $\frac{1}{z_{+-}^2}$ & 0 \\
  \hline $PT(123+-4)PT(12+-43)$ & $\frac{1}{z_{+-}^2}$ & $-\frac{0.029+0.0087 i}{ \epsilon^3}$ \\
  \hline $PT(123+-4)PT(124+-3)$ & $\frac{1}{z_{+-}^2}$ & $\frac{-0.072-0.00057 i}{ \epsilon^3}$ \\
  \hline $PT(123+-4)PT(1243+-)$ & $\frac{1}{z_{+-}^2}$ & $-\frac{0.04-0.0093 i}{ \epsilon^3}$ \\
  \hline $PT(1234+-)PT(1+-243)$ & $\frac{1}{z_{+-}^2}$ & $-\frac{0.042+0.0047 i}{ \epsilon^3}$ \\
  \hline $PT(1234+-)PT(12+-43)$ & $\frac{1}{z_{+-}^2}$ & $\frac{0.029+0.0087 i}{ \epsilon^3}$ \\
  \hline $PT(1234+-)PT(124+-3)$ & $\frac{1}{z_{+-}^2}$ & $-\frac{0.029+0.0087 i}{ \epsilon^3}$ \\
  \hline $PT(1234+-)PT(1243+-)$ & $\frac{1}{z_{+-}^2}$ & $\frac{0.042+0.0047 i}{ \epsilon^3}$ \\
  \hline

\end{tabular}
\caption{Leading order of the tree integrand}
\label{tab:he1tree6}
\end{table}

Although the divergence is not shown at the final, how the cancellation happens is not clear. To better understand, we take the second approach: for each solution, we put it to all $16$ terms and sum them up first, and then sum up six solutions. One very interesting observation is that the divergence will cancel each other after summing $16$ terms for each solution as shown in table \eref{Leading-Order-he16}. We could also verify this matter by summing $16$ terms analytically, which  is given by:
\bea
  N &= &(z_{1+} z_{2-} (z_{13} z_{2+} z_{3-} z_{4+} z_{4-}+z_{1-} z_{3+} (z_{2+}
   z_{34} z_{4-}-z_{24} z_{3-} z_{4+}))-z_{12} z_{1-} z_{2+} z_{3+} z_{3-} z_{4+}
   z_{4-})\times \nn & &  (z_{12} z_{1-} z_{2+} z_{3+} z_{3-} z_{4+} z_{4-}+z_{1+} z_{2-}
   (z_{1-} z_{4+} (z_{2+} z_{34} z_{3-}+z_{23} z_{3+} z_{4-})-z_{14} z_{2+}
   z_{3+} z_{3-} z_{4-})) \nn
   I & = & \frac{N}
   {z_{12}^2 z_{13} z_{14} z_{1+}^2 z_{1-}^2 z_{23} z_{24}
   z_{2+}^2 z_{2-}^2 z_{34}^2 z_{3+}^2 z_{3-}^2 z_{4+}^2 z_{4-}^2 z_{+-}^2}
\eea
As one can check, for each solution, the $N$ will give $\eps^a$ factor to cancel the singular from 
the denominator and the measure. 

\begin{table}[htbp]
  \centering
  \begin{tabular}{|c|c|}
  \hline
  Solution   &   Leading Order \\
  \hline
  Regular (1) & $(0.0738-0.128 i)+O\left(  \epsilon\right)$   \\
  \hline
  Regular (2) & $-(0.126+0.00619 i)+O\left(  \epsilon\right)$   \\
  \hline
  SingularI (1) & $(0.0277+0.2 i)+O\left(  \epsilon\right)$   \\
  \hline
  SingularI (2) & $-(0.055-0.214 i)+O\left(  \epsilon\right)$   \\
  \hline
  SingularII (1) & $(0.00452+0.00782 i)  \epsilon+ O\left(  \epsilon^2\right)$   \\
  \hline
  SingularII (2) & $(0.0145+0.0407 i)  \epsilon+ O\left(  \epsilon^2\right)$   \\
  \hline
  Total  & $-(0.079-0.28i) + O\left(  \epsilon\right)$   \\
  \hline

  \end{tabular}
  \caption{Leading order of the loop integrand}
  \label{Leading-Order-he16}
\end{table}

The cancelation pattern observed above for the integrand \eref{loop-tree} is not a coincident.
In fact, in the construction of the CHY-integrand, the tadpole contribution under the 
forward limit is manifest avoid,
while the massless bubble must be canceld by different terms in the combinations.  

\subsubsection{The second type of integrands}

A different idea to remove the tadpole and massless bubble singularities has been used in  \cite{Feng:2019xiq}. Based on this idea, different one-loop CHY-integrand has been constructed by multiplying each term with the proper pole-picking operators. For the example $m_{4}^{1-\text { loop }}[1234 \mid 1243]$, there are only three terms as given in the table \eref{Leading-Order-eachhu3}. By the construction,  for each term, the cancellation of divergence comes from the summing over six solutions as shown in the second column of the table \eref{Leading-Order-eachhu3}. More accurately, the cancellation happens when and only when summing over singular-I and singular-II solutions. Just summing over singular-I solutions or singular II solutions, the divergence will not be cancelled.

\begin{table}[htbp]
  \centering
  \begin{tabular}{|c|c|}
  \hline Integrand  & Leading order \\

  \hline $(1-\frac{z_{14}^2 z_{2-} z_{-+}}{z_{12} z_{1-} z_{4-}
  z_{4+}}-\frac{z_{23}^2 z_{1+} z_{-+}}{z_{12} z_{3-}
  z_{2+} z_{3+}}-\frac{z_{2-} z_{1+}}{z_{1-} z_{2+}})PT(1+-234)PT(1+-243)$ & $0 .015+0.03 i$ \\
  \hline $(1-\frac{z_{12}^2 z_{3-} z_{-+}}{z_{23} z_{1-} z_{2-}
  z_{1+}}-\frac{z_{3-} z_{2+}}{z_{2-} z_{3+}})PT(12+-34)PT(12+-43)$  & $0.00063+0.0163 i$ \\
  \hline $(1-\frac{z_{12}^2 z_{4+} z_{-+}}{z_{14} z_{2-} z_{1+}
  z_{2+}}-\frac{z_{1-} z_{4+}}{z_{4-} z_{1+}})PT(1234+-)PT(1243+-)$  & $0.00087+0.0173 i$ \\

  \hline
\end{tabular}
\caption{Leading order of the tree integrand}
\label{Leading-Order-eachhu3}
\end{table}

\section{Conclusion}

Motivated by our curiosity of the singular solutions and potential application to other frontiers of researches, for example, the construction of two-loop CHY-integrands by double forward limits, in this paper, we have initiated the systematic study of the relation between the singular kinematic configurations and the singular solutions of scattering equations. We find that the singular solutions will always lead to singular kinematics. Furthermore, the layer structure of singular solutions gives a clear picture of the structure of singular kinematics. However, the reverse is not always transparent, as shown by the soft limit and forward limit. From these examples, we guess the compatibility of various singular poles can characterize the singular kinematics. Although we can not give rigorous proof, from the examples discussed in this paper, it seems that compatible singular poles will lead to singular solutions, but non-compatible will not give singular solutions.

Although we have discussed several types of singularities in this paper, it is obvious that there are many other types of singularities. One-by-one analysis can be done using the method presented in the paper. However, it is more desirable to have a systematical algorithm so that the relation between the singular kinematic configurations and the singular solutions of scattering equations can be well established.

\section*{Acknowledgments}

We would like to thank Ye Yuan and Song He for very useful discussion. This work is supported by Qiu-Shi Funding and Chinese NSF
funding under Grant  No.11935013, No.11947301, No.12047502 (Peng Huanwu Center).

\appendix

\section{Kinematics~~~\label{kine}}

In this note, we focus on solutions of scattering equations defined by \eref{SE-def} or the equivalent polynomial equations given in  \cite{Dolan:2014ega}
\bea 0= h_m\equiv \sum_{S\in A, |S|=m} k_S^2 z_S~,~~~~2\leq m\leq
n-2~,~~~~ \label{DG-2} \eea
Where the sum is over all ${n!\over (n-m)! m!}$ subsets $S$ of
$A=\{1,2,...,n\}$ with exactly $m$ elements and $k_S=\sum_{b\in S}
k_b$ and $z_S=\prod_{b\in S} z_b$. It is worth pointing out that there are
exactly $(n-3)$ independent equations of the form \eref{DG-2}, since the cases of $m=1,n-1,n$ are trivially
true by momentum conservation plus the on-shell conditions. For general kinematics, it is
impossible to write down analytic results for $n\geq 6$, thus to study the properties of solutions,
we need to rely on the numerical method. One important step of the numerical calculation is the parametrization of kinematics, especially when we try to study different limit processes, such as the factorization limit, the forward limit and the soft limit, etc.

There are two ways to define a kinematic configuration. The first way is to use the Lorentz invariant combination $k_i\cdot k_j$. With momentum conservation and null conditions, there
are ${n(n-3)\over 2}$ independent contractions\footnote{To get this counting, we have neglected the constraints coming from the dimension of space-time.}. To see it, using momentum conservation, we can eliminate $p_n$. For remaining $(n-1)$ momenta, there are $\binom{n-1}{2}={(n-1)(n-2)\over 2}$ contractions. A further constraint comes from
\bea 0=k_n^2=(\sum_{i=1}^n k_i)^2=\sum_{i<j} 2k_i\cdot
k_j~~~\label{Eden-2-1}\eea
thus we get ${(n-1)(n-2)\over 2}-1={n(n-3)\over 2}$ contractions. Let us apply this method to
the kinematic configuration of forward limit of $(n+2)$ legs defined by
\bea L_++L_-=2 \epsilon q,~~~~L_+\cdot L_-=2 \epsilon^2 q^2,~~~q^2\neq 0 ~~\label{My-set-2-1}\eea
where  $t=0$ is the forward limit. For general $t\neq 0$, we take following data to parameterize kinematics:
\bea q^2;~~~ P_i\cdot P_j,~1\leq i<j\leq n-1;~~~L_+\cdot P_i,q\cdot P_i,~~~i=1,...,n-1~~~\label{F-para}\eea
There are ${(n-1)(n-2)\over 2}+1+ 2(n-1)={(n+2)(n-1)\over
2}+1$ combinations, thus they are not independent. One extra constraint
comes from the on-shell condition $P_n^2=0$, i.e.,
\bea 0 & = & {1\over 2}(\sum_{i=1}^{n-1} P_i+ 2\epsilon q)^2=  \sum_{1\leq i<j\leq n-1} P_i\cdot P_j+ 2\epsilon q\cdot
\sum_{i=1}^{n-1} P_i +2 \epsilon^2 q^2~~\label{My-set-2-7}\eea
Using above data, other Lorentz contractions are given by
\bea L_-\cdot P_i &= & -L_+\cdot P_i+ 2 \epsilon q\cdot
P_i;~~~~~~ P_i\cdot P_n  =  -2\epsilon q\cdot P_i-\sum_{j=1,j\neq i}^{n-1}
P_i\cdot P_j\nn
L_+\cdot P_n&= &-2 \epsilon^2 q^2-\sum_{j=1}^{n-1} L_+\cdot
P_j;~~~~
L_-\cdot P_n= -2 \epsilon^2 q^2+\sum_{j=1}^{n-1} (L_+\cdot P_j-2
\epsilon q\cdot P_j)~~\label{My-set-2-5}\eea
By momentum conservation. With the above choice, we need to determine how data in \eref{F-para}
depend on $\epsilon$ to describe the limit process. This is not a trivial task. For example,
a naive choice is to take $q^2$, $L_+\cdot P_i$, $q\cdot
P_i$ as well as all $P_i\cdot P_j$ except one by constraint \eref{My-set-2-7}
as invariant data under the limit. This naive choice will have the maximum number of combinations, which are independent of $\epsilon$. However, this choice is wrong because now we will have  $L_+$ to be invariant under the forward limit, thus
\bea L_-^2= (-L_++2\epsilon q)^2= L_+^2- 4\epsilon L_+\cdot q+4 \epsilon^2 q^2=- 4\epsilon L_+\cdot q+4 \epsilon^2 q^2 \eea
can not be zero for all value of $\epsilon$. The lesson from this example is the following.
Although taking Lorentz invariant combinations
to be the independent input data for the kinematics looks simple, it is not easy to
determine how they depend on the parameter $\epsilon$, which prescribes the limit process.

To avoid the above-mentioned subtlety, we take the second parametrization method, i.e., parametrising directly in the momentum component form. The first step to set up a general kinematic configuration is that
there is no subset of $A$ such that $S_A=0$. Starting from this choice, we can make various
limit procedure:
\begin{itemize}

\item {\bf The factorization limit:} For this case, we consider several situations. The first situation is there is one and only one $S_A\to 0$. To reach this goal, a simple way is to use
BCFW-deformation, i.e., while keeping all other $p_i$'s invariant, we deform
\bea p_1(z)=p_1+z q,~~~p_n(z)=p_n-zq,~~~q^2=q\cdot p_1=q\cdot
p_n=0~~~\label{One-deform-1-1}\eea
For any subset $1\in A, n\not \in A$, it is easy to know that when
\bea z_A=-{P_A^2\over 2P_A\cdot q}\Longrightarrow S_A(z_A)=0~~~\label{One-deform-1-2}\eea
Now we parameterize the kinematic as
\bea p_1(\eps)\equiv p_1+z_A q+ \eps q,~~~p_n(\eps)=p_n-z_Aq-\eps q,~~~q^2=q\cdot p_1=q\cdot
p_n=0~~~\label{One-deform-1-4}\eea
so when $\eps\to 0$  the factorization limit has been reached. Similar idea can be used to putting several $S_A\to 0$ at the same time. For example, using two different pairs of BCFW deformations
\bea p_1(w_1)=p_1+w_1 q_{1n},~~~p_n(w_1)=p_n-w_1 q_{1n},~~~q_{1n}^2=q_{1n}\cdot p_1=q_{1n}\cdot
p_n=0~~~\label{One-deform-2-1}\eea
\bea p_2(w_2)=p_2+w_2 q_{23},~~~p_3(w_2)=p_3-w_2 q_{23},~~~q_{23}^2=q_{23}\cdot p_1=q_{23}\cdot
p_3=0~~~\label{One-deform-2-2}\eea
we can set two poles $S_A,S_B\to 0$ with $1\in A, n\not\in A$ and $2\in B, 3\not\in B$ at the same time. We want to emphasise that the poles $S_A, S_B$ can be compatible or not compatible. Different patterns will give different behaviours of singular solutions. A special variation of two BCFW-deformations is that they share one common node, for example, $p_1$, i.e.,
\bea & & p_1(w_1)=p_1+w_1 q_{1n}+w_2 q_{12},~~~p_n(w_1)=p_n-w_1q_{1n},~~~p_2(w_2)=p_2-w_2 q_{12},~~~ \nn & & q_{1n}^2=q_{1n}\cdot p_1=q_{1n}\cdot
p_n=0,~~~~q_{12}^2=q_{12}\cdot p_1=q_{12}\cdot
p_2=0~~q_{1n}\cdot q_{12}=0,~~~\label{One-deform-3-1}\eea

\item {\bf The forward limit:} For the pair of the forward limit, we take them as
\bea L_+=(L+\eps q,\mu),~~~~L_-=(-L+\eps q,-\mu)~~~~q^2,L^2\neq 0,~~~\label{My-set-1-1}\eea
where we have lifted $L_{\pm}$ to be null in higher dimension. To make sure the null condition
for any value of $\eps$, we need to impose conditions
\bea  q\cdot L=0,~~~~-\mu^2= L^2+\epsilon^2 q^2~~\label{My-set-1-2}\eea
For other $n$ momenta, we find $n$ pairs of $p_i, q_i$, $i=1,...,n$ such that
\bea p_i^2=q_i^2=p_i\cdot q_i=0,~~~i=1,...,n;~~~\sum_{i=1}^n p_i=0,~~\sum_{i=1}^n q_i=2 q \eea
and use them to define
\bea P_i=p_i-\eps q_i~~~~~~\label{My-set-1-4}\eea
Above choice of the forward limit is very general. We can make a simple choice with the minimum numbers of $P_i$ depending on $\eps$. The procedure is following:
\begin{itemize}

\item (a1) First  setting $q_j=0,j=1,...,n-2$, so $P_j=p_j$.

\item (a2) Taking a null momentum $k_{n-1}$, we construct
    \bea p_{n-1}= -{K^2\over 2K\cdot k_{n-1}} k_{n-1},~~~p_n= -K+{K^2\over 2K\cdot k_{n-1}} k_{n-1},~~~~K=\sum_{j=1}^{n-2} p_j~~~~~\label{My-set-3-3}\eea

\item (a3) Taking null momenta $q_{n-1},q_n$ such that $p_a\cdot q_a=0,a=n-1,n$ and $q_n\cdot q_{n-1}\neq 0$ and choosing $b_{n-1}, \b_{n}$ such that $L\cdot (\b_{n-1} q_{n-1}+\b_n q_{n})=0$, we can write down
\bea
     P_{n-1} & = & p_{n-1}-2\eps \b_{n-1} q_{n-1},~~~
     P_n = p_n-2\eps \b_n q_n,~~~q=\b_{n-1} q_{n-1}+\b_n q_{n},~~~~~~\label{My-set-3-5}\eea

\end{itemize}
Above construction \eref{My-set-3-3} and \eref{My-set-3-5} have two exceptions. The
first case is $K=0$, i.e., the case $n=2$. For this one, we can trivially take
 $p_n=-p_{n-1}$ in \eref{My-set-3-3}. The second case is that $K\neq 0$, but
$K^2=0$, which will happen when $n=3$  (for  general momentum configuration, it could not be true for $n\geq 4$). For this one, we can take
\bea P_1& = & - p_2 - p_3,~~~P_2=p_2-2 \epsilon q_2,~~~~P_3=p_3-2\epsilon q_3,
\nn
L_+ & = & (L+\epsilon q,\mu),~~~L_-=(-L+\epsilon q,
-\mu),~~~q=q_2+q_3,~~~\mu^2=L^2+\epsilon^2 q^2~~\label{My-set-3-6}\eea
%
with conditions
\bea & & p_i^2=q_i^2=0=p_3\cdot p_2=L\cdot q,  ~~~i=2,3
\nn
& & q_2\cdot p_2=0=q_3\cdot p_3,~~~q_2\cdot q_3,q_2\cdot p_3, q_3\cdot
p_2\neq 0~~\label{My-set-3-7}\eea

\item {\bf The soft limit:} The kinematic parametrization is following \cite{Cachazo:2014fwa}. Assuming the $p_s$ is the
soft particle, we take two arbitrary momenta, for example, $a,n$ and make the shifting
\bea \W\la_a(\eps)=  \W\la_a+ (1-\eps){\Spaa{n|s}\over \Spaa{n|a}}\W\la_s,~~~\W\la_n(\eps)=
\W\la_n+ (1-\eps){\Spaa{a|s}\over \Spaa{a|n}}\W\la_s,~~~\W\la_s(\eps)=\eps \W\la_s~~~~\label{soft-shift}\eea
It is easy to see that
\bea \la_a \W \la_a+ \la_n \W \la_n+\la_s \W \la_s=\la_a \W\la_a(\eps)+\la_n \W\la_n(\eps)+
\la_s \W\la_s(\eps)~~~~\label{soft-shift-1} \eea
and $\eps\to 0$ gives the soft limit. Although we have used the spinor formalism special for $D=4$, for general $D$-dimension, we can make the Lorenz transformation to put $p_n, p_a, p_s$ to the $4$-dimensional subspace, thus above construction is general.

\end{itemize}
%


\end{document}